\documentclass[utf8]{frontiersSCNS} 

\usepackage{url,hyperref,lineno,microtype,subcaption}
\usepackage[onehalfspacing]{setspace}
\usepackage[british,UKenglish,USenglish,english,american]{babel}

\def\keyFont{\fontsize{8}{11}\helveticabold }
\def\firstAuthorLast{Buldgen {et~al.}} 
\def\Authors{Gaël Buldgen\,$^{1,2}$, Sébastien Salmon\,$^{3}$ and Arlette Noels\,$^{3}$}
\def\Address{$^{1}$School of Physics and Astronomy, University of Birmingham, Edgbaston, Birmingham B15 2TT, UK.\\
$^{2}$Observatoire de Gen\`eve, Universit\'e de Gen\`eve, 51 Ch. Des Maillettes, CH-1290 Sauverny, Suisse.\\
$^{3}$STAR Institute, Universit\'e de Li\`ege, All\'ee du Six Ao\^ut 19C, B-4000 Li\`ege, Belgium.}
\def\corrAuthor{Gaël Buldgen}

\def\corrEmail{Gael.Buldgen@unige.ch}

\begin{document}
\onecolumn
\firstpage{1}

\title[Progress in global helioseismology]{Progress in global helioseismology: a new light on the solar modelling problem and its implications for solar-like stars} 

\author[\firstAuthorLast ]{\Authors} 
\address{} 
\correspondance{} 

\extraAuth{}

\maketitle

\begin{abstract}
Since the first observations of solar oscillations in 1960, helioseismology has probably been one of the most successful fields of astrophysics. Data of unprecedented quality were obtained through the implementation of networks of ground-based observatories such as the GONG project or the BiSON network, coupled with space-based telescopes such as SOHO and SDO missions and more data is expected from the Solar Orbiter mission. Besides the improvement of observational data, solar seismologists developed sophisticated techniques to infer the internal structure of the Sun from its eigenfrequencies. These methods, then already extensively used in the field of Geophysics, are called inversion techniques. They allowed to precisely determine the position of the solar convective envelope, the helium abundance in this region and the internal radial profiles of given thermodynamic quantities. Back in 1990s these comparisons showed a very high agreement between solar models and the Sun. However, the downward revision of the CNO surface abundances in the Sun in 2005, confirmed in 2009, induced a drastic reduction of this agreement leading to the so-called solar modelling problem. More than ten years later, in the era of the space-based photometry missions which have established asteroseismology of solar-like stars as a standard approach to obtain their masses, radii and ages, the solar modelling problem still awaits a solution. In this paper, we will present the results of new helioseismic inversions, discuss the current uncertainties of solar models as well as some possible solutions to the solar modelling problem. We will show how helioseismology can help us grasp what is amiss in our solar models. We will also show that, far from being an argument about details of solar models, the solar problem has significant implications for seismology of solar-like stars, on  the main sequence and beyond, impacting asteroseismology as a whole as well as the fields requiring precise and accurate knowledge of stellar masses, radii and ages, such as Galactic archaeology and exoplanetology.

\tiny
 \keyFont{ \section{Keywords:} The Sun, Helioseismology, Asteroseismology, Solar Abundances, Stellar Structure and Evolution, Solar-like Stars} 
\end{abstract}

\section{Introduction}

For the past decades, helioseismology has been a thriving field, enjoying numerous successes and paving the way for asteroseismology of solar-like oscillators. These achievements are a consequence of the very high-quality seismic data obtained thanks to ground-based observation networks \citep{Brookes1978,Harvey1988, Isaak1989} and space based observatories such as the SOHO satellite \citep{Domingo1995}. 

These excellent data enabled the precise determination of the position of the base of the convective envelope \citep{KosovBCZ,JCD91Conv,Basu97BCZ}, the determination of the solar rotation \citep{BrownRota,KosovichevRota,SchouRota,JCD2007,Garcia2007}, density and sound speed profile \citep{Antia94} as well as an estimation of the helium abundance in the convective envelope through the use of sophisticated seismic analysis techniques \citep{Vorontsov91,Antia94,BasuYSun,RichardY}. The importance of helioseismology as a test of fundamental physics was also highlighted with the so-called solar neutrino problem, which was first thought to stem from inaccurate modelling of the solar core but was ultimately solved with the discovery of neutrino oscillations \citep{Fukuda1999,Ahmad2002,Eguchi2003} and its impact for solar models \citep{TC1988,Elsworth1990,TC2005, BahcallNeutrino,TC2011, Haxton2013}. These successes led to the elaboration of stellar models well-suited for helioseismic studies and valided by seismic inversions \citep[see for example][]{MODELS} and served as a validation of the depiction of the solar structure and evolution to an excellent degree of accuracy. 

However, the downward revision by about $30\%$ of the abundances of carbon, nitrogen and oxygen determined in \citet{AGS04O}, \citet{AGS05C} revealed a new problem for standard models, the so-called solar metallicity, or solar modelling problem. This revision stems from the use of $3$D atmospheric models instead of outdated $1$D empirical models, from the inclusion of NLTE effects and from a careful selection of spectral lines. As C, N and O are key contributors to the opacity in solar conditions, standard solar models built using the revised abundances from \citet{AGS05} strongly disagreed with helioseismology \citep[e.g.][]{TC2004, Guzik2008}. Further studies were performed in $2009$, $2011$ and $2015$ \citep{AGSS09,Caffau,Scott2015I, Scott2015II, Grevesse2015}, showing that the $3$D models agreed with each other and that the remaining differences were due to line selection effects. Recent re-investigations using spectroscopy further confirmed the results of $2009$ and recent helioseismic determinations of the solar metallicity also agreed with a rather ``low'' value \citep{Vorontsov13,BuldgenZ}.

Quickly, it became clear that the solution to the solar metallicity problem was not purely a question of chemical abundances but could also be linked to other ingredients of the models. Investigations on various possible modifications to the solar models were rapidly performed following the publication of the revised abundances \citep[see e.g.][]{Bahcall2005a,Bahcall2005b,Guzik05,Bahcall05,Bahcall2006, Delahaye2006, Guzik06, Montalban06,SerenelliComp,Basu08,Delahaye2009}. These studies showed that a higher opacity could help solving the current discrepancies between solar models and helioseismology. This hypothesis gained some credence with the first experimental measurements of iron opacity in conditions close to those of the base of the solar convective envelope, showing strong disagreement with theoretical opacity computations \citep{Bailey}.

At the same time, new theoretical opacity computations became available for solar and stellar modelling \citep{Colgan,Mondet,LePennec}. Various groups showed that the modifications stemming from these recomputations were insufficient to solve the solar modelling problem and could even lead to larger discrepancies than before. Recently, \citet{Pradhan} and \citet{Zhao} presented new computations of iron opacity showing an increase compatible with experimental measurements. Yet, an opacity increase for a specific element will not necessarily be sufficient to solve the solar problem \citep{Iglesias2017} and other sources of uncertainties are present in the models. 

Ultimately, the solar problem encompasses a wide variety of uncertain physical processes and key ingredients whose impacts are often neglected in standard stellar modelling. In this study, we will list some of the key contributors to the solar issue in section \ref{Sec:solarproblem} and illustrate their impact on helioseismic constraints in section \ref{Sec:Inversions}. We discuss the usual suspects of the micro- and macrophysics of the solar models but also briefly present some non-standard scenarios including accretion of material during the early stages of solar evolution as well as the hypothesis of an initial solar mass higher than the currently measured value. In addition, the solar modelling problem, despite its very specific nature, cannot be easily overlooked by stellar modellers. In the current era of high-quality asteroseismic data, stellar modellers are asked to provide very precise and accurate fundamental parameters for solar-like stars. This race to precision is however meaningless if the accuracy of the stellar models is not ensured. Currently, it is well known that the main limitations of stellar physics are the shortcomings of the theoretical models. Consequently, the solar modelling problem is still a very timely issue, as the recipe applied for the Sun is replicated for most of the solar-like oscillators. To illustrate the revelance of the solar modelling problem in asteroseismology, we briefly discuss in section \ref{Sec:Cygni} the impact of a potential solution to the solar problem on the seismic parameters of the $16$Cyg binary system, one of the most observationally constrained solar-like oscillator, and discuss the potential use of seismic inversions to further constrain this system using asteroseismology. In section \ref{Sec:Discussion}, we discuss some future prospects for solar models and further tracks for improving the physical accuracy of solar and stellar models. This discussion is then followed by a brief conclusion in section \ref{Sec:Conclusion}.

\section{The solar modelling problem and its various contributors}\label{Sec:solarproblem}
While the solar modelling problem has at first been linked to the revision of the solar metallicity, its clear origin is still disputed and could well be the result of multiple small contributions from various micro- and macrophysical ingredients of the solar models. As such, the definition of the standard solar models derived by \citet{Bahcall82} 30 years ago imposes a strict framework which does not take into account all the information we have on the solar structure. 

A standard solar model is a one solar mass model, evolved to the solar age, taking microscopic diffusion into account and reproducing the current photospheric ratio of heavy elements over hydrogen, the current solar luminosity (or effective temperature) and the solar photospheric radius. To fulfill these constraints, the models are built using the initial abundance ratio of the heavy elements to hydrogen and the mixing length parameter for convection as free parameters of a minimization process. With this definition, the mathematical problem of reproducing the Sun is well-posed for a given set of constraints.

While this methodology leads to a simple approach for producing solar models using standard stellar evolution codes, it does not take into account all observational constraints. For example, standard solar models do not reproduce neither the rotation profile inside the Sun, nor the photospheric lithium abundance. Both constraints are well determined and point towards the absence or inaccurate implementation of various transport processes of both angular momentum and chemical elements in the current solar models. 

Moreover, it is also clear that the mixing-length formalism of convection is inherently flawed and leads to an inaccurate depiction of both the upper layer of the solar convective envelope and its lower boundary, where additional chemical mixing is supposed to occur. 

Finally, it should also be pointed out that standard solar model properties are strongly dependent on fundamental physical ingredients such as nuclear reaction rates, radiative opacities, chemical abundances and the equation of state used for the stellar material. 

Consequently, when discussing the inadequacy between standard solar models and helioseismic constraints, various contributors can be listed and could be held responsible for the observed disagreements. In the next sections, we will briefly discuss some of these contributors and their potential impact.  

\subsection{Chemical abundances}\label{Sec:ChemAbund}

The chemical abundances are the first and perhaps most important contributors to the solar modelling problem. The determination of the photospheric abundances of most elements heavier than helium is performed using spectroscopic data. For decades, spectroscopists used $1$D empirical models of the solar atmosphere to determine the solar metallicity \citep[e.g.][]{Holweger1974,Vernazza1976}. These abundance tables are the so-called GN$93$ abundances from \citet{GrevNoels}, which were used in the standard solar models of the $90$s and led to the tremendous successes of helioseismology. They were slightly revised a couple of years later and recompiled in the so-called GS$98$ abundance tables \citep{GreSauv} still used today in helio- and asteroseismology. 

The first solar abundance tables using $3$D atmospheric models were the so-called AGS$05$ abundances \citep{AGS05} which initiated the solar modelling problem. These tables were revised in $2009$ and became the AGSS$09$ abundance tables. Further determinations were made in $2011$ and $2015$ \citep{Caffau,Scott2015I, Scott2015II, Grevesse2015}, one leading to an intermediate value between the GS$98$ and AGSS$09$ and the most recent confirming the results of $2009$. Ultimately the remaining differences are related to the important aspects of line selection and blends \citep{Allende2001}, which can lead to different values for key chemical elements.

In a series of paper \citet{Bahcall2005b,Bahcall2006,SerenelliComp,Vinyoles} discussed comparisons between standard solar models using photospheric and meteoritic abundances for the refractory elements. Comparisons between photospheric and meteoritic values for these elements have shown slight differences. \citet{Vinyoles} suggest that the meteoritic scale could be used as a higher precision substitute to the solar photospheric values. The main argument is that in recent revisions of solar abundances by \citet{Scott2015I,Scott2015II,Grevesse2015}, the differences between photospheric and meteoritic values have been further reduced. However, this approach makes the assumption that the CI chondrites used to infer the meteoritic scale have not undergone any differentiation and represent a realistic sample of mean solar system materials. Recent investigations seem to indicate that this is not the case and that meteoritic abundances cannot be used as such substitutes for solar materials (N. Grevesse, private communication).  

While spectroscopy is the most famous approach to determine the solar metallicity, helioseismology has also been used to derive this key ingredient of solar models. The first of such studies was performed by \citet{Takata2001}, who favoured a low value for the metallicity, in agreement with the results of AGS$05$. However, the precision of these results did not allow them to conclude, as the uncertainties were large enough to agree with all abundance tables. \citet{Antia2006} used a different seismic technique and found an agreement with the GS$98$ abundance tables whereas \citet{Houdek2011} found an intermediate value. Recently, \citet{Vorontsov13} and \citet{BuldgenZ} used different techniques and concluded that helioseismic methods favoured a low metallicity in the solar envelope, more in agreement with the AGSS$09$ determination. Both studies stressed the strong dependency of these inferences on the equation of state, which dominates the uncertainties. 

It is also worth noticing that the abundance of some elements cannot be directly inferred from spectroscopy of the solar photosphere. One of such elements is neon, which is derived from quiet regions of the solar corona \citep[see][and references therein]{Young}. Varying the neon abundance has a significant impact on opacity. Quickly after the revision of the solar abundances, \citet{Antia05} and \citet{Bahcall05} investigated the impact of changing the neon abundances to reconcile the AGS05 models with helioseismology. They found that a large increase was required. Recently, two independent studies \citep{Landi,Young} have demonstrated that the abundance ratio of neon over oxygen should be increased by $40\%$, which leads to significant changes in solar models, but still well below the values found by previous studies. 

The reason for this large impact of the abundances of elements heavier than helium is due to their large contribution to the radiative opacity inside the Sun \citep{BlancardOpacDetail, Mondet}. Despite their low abundance, they significantly shape the transport of energy in the radiative layers of the Sun, which represent most of its structure. This implies that they have a significant impact on the stratification of solar models and therefore on their (dis)agreement with observational constraints.

\subsection{Opacity tables}\label{Sec:Opacity}

Since the transport of energy in most of the solar structure is carried out by radiation, it is unsurprising that the radiative opacities have a large impact on solar models. As the solar modelling problem was unveiled in $2004$, the opacities were quickly pointed out as one of the potential causes of the discrepancies between the models and helioseismology \citep{Basu2004}. 

Today, they remain one of the most uncertain elements of the solar models. Indeed, various tables disagree with each other and lead to significantly different solar models at the level of precision of helioseismic constraints. Moreover, none of the current tables provides a satisfactory agreement with helioseismic constraints with recent abundances. In this paper, we will present results using the OPAL \citep{OPAL}, OP \citep{Badnell}, OPAS \citep{Mondet} and OPLIB \citep{Colgan} tables which have been computed by different groups at different times. Moreover, purely numerical considerations are also relevant, related to the various approaches chosen for the interpolation procedure of the opacity tables \citep{Houdek1996}.

These disagreements have motivated attempts to measure experimentally the opacity of key elements in physical conditions as close to solar as possible. The first of such measurements using a Z machine at the Sandia National Laboratories have been recently published for iron \citep{Bailey} and showed large discrepancies with theoretical calculations of iron spectral opacities, between $30$ and $400\%$. The origin of these discrepancies is still unclear and these experimental results still await independent confirmation. Nevertheless, various studies have been carried out to try to close the gap between theoretical calculations and the experiments \citep{Nahar,Iglesias2015, Blancard2016,Iglesias2017, Pradhan, Zhao, Pain2018}, some of which finding opacity increases compatible with the experimental results \citep{Bailey}. The debate is, however, still very much open and will probably require further extensive theoretical computations and comparisons with experiments.

\subsection{Equation of state}\label{Sec:EOS}

Another key elements of solar models is the equation of state. Throughout the years, refinements to the equation of state have also contributed to improve the agreements of solar models with helioseismic constraints. 

Two different approaches are used to compute an equation of state for stellar models. The first and most common approach is the so-called ``chemical picture'', where the thermodynamical quantities are computed from a free-energy minimization approach. The chemical picture has been used in the computation of the CEFF \citep{CEFF}, FreeEOS \citep{Irwin}, SAHA-S \citep{Gryaznov04,Gryaznov06, Baturin, Gryaznov13} and MHD \citep{MHDI,MHDII,MHDIII,MHDIV} equations of state. Moreover, in the regimes of astrophysical applications, effects as those of radiation pressure, relativistic corrections and electron degeneracy, amongst other, have to be included in the free energy and included consistently in the equation of state. Slight differences between various equations of state using the ``chemical picture'' might however result from different hypotheses made when taking into account these effects.  

The other approach used in equation of state calculations is the so-called ``physical picture'', which uses fundamental constituents and computes their interactions ab initio. Namely, this formalism considers separately atomic nuclei and electrons and describes their states using quantum wavefunctions. Again, additional corrections are included for astrophysical considerations. This approach has been used to compute the OPAL equation of state \citep{Rogers1996,Rogerseos}. 

The equation of state is a fundamental constituent of solar models, as it impacts indirectly multiple processes acting in solar and stellar interiors. For example, it influences the ionization levels of the chemical elements, which impacts the opacity at various temperatures. In some cases, differences between opacity tables do actually stem from the fact that a different equation of state has been associated with the computations. Consequently, one should in principle use opacity tables with the same equation of state employed in their computation. This is however unfortunately not always possible. The impact of the equation of state can also be directly seen in the sound-speed profile of solar models. This is particularly important when comparing models with various constitutents with helioseismic inferences, as the equation of state will impact the results in a significant manner. The induced variations are such that it is often stated that inversions of density profile should not be done using kernels such as the $(\rho,Y)$ kernels, as they lead to biases in the inferred profiles \citep[see][for a discussion and an illustration of this effect]{BasuSun}.

Moreover, differences in ionization level will impact diffusion velocities and hence the transport of chemicals during the evolution of the Sun. The equation of state also affects the adiabatic temperature gradient, which will influence the onset of convective transport and hence macroscopic mixing in solar and stellar models.   

Various studies have been performed to improve the current equation of state in the solar models by carrying out inversions of the profile of the adiabatic exponent, $\Gamma_{1}=\frac{\partial \ln P}{\partial \ln \rho}\vert_{S}$ \citep[e.g.][]{Elliott1996, Basu1997, Vorontsov13}.  

\subsection{Mixing of chemical elements}\label{Sec:Chemix}
\subsubsection{Microscopic diffusion}\label{Sec:MicroMix}

It is well known from first principle that a slow transport of the chemical elements is present in stellar radiative layers. This transport process is called microscopic diffusion and is linked to the various effects of temperature, pressure and composition gradients as well as the effects of ionization and radiation pressure with the various chemical elements of the stellar plasma. These effects induce chemical composition gradients in the stellar radiative regions and thus drastically change the expected initial chemical composition of solar models and their structure. In terms of nomenclature, solar models including the effects of microscopic diffusion are called ``standard solar models'' whereas models not including this transport process are called ``classical solar models''. It was one of the big successes in the early days of helioseismology to show that diffusion was acting in the Sun and thus had to be included in stellar model computations \citep[e.g.][]{JCD1993Diff,Basu1994, Basu1996}.

While it has been proven that solar models including microscopic diffusion are by far superior to models neglecting it, there are still some uncertainties linked to details in the physical processes underlying the generic term ``microscopic diffusion'' described in textbooks such as \citet{Burgers1969,Chapman1970, Ferziger1972,Michaud2015}. As such, various approaches for its implementation exist in the litterature \citep{Michaud1976,Noerdlinger1977,Paquette,Michaud1993,Thoul}, with various hypotheses linked to the components of the stellar plasma and the physical processes considered. It is also worth noticing that many standard solar models do not consider the effects of partial ionization nor the effects of radiation pressure when computing the transport of chemicals by microscopic diffusion. Including radiation pressure can be done in various ways; a simple approximate formula has been derived by \citet{Alecian2002}, to avoid the full computation of radiative accelerations for each element, which is very expensive numerically. Indeed, computing the effects of radiation in a fully consistent manner requires to compute the opacities for each chemical element on the fly at the given conditions of the layer of stellar material. This requires to interpolate in the individual opacity tables whenever these are made available and is computationally very expensive. In the solar case, \citet{Turcotte} have demonstrated that these effects are negligible for the solar case. However, other studies have shown that slight modifications should be expected \citep{Schlattl2002,Gorshkov2008,Gorshkov10} while the radiative accelerations for certain elements will of course be ultimately influenced by potential significant opacity modifications. Other effects, such as quantum corrections on diffusion coefficients will also slightly affect the transport of chemical elements in the Sun and thus alter the (dis)agreement with helioseismic constraints \citep{Schlattl2003}. Recently, careful investigations of the numerical integrations of the resistance coefficients have also been undertaken by \citet{Zhang2017}. This study found slight but significant modifications to the properties of solar models, resulting from singularities in the case of an attractive screened Coulomb potential. In addition, while many of these effects might well be of small importance, when not completely negligible for the Sun, this hypothesis does not hold for other stars \citep[see e.g.][]{Richard2002a,VandenBerg2002,Richard2002b,Theado2005,Michaud2008,Theado2010,Deal2018}.   

\subsubsection{Macroscopic chemical mixing at the base of the convective zone}\label{Sec:MacroMix}

Besides microscopic diffusion, macroscopic motions of the solar plasma are also responsible for alterations of the chemical stratification inside the Sun. The most well-known process is turbulent convection, which occurs in the upper layers of the solar envelope. The modelling of convection is one of the most central problem in stellar astrophysics, as most of the current stellar evolution codes use the so-called mixing length theory (MLT) which is a very crude representation of the turbulent motions occuring in stellar conditions \citep{Bohm, Cox}. For the solar modelling problem, the shortcomings of the MLT are especially crucial for the positioning of the base of the convective zone and the transition from convective to radiative transport of energy. Indeed, the largest differences between the Sun and standard models are found right below the base of the convective zone. 

The problem is linked to the criterion used to determine the extension of convective region, the so-called Schwarzschild criterion \citep{Schwarzschild1906}. This criterion is based on the cancellation of the convective flux, which translates into a local criterion for the temperature gradients inside the star. However, the cancellation of the flux does not necessarily imply a cancellation of the velocity of the convective elements, which is the parameter determining the extent of the mixed region. This extra-mixed region and its thermal stratification are still uncertain, although hydrodynamical simulations can provide some guidelines in the computation of this so-called ``overshooting'' or ``penetrative convection'' at the base of the solar convective zone \citep{Xiong01, Rempel04,YangI, YangII, Viallet, Hotta}. Helioseismology can also be used to provide some insights on the transition of the temperature gradient from adiabatic to radiative in this region \citep{Monteiro94,JCDOV} but unfortunately, it is difficult to disentangle the effects of overshoot from the effects of opacities which can also alter the temperature gradient in these layers.

Besides the effects of overshooting, the base of the solar convective zone is also affected by the effects of rotation in a thin region called the tachocline \citep{SpiegelZahn1992}. In this region of around $0.04R_{\odot}$ wide \citep{Elliott99Tacho,corbard99}, the rotational profile of the Sun changes from differentally rotating in latitude to solid body rotating. This transition implies shear-induced mixing of the chemical elements. However, comparisons of helioseismic inversions of the solar rotation profile to rotating models have shown that the effects of meridional circulation and shear-induced turbulence were insufficient to reproduce the inferred properties. Hence, additional processes linked to magnetism or internal gravity waves have to be invoked to reproduce the solar rotation profile \citep{Gough1998, Charbonnel, Eggenberger}. These effects impact the chemical evolution of the Sun, being for example thought to be responsible for the observed lithium depletion and influencing the evolution of the solar convective zone. These effects, while localized, also slightly influence the calibration procedure and hence the initial chemical composition of the standard solar models \citep{Proffitt,Richard96Sun,Gabriel97,Brun02}. It should be noted, however, that including these processes in a calibration procedure is extremely difficult and somewhat dangerous as they introduce additional parameters which are not constrained from first principles. Hence, further theoretical work is required to avoid the artificial fine-tuning of correlated parameters which could lead to spurious solutions. The recent detection of gravity modes by \citet{Fossat2017} could prove to be a game changer in that respect, by providing an average rotation of the solar core. This would provide a link between mean molecular weight and potential rotation gradients, providing very stringent constraints on the nature of the physical process responsible for the flat rotation profile of the upper radiative layers \citep{Eggenberger}. This detection, however, still needs to be confirmed independently as it has already triggered some controversy \citep{Schunker2018}.

The extra-mixing below the envelope is often treated in a parametric way, by introducing an additional turbulent diffusion coefficient depending on various parameters. In our study, we parametrize this diffusion coefficient as a function of $\rho_{\mathrm{cz}}$, the density value at the base of the convective zone
\begin{align}
D_{Turb}=D\left( \frac{\rho_{\mathrm{cz}}}{\rho(r)} \right)^{N}, \label{eqDTURB}
\end{align}
with the free parameters $D$ $\left[ \mathrm{cm}^{2}\mathrm{s}^{-1} \right]$, and $N$ which were fixed to $7500$ and $3$ respectively in the work of \citet{Proffitt}.

\subsection{Early evolution}\label{Sec:SolarPMS}

In the previous sections, we discussed mainly effects that occured largely on the main sequence and consisted in the ``usual suspects'' of the solar modelling problem. There are, however, other sources of uncertainties in the early solar evolution that could have an impact on the present-day solar structure as seen from helioseismic constraints.

These include accretion of material during the early stages of the formation of the solar system. This would lead to a contrast in the models, where the internal structure would behave as if the model had a high metallicity, whereas the upper layers would have the observed photospheric abundances. Accretion of low metallicity material was considered by \citet{Winnick2002}, \citet{Guzik06} or \citet{Castro2007}. The proposed scenario was that $98\%$ in mass of the Sun could have formed from metal-rich material, in agreement with the GS98 or GN93 abundances, while the last $2\%$ of material would be metal-poor or metal-free and would have been accreted after the apparition of the radiative core of the Sun, to avoid a full mixing of the elements. This scenario provided some improvement in the position of the base of the convective envelope, the helium abundance in the convective zone and to some extent in the sound speed profile (at least in \citet{Guzik06}, whereas \citet{Castro2007} still find large discrepancies just below the convective zone).

\citet{Serenelli2011} have tested the accretion scenario using various metallicities, masses and times at which accretion took place. They found that accretion alone could not solve the solar problem, as metal-rich accretion led to a good agreement in the position of the base of the convective zone and sound speed profile, but reduced the agreement in helium abundance. Metal-poor accretion only provided a good agreement in helium abundance in the convective zone in their tests. They also noted that accretion of material could easily lead to a strong disagreement in lithium abundances, implying that at least additional mixing would be required to reproduce the proper lithium depletion.

Besides accretion, the so-called "faint young Sun paradox" has also motivated non-standard computations of the evolution of the Sun, including exponentially decaying mass loss on the main-sequence. The paradox resides in the fact the solar luminosity on the zero-age main sequence, according to a standard model evolution, would be around $70\%$ of its current luminosity, which is insufficient to explain the presence of liquid water on Mars and the Earth at an early stage of the evolution of the solar system. Other solutions have been suggested to explain these discrepancies, such as greenhouse gases \citep[see][]{Forget2013,Airapetian2016,Wordsworth2016,Bristow2017,Turbet2017}, a revision of the carbon cycle in the early Earth's atmopshere \citep{Charnay2017} or a slightly more massive young Sun \citep{Sackmann2003,Minton2007,Turck2011,Weiss2013}.

Physically, one makes the hypothesis that large mass loss on the pre-main sequence could still be present at the very beginning of the main-sequence. Indeed \citet{Wood2005} have observed large winds on young solar-like stars.  Increasing the mass loss on the early main sequence implies that the mass and hence the solar luminosity at the zero-age main sequence would be slightly higher and could then provide the physical conditions required for the presence of liquid water. Typically, this effect is erased as the models including mass loss recover the standard evolution of luminosity at about $2$Gy. Of course, such a non-standard evolution leaves traces on observational constraints. Early works by \citet{Guzik1987} and \citet{Graedel1991} studied its impact on the lithium depletion problem and recently, \citet{Guzik2010} and \citet{Wood2018} investigated its impact on seismic properties and neutrino fluxes. It appears that such massive models improve the agreement of low metallicity models in the upper radiative layers but not in the core. The disagreement in the central regions has to be mitigated by modifying other physical ingredients such as the screening factors of nuclear reaction. Following \citet{Wood2018}, this can be done using the dynamical screening factor of \citet{Mussack2011}. As for the rotational profile of the Sun, the potential detection of solar gravity modes would provide stringent constraints on the solar core, which could eventually require to question key ingredients linked to the nuclear reactions. \citet{Spalding2018} suggest another way to test this hypothesis by analysing terrestial or martian sediments to look for traces of specific Milankovitch cycle imprints scaling with the solar mass. Detecting such frequencies at different epochs could provide a direct hint at the history of the Sun and thus insights on the ``young massive Sun hypothesis''. 

Following the neutrino measurements by \cite{Davis1968} and their disagreement with the solar models of the time, \citet{Dilke1972}, adapting the formalism of \citep{Defouw1970}, suggested a mechanism that could alter the core properties of the Sun in its early evolution and provide a solar explanation for the some climate cycles on Earth. The mechanism received some criticism by \citet{Ulrich1973}, \citet{Ulrich1974} and \cite{Ulrich1975} and was further investigated by \citet{Unno1975} and also discussed by various other authors \citep{Ledoux1974,JCD1974,Shibahashi1975,Boury1975,Scuflaire1975,Gabriel1976,Noels1976}. The original idea was called the ``solar spoon'' and was linked to the potential intermittent mixing of the solar core as a result of gravity modes, which would be excited by a form of $\epsilon$ mechanism due to $^{3}\mathrm{He}$ burning. In practice, the first appearance of overstability is favoured by some form of mixing such as the aftermath of the intermittent convective core at early stages of solar evolution, some amount of rotational mixing or other unknown processes such as magnetic convection \citep{Schatten1973}. Once a favourable condition for overstability is provided at some point during the solar evolution, the oscillations can be excited by the intermittent burning of $^{3}\mathrm{He}$, which starts once the first oscillations have grown large enough. However, since the gravity modes are stabilized by radiative damping, a trapping condition has to be ensured so that they can grow large enough in the deep layers to trigger the intermittent burning and self-sustain the cycle.

Provided an adequate trapping of the modes, mixing of $^{3}\mathrm{He}$ will occur as a result of the oscillation and the Brunt-Väisälä profile of the solar model will be altered. After a sufficient nuclear time linked to the $^{3}\mathrm{He}$, the depletion of nuclear fuel will induce the disappearance of the overstability. However, the overstability will propagate towards lower temperatures and thus subsist in regions where its timescale will be greater, provided that the $^{3}\mathrm{He}$ profile is adequate for its development. The whole process will thus be quasi-periodic, as the different timescales involved will change over the course of the solar evolution.

In a recent paper \citet{Gough2015} discussed the process and considered it ruled out. However, a steep $^{3}\mathrm{He}$ is suggested by non-linear inversions of the solar core \citep{Marchenkov}, which could drive the overstable oscillations described by \cite{Dilke1972} and \citet{Unno1975}. \citet{Roxburgh1976,Roxburgh1984} suggested that the instability would break down into mild turbulence and locally modify the sound-speed gradient. Various computations have been undertaken to investigate the stability of g modes to this form of $\epsilon$ mechanism. \citet{JCD1974,Shibahashi1975,Boury1975} found, using the quasi-adiabatic approximation, that some low order g modes could be unstable. \citet{JCD1975} investigated the issue using fully non-adiabatic computations and found the modes to be likely stable, as a result of significant damping in the upper layers of the convective envelope. \citet{Saio80}, using a linear non-adiabatic analysis taking into account time-dependent convection, confirmed the instability of some g modes in early stages of the solar evolution. Moreover, \citet{Saio80} confirmed the potential instability of the $g_{2}$ $\ell$=1 mode in the present Sun and suggested that some higher degree modes could also be non-linearly coupled with the $g_{2}$ $\ell$=1 mode. The issue was later investigated by \citet{Kosovichev1985}, which confirmed that mixing and a low-metallicity of the solar models would enhance the instability of the gravity modes with respect to the $\epsilon$ mechanism. The main difficulty in reaching a definitive answer on the issue is linked to the treatment of the behaviour of the convective envelope in the stability analysis. Moreover, a fully non-linear analysis of the development of the instability is required to prove that it would lead to a significant transport of chemical elements which would self-sustain the process. \citet{Ulrich1973} and \citet{Ulrich1974} have stated that an additional agent was required to provide the necessary chemical mixing, as the non-radial oscillation would be insufficient to do so. Non-linear calculations of resonant coupling of gravity modes by \citet{Dziembowski1983}, using the approach of \citet{Dziembowski1982}, confirmed this criticism of the original formalism of \citet{Dilke1972}. Finally, the absence of undisputed detection and identification of gravity modes does not allow to close the debate. In the quest for solar g-modes, other excitation mechanisms have been suggested and investigated \citep[see][for a review on solar gravity modes]{Appourchaux2010}, predicting various detectability levels for these highly-sought pulsations.

In recent years, the $\epsilon$ mechanism has been reinvestigated in metal-poor low-mass main-sequence stars \citep{Sonoi2012a,Sonoi2012b}. In this case, the reduced size of the outer convective zone simplifies the treatement of the stability analysis, as it is thought to play a minor role in the total energy budget.

Other more subtle effects, like the low-temperature opacities \citep{Guzik06}, the equation of state or the properties of the chemical mixing at the base of the convective zone in the early phases of the solar evolution \citep{Baturin2015} could affect the observed properties of the current Sun and the conclusion we may draw from them. At first, these effects may seem negligible but they would actually impact the initial conditions of a solar calibration, hence leading to overall changes in the structure that cannot be fully neglected. A very stringent constraint on such effects is the lithium depletion observed in the solar photosphere, which is strongly affected by micro- and macrophysical effects in the solar models. 

\section{Combined structural inversions and structural diagnostics}\label{Sec:Inversions}

In this section, we will present inversion results of solar models built with various physical ingredients. All models have been computed with the Liège stellar evolution code \citep[CLES,][]{ScuflaireCles}. Their oscillations have been computed using the Liège adiabatic oscillation code \citep[LOSC,][]{ScuflaireOsc} and the inversions have been carried out using the SOLA method \citep{Pijpers} implemented in the InversionKit software \citep{ReeseDens}. 

We followed the guidelines of \citet{RabelloParam} to adjust the trade-off parameters of the inversion techniques and used the data of \citet{BasuSun} supplemented by an extension of BiSON observations of \citet{Davies} \citep[as used in][]{Buldgen2018}. We computed inversions of the squared adiabatic sound speed $c^{2}=\frac{\Gamma_{1}P}{\rho}$, an entropy proxy, denoted $S_{5/3}=\frac{P}{\rho^{5/3}}$ presented in \citet{BuldgenS}; and the Ledoux discriminant, defined as $A=\frac{d \ln \rho}{d \ln r}-\frac{1}{\Gamma_{1}}\frac{d \ln P}{d \ln r}$ as in \citet{BuldgenA}.

We start in sections \ref{Sec:SoundSpeed}, \ref{Sec:EntropyProxy} and \ref{Sec:Ledoux} by presenting inversion results for solar models built using different physical ingredients. To test the dependency of standard solar models on chemical compositon and opacities, we used models built using the AGSS09 and GS98 abundances tables, the OPAS, OPAL and OPLIB opacity tables and models including the revision of the neon abundance found in \citet{Landi} and \citet{Young}, hereafter AGSS09Ne. We also present results for various implementations of the mixing of chemical elements, namely the use of the \citet{Paquette} collision integrals in the diffusion coefficients, the effects of considering the partial ionization of the heavy elements in the computation of microscopic diffusion. Besides microscopic effects, we also consider macroscopic mixing, in the form of an adiabatic overshoot and in the form of turbulent diffusion. All models presented here have been built using the FreeEOS equation of state and the \citet{Adelberger} nuclear reaction rates, except for the model taking into account partial ionization in the computation of microscopic diffusion which used the SAHA-S equation of state \citep{Gryaznov04, Baturin, Gryaznov13}. 

\begin{table*}[t]
\caption{Physical ingredients of the standard solar models used in this study}
\label{tabSTDParamModels}
  \centering
\begin{tabular}{r | c | c | c | c | c }
\hline \hline
\textbf{Name}&\textbf{EOS}&\textbf{Opacity}&\textbf{Abundances} & \textbf{Diffusion} & \textbf{Convection}\\ \hline
 AGSS09-OPAL & FreeEOS & OPAL & AGSS09 & Thoul & MLT\\
 AGSS09-OPLIB & FreeEOS & OPLIB & AGSS09 & Thoul & MLT\\ 
 AGSS09-OPAS & FreeEOS & OPAS & AGSS09 & Thoul & MLT\\
 AGSS09-OPAL-Paquette & FreeEOS & OPAL & AGSS09 & Paquette & MLT\\
 GS98-OPAL & FreeEOS & OPAL & GS98 & Thoul & MLT\\
 AGSS09Ne-OPAL & FreeEOS & OPAL & AGSS09Ne & Thoul & MLT\\
 AGSS09-OPAL-PartIon & SAHA-S & OPAL & AGSS09 & Thoul + PartIon & MLT\\
 AGSS09-OPAL-OvAd & FreeEOS & OPAL & AGSS09 & Thoul + OvAd $(0.1H_{P})$ & MLT\\
 AGSS09-OPAL-DT & FreeEOS & OPAL & AGSS09 & Thoul + DT& MLT\\
 AGSS09-OPAL-Proffitt & FreeEOS & OPAL & AGSS09 & Thoul + Proffitt& MLT\\
\hline
\end{tabular}
\end{table*}

\begin{figure}[h!]
\begin{center}
\includegraphics[width=12cm]{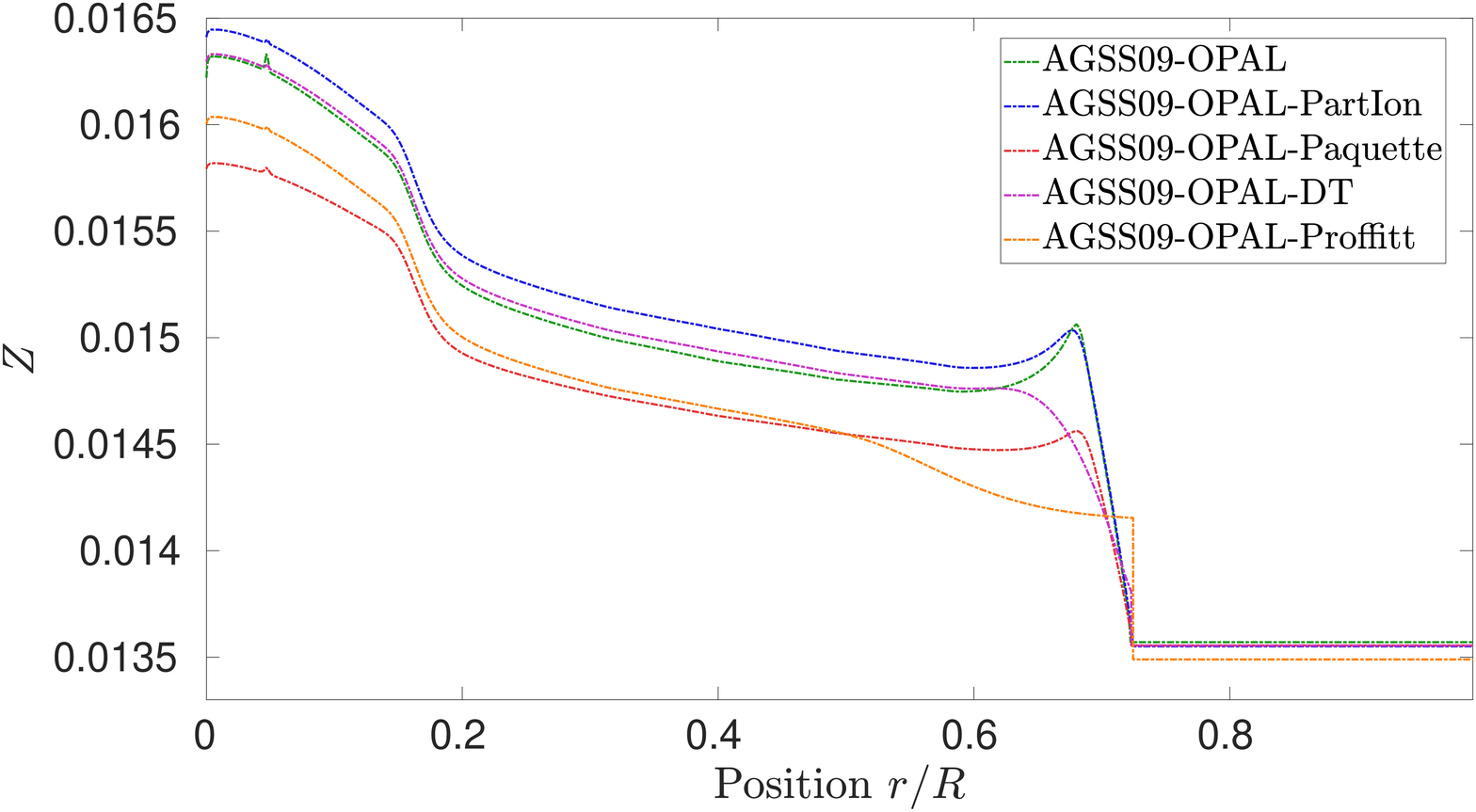}
\end{center}
\caption{Metallicity profile of the standard solar models of table \ref{tabSTDParamModels} including various prescriptions for the transport of chemicals.}\label{fig:ZDiff}
\end{figure}

In Fig. \ref{fig:ZDiff}, we illustrate the effect of the changes in the properties of the chemical mixing on the metallicity profile of the solar models. Most of the trends can be easily understood. For example, taking into account partial ionization of the metals when computing microscopic diffusion as in the AGSS09-OPAL-PartIon model (blue) will lead to a slightly more efficient diffusion of these elements, as they encounter less repulsion near the base of the convective zone and thus will more easily fall down towards central layers. Using the screened Coulomb potentials in the diffusion coefficients, as in \citet{Paquette}, in the AGSS09-OPAL-Paquette model, leads to a less efficient diffusion during the evolution, as the ions will experience more repulsion than in the case of the cut-off hypothesis over a Debye sphere used in the original \citet{Thoul} formalism. The models including turbulent diffusion, denoted ``AGSS09-OPAL-DT'' and ``AGSS09-OPAL-Proffitt'' in table \ref{tabSTDParamModels} , show a much more different behaviour. In the ``AGSS09-OPAL-DT'' model, we have fixed the $D$ and $N$ parameter of equation \ref{eqDTURB} to respectively $50$ and $2$ and to $7500$ and $3$ the ``AGSS09-OPAL-Proffitt'' model. The peak stemming from the variations of diffusion velocity near the base of the convective envelope is erased by the turbulent mixing, which induces a very different metallicity profile. The disappearance of this metal-peak is actually seen in the Ledoux discriminant inversion through its impact on the temperature gradient around $0.65$ solar radii (see Sect. \ref{Sec:Ledoux}). It is also worth noticing that including the prescription of \citet{Proffitt} for turbulent diffuson has a sufficiently large impact on the calibration to alter the initial chemical composition of model `AGSS09-OPAL-Paquette', as can also be seen from table \ref{tabSTDParamModels}, whereas the coefficients used in \citet{BuldgenA} have a negligible impact on the initial conditions. 

\subsection{Sound speed inversions}\label{Sec:SoundSpeed}

\begin{figure}[h!]
\begin{center}
\includegraphics[width=17cm]{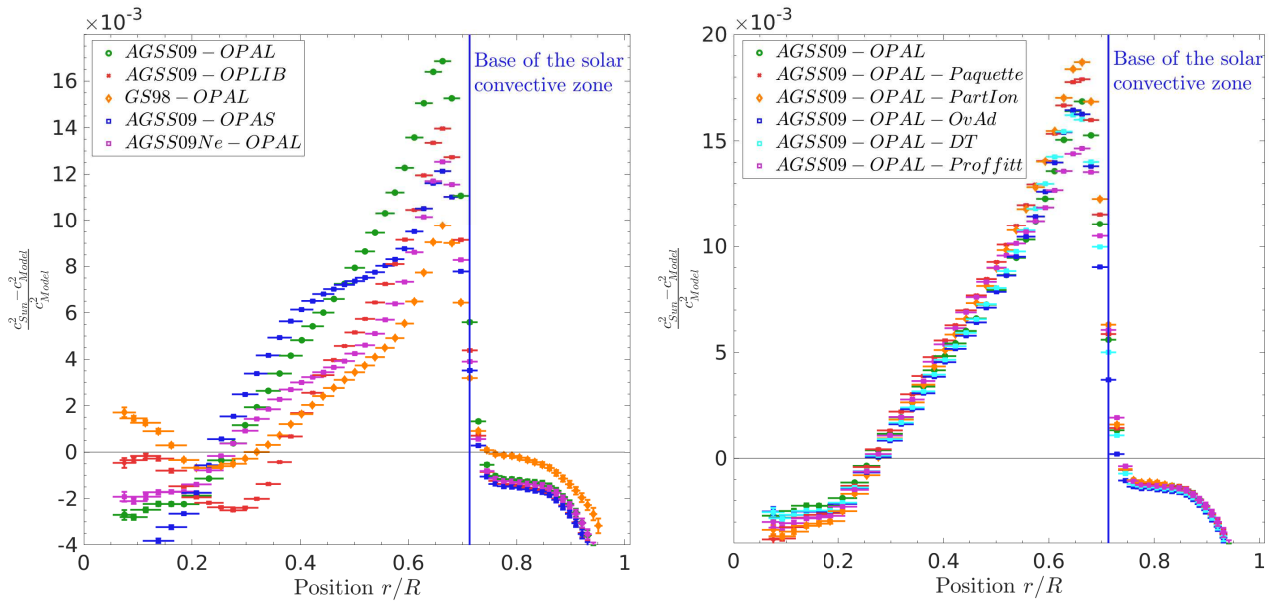}
\end{center}
\caption{Left panel: relative squared sound speed differences between standard solar models using various abundance and opacity tables and helioseismic results. Right panel: relative squared sound speed differences between models including various prescriptions for the mixing of the chemical elements and helioseismic results.}\label{fig:c2STD}
\end{figure}

We start with classical sound speed inversions, presented in Fig. \ref{fig:c2STD}. In the left panel of Fig. \ref{fig:c2STD}, we present results for standard solar models built with various abundances and opacities. We can see the illustration of the well-known solar modelling problem when comparing the standard AGSS09 models, in green, with the GS98 standard model, in orange. However, it appears that considering the $40\%$ increase of the Ne/O ratio derived independently by \citet{Landi} and \citet{Young} provides a significant improvement of the agreement between AGSS09 models and helioseismic inversions. This is not a surprise, since a neon increase, although much larger, was already suggested by \citet{Antia05}, \citet{Zaatri2007} and \citet{Basu08} as a potential solution to the solar modelling problem. Similarly, using the more recent OPAS or OPLIB opacity tables also leads to a non-negligible improvement of the agreement of low-metallicity models and helioseismic results. However, this significant improvement is restricted to the radiative layers. Indeed, large discrepancies in sound speed in the convective envelope are still present for all the AGSS09 models. This is likely due to the large discrepancies in helium in the convective envelope, since, as we will see in Sect. \ref{Sec:OtherConstraints}, none of the models presented in Fig. \ref{fig:c2STD} shows a good agreement with the helioseismic helium abundance. 

In the right panel of Fig. \ref{fig:c2STD}, we illustrate squared adiabatic sound speed inversions for models including various prescriptions for the mixing of the chemical elements. Using the \citet{Paquette} collision integrals or considering partial ionization in the computation of microscopic diffusion leads to an increase of the disagreements in the sound speed profile just below the convective envelope (orange and red symbols in the right panel of Fig. \ref{fig:c2STD}). Adding a form of macroscopic mixing improves the agreement of AGSS09 models and helioseismic inversions, as can be seen from the models including either turbulent diffusion or a form of overshooting. The best agreement is found for the polynomial formulation of turbulent diffusion used in \citet{Proffitt} to reproduce the solar lithium abundances (purple symbols in the right panel of Fig. \ref{fig:c2STD}). However, the improvement is very localized and the mixing has little to no impact on the deeper radiative layers. This demonstrates, as is now well-known, that the solar modelling problem cannot stem only from an inaccuracy of the mixing of the chemical elements, but that other ingredients such as the radiative opacities, may be partially responsible for the discrepancies.   

\subsection{Entropy proxy inversions}\label{Sec:EntropyProxy}

\begin{figure}[h!]
\begin{center}
\includegraphics[width=17cm]{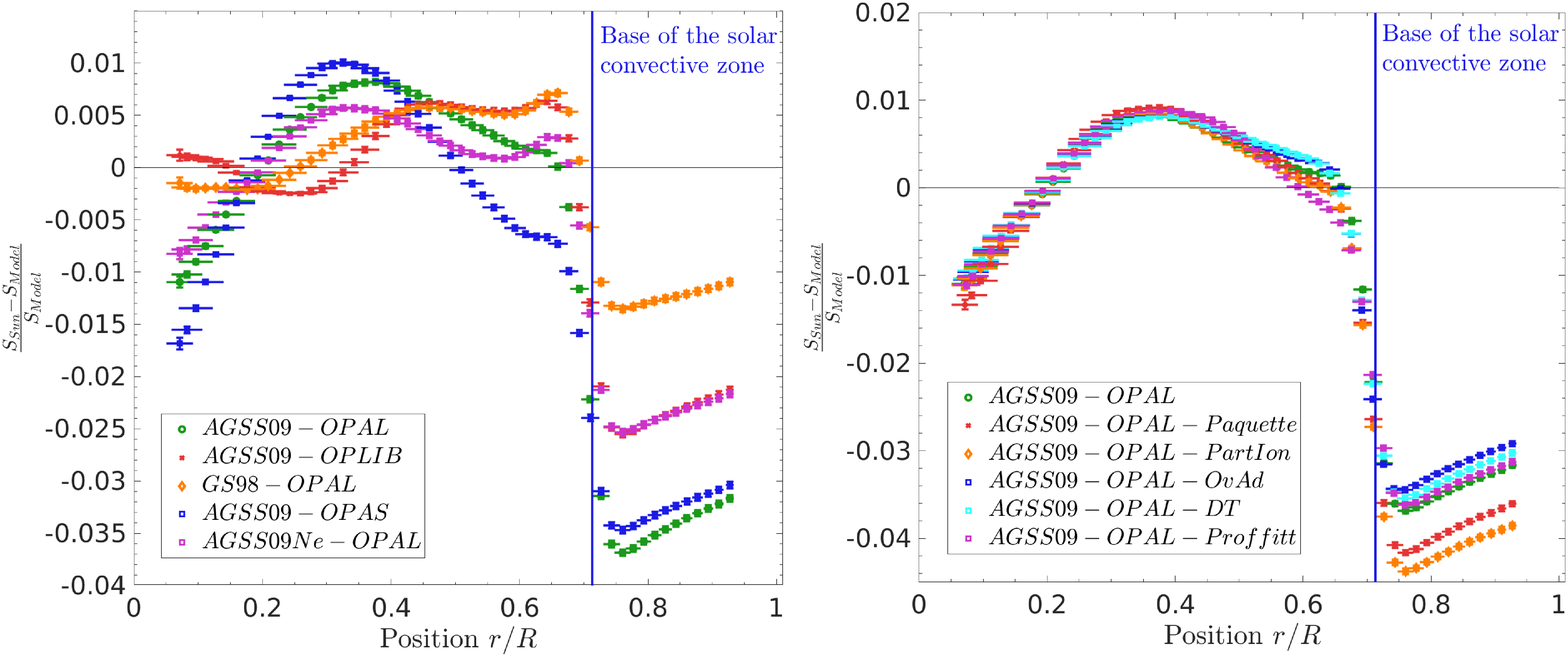}
\end{center}
\caption{Left panel: relative entropy proxy differences between standard solar models using various abundance and opacity tables and helioseismic results. Right panel: relative entropy proxy differences between models including various prescriptions for the mixing of the chemical elements and helioseismic results.}\label{fig:SSTD}
\end{figure}

In addition to squared adiabatic sound speed, other structural quantities can be inverted, such as for exemple the density, using the $\left( \rho, \Gamma_{1} \right)$ structural pair \citep[see e.g.][]{Antia94} or the squared isothermal sound speed \citep[see e.g.][]{Dziemboswki90,Gough1991}. Recently, we presented in \citet{BuldgenKer} approaches to change the structural variables of the variational equations which could in turn be used in helio- and asteroseismology. In \citet{BuldgenS}, we presented inversion results of an entropy proxy, denoted $S_{5/3}=\frac{P}{\rho^{5/3}}$ which provides interesting insights on the solar structure. In Fig. \ref{fig:SSTD}, we show the inversion results of this structural quantity for the models discussed in Sect. \ref{Sec:SoundSpeed}. 

From Fig. \ref{fig:SSTD}, a slightly different picture of the problem is drawn. In the left panel, the model built with the OPAS opacity tables which performed quite well in the sound speed inversion does not provide a good agreement in the entropy proxy, whereas the OPLIB opacities provide a very significant improvement over the OPAL opacities, similar to the effect of the neon revision. Nevertheless, it is clear that the performance of the AGSS09 models is still very far from the agreement obtained using the GS98 abundances. The performance of the models built with the OPLIB opacities and the revision of the neon abundance is due to the steeper temperature gradient of these models below the convective envelope due to either the behaviour of the opacity profile \citep[see][for a discussion]{Colgan,Guzik16} or simply the increase of neon which leads to an increased opacity. Indeed, from \citet{BlancardOpacDetail}, it appears that neon is the third most important contributor to the opacity at the base of the solar convective zone. 

In the right panel of Fig. \ref{fig:SSTD}, we can see again that none of the modifications of the transport of chemical elements have led to a large improvement of the performance of low-metallicity solar models. Slight modifications to the $S_{5/3}$ profile are seen, with the model including adiabatic overshooting performing slightly better than the models including turbulent diffusion. Again, we also see that the models including the \citet{Paquette} collision integrals or partial ionization when computing microscopic diffusion lead to an increase of the disagreements with helioseismic results. Overall, this inversion confirms that the solution to the solar modelling problem is not to be found from the mixing of the chemical elements alone, but also that some distinction can be made over the type of mixing if one refines the diagnostic by combining it to a quantity more sensitive to local variations. This will be further discussed in Sect. \ref{Sec:Ledoux}, when presenting the results of the Ledoux discriminant inversions.

\subsection{Ledoux Discriminant inversions}\label{Sec:Ledoux}

In Sect. \ref{Sec:EntropyProxy}, we discussed the results of inversions of an entropy proxy and showed the importance of combining the information from various inversion techniques to lift potential degeneracies that could hinder our understanding of the solar modelling problem. This thinking can be pushed even further by carrying out inversions of the Ledoux discriminant. These inversions were already presented in \citet{Gough1993, Elliott1996, Takata2002, Kosovichev} but have not been exploited to analyse the discrepancies found for models built with the recent abundance tables of \citet{AGSS09}. This analysis was carried out in \citet{BuldgenS} and \citet{Buldgen2018}, where in this last paper, an extended set of models is analysed. 

\begin{figure}[h!]
\begin{center}
\includegraphics[width=17cm]{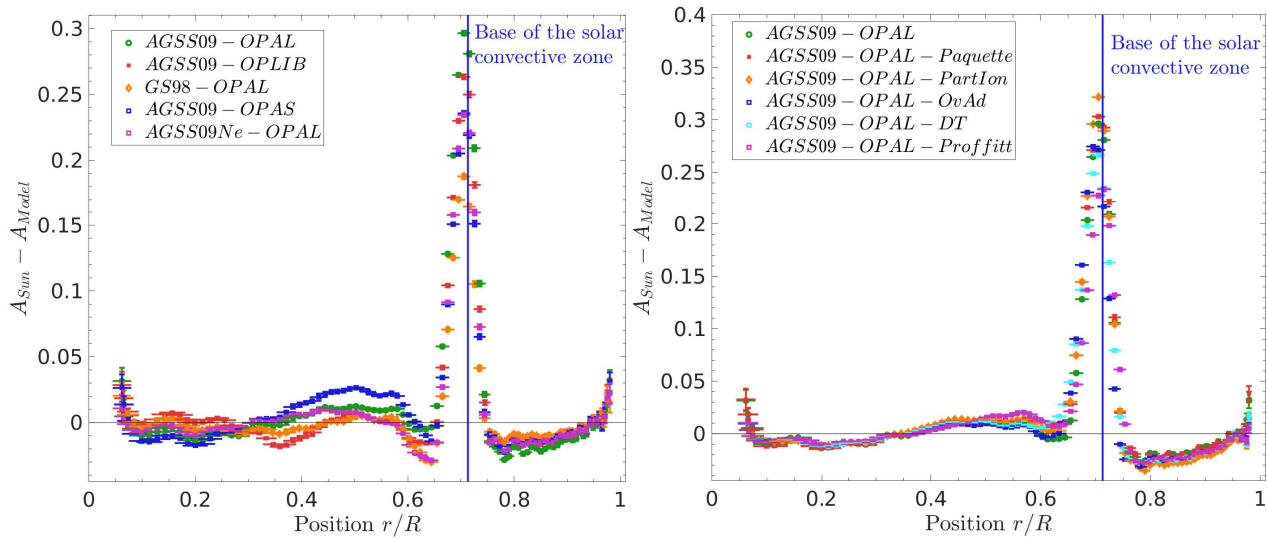}
\end{center}
\caption{Left panel: Ledoux discriminant differences between standard solar models using various abundance and opacity tables and helioseismic results. Right panel: Ledoux discriminant differences between models including various prescriptions for the mixing of the chemical elements and helioseismic results.}\label{fig:ASTD}
\end{figure}

In Fig. \ref{fig:ASTD}, we present inversion results for the models of Sect. \ref{Sec:SoundSpeed} and \ref{Sec:EntropyProxy}. The first striking feature of these inversions is the large disagreements at the base of the convective zone which is found for any opacity tables, chemical abundances and mixing considered. These discrepancies illustrate clearly the fact that the standard solar models are unable to reproduce the transition in both temperature and chemical composition gradient at the base of the convective envelope. The main difficulty is to separate each of their contributions to the $A$ inversion.

Overall, the results are again quite mixed. In the left panel of Fig. \ref{fig:ASTD}, we find that the increase in neon provides the largest improvement, bringing the models to an agreement nearly as good as that found in the GS98 models. The OPAS and OPLIB opacity tables also significantly improve the behaviour of the AGSS09 models. Nevertheless, the results are far from convincing. Moreover, even the GS98 models show large deviations below the convective zone, as deep as $0.6$ solar radii, thus in a region supposedly fully radiative. This emphasizes that while the potentially missing macroscopic mixing process is certainly very localized, it can still have an impact in deeper radiative layers. Indeed, it will influence the initial chemical abundances required to reproduce the solar surface metallicity, luminosity and temperature (or radius) at a solar age and thus the whole structure to a level that is detectable with helioseismic data.

This is confirmed by the right panel of Fig. \ref{fig:ASTD}, where we can see the impact of extra-mixing below the convective zone. Again, the parametrization of \citet{Proffitt} provides the largest improvement for the low-metallicity models, while the second parametrization used in \citet{BuldgenA} provides a similar agreement to that of using a convective overshoot and the use of \citet{Paquette} collision integrals and considering partial ionization in the computation of microscopic diffusion leads to larger deviations. This does not mean, however, that turbulent mixing is not occuring at the base of the convective zone and that such approaches should not be explored. However, it clearly shows that mixing alone is not sufficient to solve the solar modelling problem and other ingredients have to be revised. Hence, it is of crucial importance to compare physical ingredients, formalisms and numerical techniques to fully assess their importance for the current issue, in a similar fashion to what has been done in \citet{Boothroyd03}, \citet{ESTA1} and \citet{ESTA2}. 

On a sidenote, we would like to emphasize the degeneracy at play in the analysis of helioseismic inversions. Even when combining the results of sound speed, entropy proxy, and Ledoux discriminant, we cannot fully distinguish between thermal and compositional effects. Moreover, the inverted results are not independent. They could in principle all be deduced from the solar density profile. In that sense, they all provide the same information about the solar structure. For example, if one uses the Ledoux discriminant inversion to correct the $A$ profile of a standard solar model and integrate the other variables, assuming $\Gamma_{1}$ known, the agreement in both sound speed and entropy proxy is very significantly improved. However, combining the inversions is useful when trying to link an improvement with respect to the helioseismic inversions to a change in the physical ingredients of the models, as the degeneracy at play between compositional and thermal effects will not act in the same way for all structural variables. 

Despite these differences in their behaviours, one cannot fully separate thermal and compositional effects without further assumptions linked to the equation of state of stellar material and the chemical composition or the temperature stratification. This degeneracy is a consequence of the physical dependencies of the quantities for which inversions can be carried out, and of the fact that the changes considered in the models will always impact both temperature and chemical gradients. For example, changing the opacity tables will alter the position of the base of the convective zone in the models and thus alter the diffusion history, thus the chemical gradients. Similarly, adding an additional mixing at the base of the convective zone will alter the chemical history of the model and thus the temperature gradients, near the base of the convective zone but also deeper, as seen from the $A$ inversion of the model including turbulent diffusion following \citet{Proffitt}. In that sense, one should emphasize the value of independent constraints such as neutrino measurements, which provide crucial additional informations about the deep layers of the Sun, particularly the temperature. 

\subsection{Convective envelope properties and frequency separation ratios}\label{Sec:OtherConstraints}

\begin{figure}[h!]
\begin{center}
\includegraphics[width=13cm]{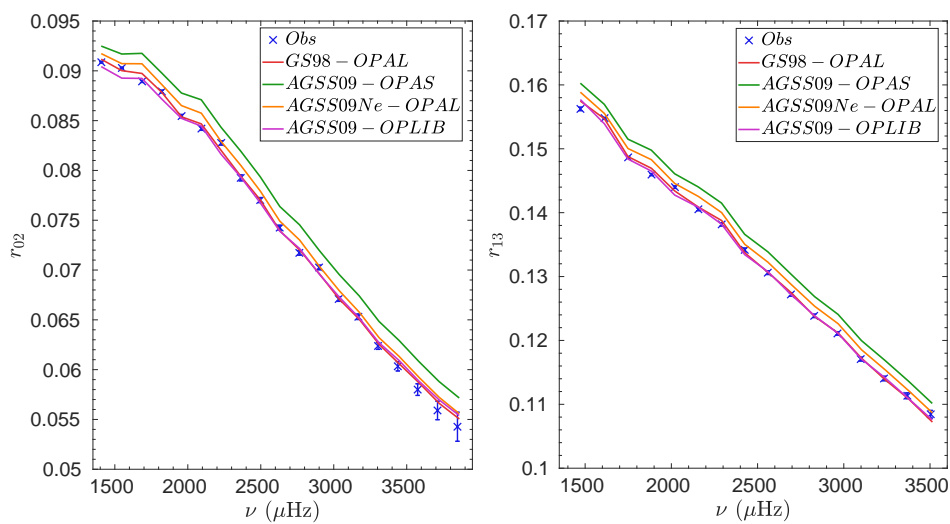}
\end{center}
\caption{Comparison between the observed frequency separation ratios $r_{0,2}$ and $r_{1,3}$ from the BiSON data and those of solar models built with various abondance and opacity tables.}\label{fig:ratiosSTD}
\end{figure}

Besides looking at seismic inversions, it is also interesting to analyse other complementary constraints. Indeed, solar models do not have only to show a good agreement in terms of inferred quantities, but should also reproduce the correct position of the base of the convective zone  \citep{KosovBCZ,JCD91Conv,Basu97BCZ}, determined to be around $0.713$ solar radii and the helium abundance in the convective envelope, determined by \citep{Vorontsov91, Vorontsov13} and found to be above $0.245$\footnote{We consider here a conservative approach given the differences in the precision of the helium determination found by various studies.}. In addition, a classical seismic diagnostic of the solar models are the so-called frequency separation ratios
\begin{align}
r_{0,2}=\frac{\nu_{n,0}-\nu_{n-1,2}}{\nu_{n,1}-\nu_{n-1,1}},\label{eqratios1} \\
r_{1,3}=\frac{\nu_{n,1}-\nu_{n-1,3}}{\nu_{n+1,0}-\nu_{n,0}}, \label{eqratios2}
\end{align} 
following the definitions of \citet{RoxburghRatios}. They showed that these ratios are very sensitive to the deep layers of stellar structure. In this section, we present in table \ref{tabSTDModels} the above properties of the solar models presented in the preceeding sections and illustrate in Fig. \ref{fig:ratiosSTD} the frequency separation ratios of some of our models and those obtained from BiSON data. 

From table \ref{tabSTDModels}, we can see that models with the AGSS09 abundances all have a surface helium abundance well below the seismically determined intervals and a too shallow convective envelope. A more worrying result is found for the models built with the latest OPLIB and OPAS opacity tables, as their helium abundance is even lower than those built using the OPAL tables. This is a consequence of the reduction of the opacity in an extended part of the solar radiative zone, which implies a higher initial hydrogen abundance of the model to allow them to reproduce the solar luminosity at the solar age. 
 
\begin{table*}[t]
\caption{Parameters of the standard solar models used in this study}
\label{tabSTDModels}
  \centering
\begin{tabular}{r | c | c | c | c | c | c }
\hline \hline
\textbf{Name}&\textbf{$\left(r/R\right)_{BCZ}$}&\textbf{$\left( m/M \right)_{CZ}$} &\textbf{$Y_{CZ}$}&\textbf{$Z_{CZ}$}&\textbf{$Y_{0}$}&\textbf{$Z_{0}$}\\ \hline
AGSS09-OPAL& $0.7224$&$0.9785$&$0.2363$&$0.01361$&$0.2664$& $0.01511$\\
AGSS09-OPLIB & $0.7205$&$0.9777$&$0.2300$&$0.01372$&$0.2588$& $0.01520$\\ 
AGSS09-OPAS & $0.7196$&$0.9779$&$0.2322$&$0.01368$&$0.2614$& $0.01516$\\
AGSS09-OPAL-Paquette & $0.7235$&$0.9788$&$0.2373$&$0.01359$&$0.2648$& $0.01480$\\
GS98-OPAL & $0.7157$&$0.9764$&$0.2465$&$0.01706$&$0.2765$& $0.01887$\\
AGSS09Ne-OPAL & $0.7207$&$0.9780$&$0.2373$&$0.01393$&$0.2655$& $0.01547$\\
AGSS09-OPAL-PartIon & $0.7240$&$0.9790$&$0.2378$&$0.01355$&$0.2690$& $0.01524$\\
AGSS09-OPAL-OvAd & $0.7207$&$0.9780$&$0.2372$&$0.01356$&$0.2666$& $0.01514$\\
AGSS09-OPAL-DT & $0.7230$&$0.9786$&$0.2375$&$0.01355$&$0.2666$& $0.01514$\\
AGSS09-OPAL-Proffitt & $0.7244$&$0.9790$&$0.2411$&$0.01349$&$0.2650$& $0.01486$\\
\hline
\end{tabular}
\end{table*}

The only AGSS09 model to show a significant improvement in the helium abundance is the one using the \citet{Proffitt} parametrization of turbulent diffusion. However, this is made at the expense of a larger disagreement of the position of the base of the convective zone. The model including the revised neon abundance, for example, does not significantly improve the helium abundance problem while it reduced the discrepancies observed in structural inversions. All other modifications lead to somewhat similar conclusions, with neither the macroscopic mixing nor the modifications to microscopic diffusion implying a decisive improvement of the models. However, the similarities between the parameters of the models including partial ionization in the computation of microscopic diffuson and those including macroscopic mixing illustrate the importance of combining the structural inversions, since they could differentiate between both effects. 

In Fig. \ref{fig:ratiosSTD}, we compare the frequency separation ratios of theoretical models including revised abundance and opacity tables. In previous papers, the good agreement between the frequency separation ratios of the GS98 and the solar data was considered as a strong argument against the revised abundances. We see in Fig. \ref{fig:ratiosSTD} that a similar agreement can be obtained by using the OPLIB opacities in AGSS09 models and that the neon revision also provided a significant improvement of the agreement. This results from the fact that the frequency separation ratios are sensitive to the sound-speed derivative. Hence, they are sensitive to both the temperature and chemical composition gradients and not only to the chemical composition of the solar radiative layers. 

Consequently, the frequency separation ratios cannot be used as a direct constraint on the solar chemical composition. However, they certainly provide some additional information to dissect the current solar modelling problem. For example, the fact that the model built with the OPAS opacity tables, while it provided a quite good improvement in the squared sound speed inversion, demonstrates that there is a clear issue. Similarly, since the AGSS09 OPLIB model reproduces quite well the ratios implies that the gradient of the ratio of temperature over mean molecular weight must be quite close to the solar one, but clearly fails at reproducing the mean molecular weight itself, since the helium abundance in the convective envelope is far too low. 

\subsection{Modified Solar Models}\label{Sec:ModifiedModels}

In addition to the models presented in the previous sections, we also carried out inversions for models built using a modified profile of the mean Rosseland opacity and taking into account the recent revision of the neon abundance. The modification is implemented as a combination of a polynomial and a Gaussian peaked around $\log T=6.35$. The general behaviour of the considered alteration of the opacity profile is motivated by the current discussions in the opacity community regarding uncertainties in conditions similar to those of the base of the solar convective envelope. These models also include an additional macroscopic mixing of the chemical elements at the base of the convective zone in the form of either turbulent diffusion or overshoot. 

\begin{figure}[h!]
\begin{center}
\includegraphics[width=9cm]{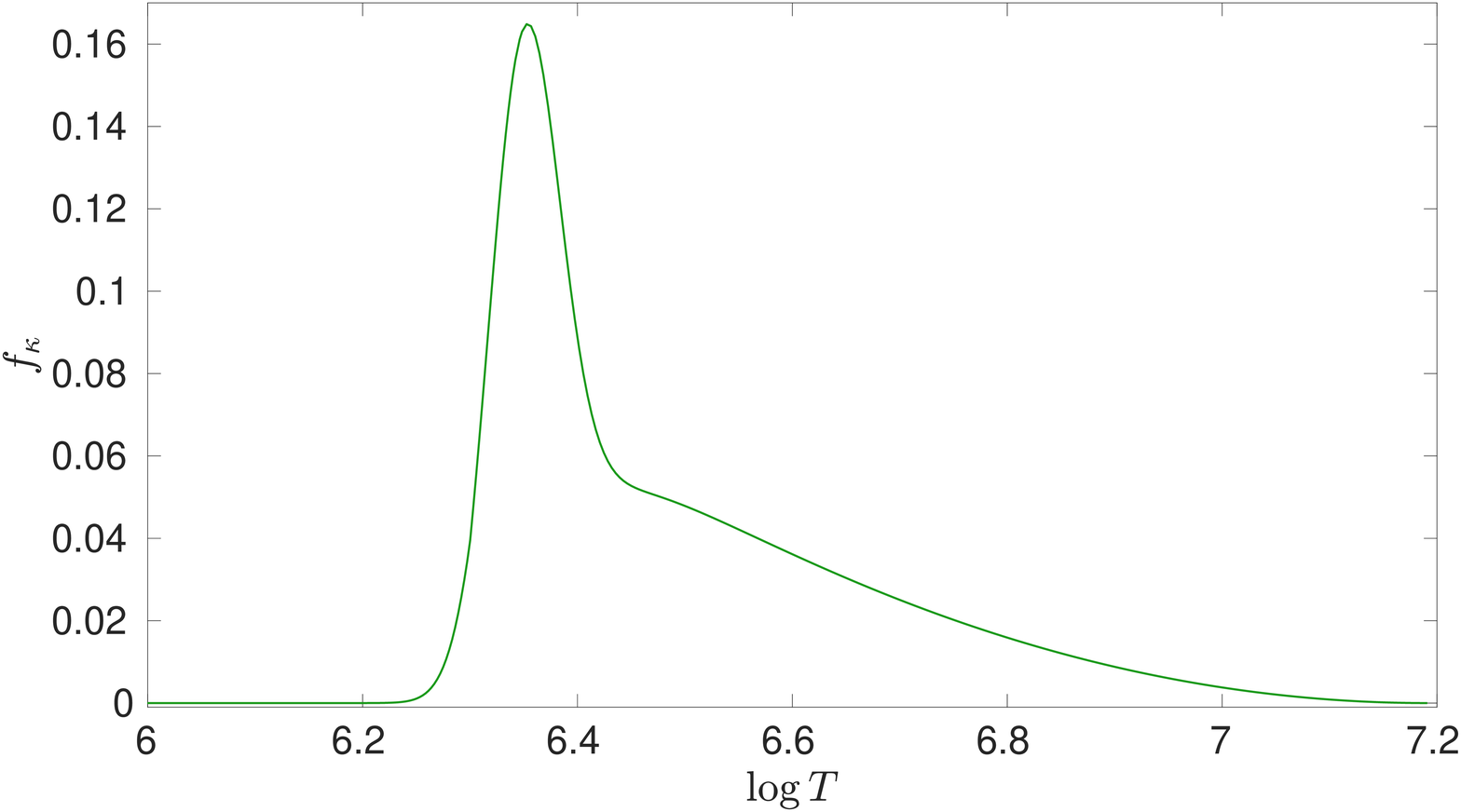}
\end{center}
\caption{Modification to the opacity profile used in the solar models denoted as `$Poly$'. $f_{\kappa}(T)$ is the increase in relative opacity applied during the evolution.}\label{fig:kappaModif}
\end{figure}

The opacity modification is implemented as a multiplicative factor to the mean Rosseland opacity 
\begin{align}
\kappa^{'} = \left( 1+f_{\kappa}(T) \right) \kappa ,
\end{align}
with $\kappa$ the original value of the mean Rosseland opacity, $\kappa^{'}$ the modified value and $f_{\kappa}(T)$ the parametric function considered. An illustration of $f_{\kappa}(T)$ is provided in Fig. \ref{fig:kappaModif}, the modification is cut at lower temperatures than those of the position of the base of the convective zone, as these regions will not affect the solar modelling problem. However, in stars other than the Sun, modifications can also be expected in other regimes and their amplitude might be higher than what is found in the solar case. As can be seen, most of the alteration is localized below the base of the solar convective zone, and the order of magnitude is similar to the value given by \citet{Zhao} and \citet{Pradhan}\footnote{A. Pradhan, private communication.} whereas at higher temperatures, the modification quickly drops to values of the same order of magnitude as the various standard opacity tables. From a physical point of view, the sharp decrease in opacity uncertainties at higher temperatures due to the higher ionization state of the various chemical elements and the reduced contribution of photon absorption to the total opacity budget. In our study, the opacity modification is applied throughout the evolution and each of these ``corrected'' models is recalibrated individually.

We considered models built with the AGSS09 abundance tables, including the corrected neon abundance, the OPAL opacity tables and the SAHA-S equation of state. The motivation behind the use of the SAHA-S EOS was to include the most recent version of an EOS relying on the chemical picture. All these models have been built with the idea of seeing how well the agreement with all the seismic diagnostics could be improved and what we could learn about the degeneracies of the solar modelling problem. A more extended study can be found in \citet{Buldgen2018} where we have investigated various modifications to the opacity profile using various standard opacity tables as a starting point. Similar studies using modified models can also be found in \citet{Montalban06}, \citet{JCD2009}, \citet{JCD2010}, \citet{Ayukov2011}, \citet{Ayukov2017} and \citet{JCD18}.

\begin{table*}[t]
\caption{Physical ingredients of the solar models with modified opacities and additional mixing used in this study}
\label{tabParamMixModels}
  \centering
\begin{tabular}{r | c | c | c | c | c }
\hline \hline
\textbf{Name}&\textbf{EOS}&\textbf{Opacity}&\textbf{Abundances} & \textbf{Diffusion} & \textbf{Convection}\\ \hline
AGSS09Ne-Poly & SAHA-S & OPAL+Poly & AGSS09Ne & Thoul & MLT\\
AGSS09Ne-Poly-DT & SAHA-S & OPAL+Poly & AGSS09Ne & Thoul+$D_{Turb}$ & MLT\\
AGSS09Ne-Poly-Prof & SAHA-S & OPAL+Poly & AGSS09Ne & Thoul+$D_{Turb}-\mathrm{Prof}$ & MLT\\
AGSS09Ne-Poly-Rad & SAHA-S & OPAL+Poly & AGSS09Ne &  Thoul+$\mathrm{Ov}-\mathrm{Rad}$ $(0.3H_{P})$& MLT\\
AGSS09Ne-Poly-Ad & SAHA-S & OPAL+Poly & AGSS09Ne & Thoul+$\mathrm{Ov}-\mathrm{Ad}$ $(0.3H_{P})$ & MLT\\
\hline
\end{tabular}
\end{table*}

\begin{figure}[h!]
\begin{center}
\includegraphics[width=17cm]{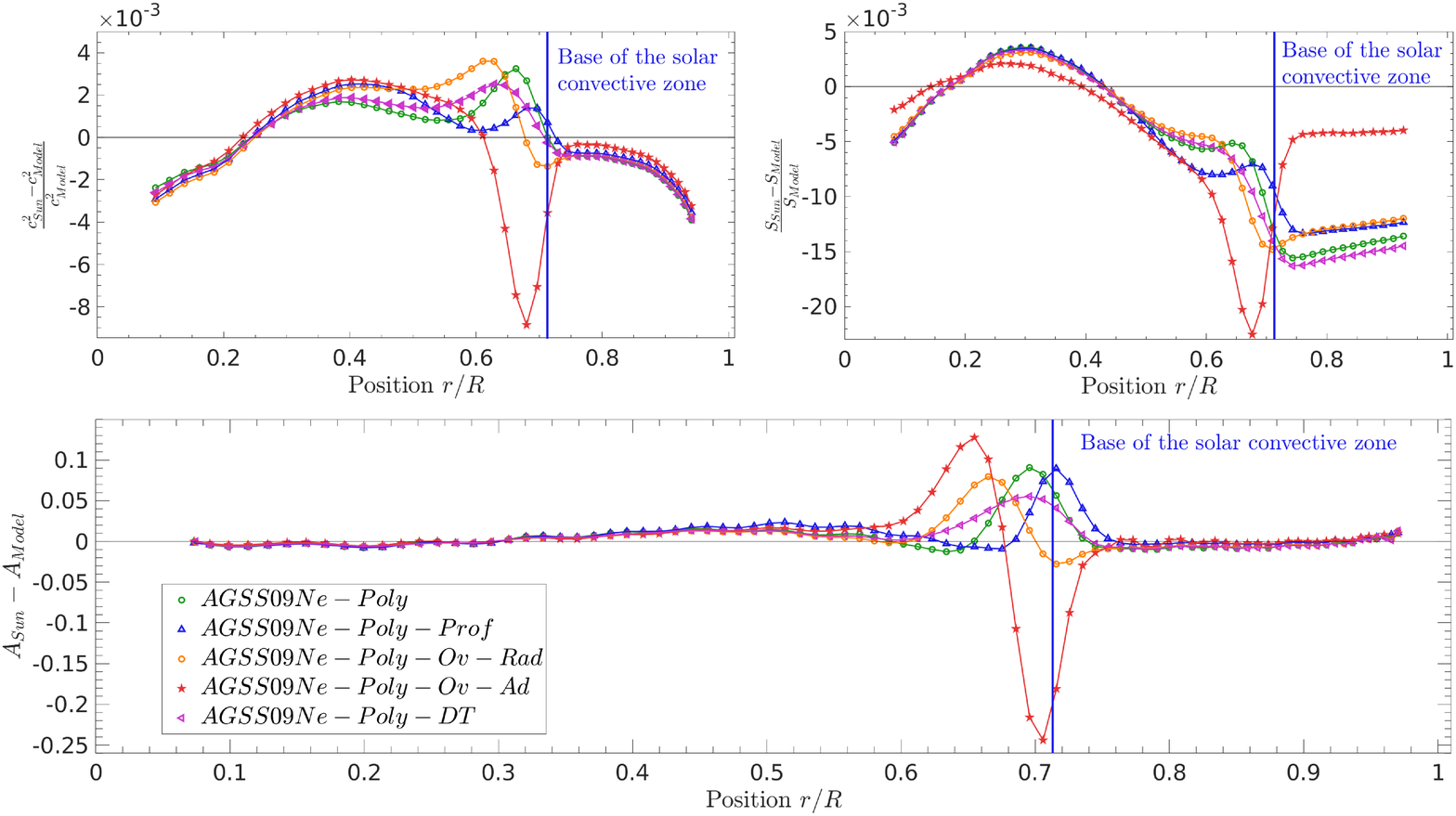}
\end{center}
\caption{Upper-left panel: squared adiabatic sound speed inversions for the solar models including a modified mean Rosseland opacity and additional macroscopic mixing. Upper-right: entropy proxy inversions for the solar models including a modified mean Rosseland opacity and additional macroscopic mixing. Lower panel: Ledoux discriminant inversions for the solar models including a modified mean Rosseland opacity and additional macroscopic mixing. The error bars have the same amplitude as for the standard models but were left out to ease of readibility of the figure.}\label{fig:OpacPolyInv}
\end{figure}

\begin{figure}[h!]
\begin{center}
\includegraphics[width=13cm]{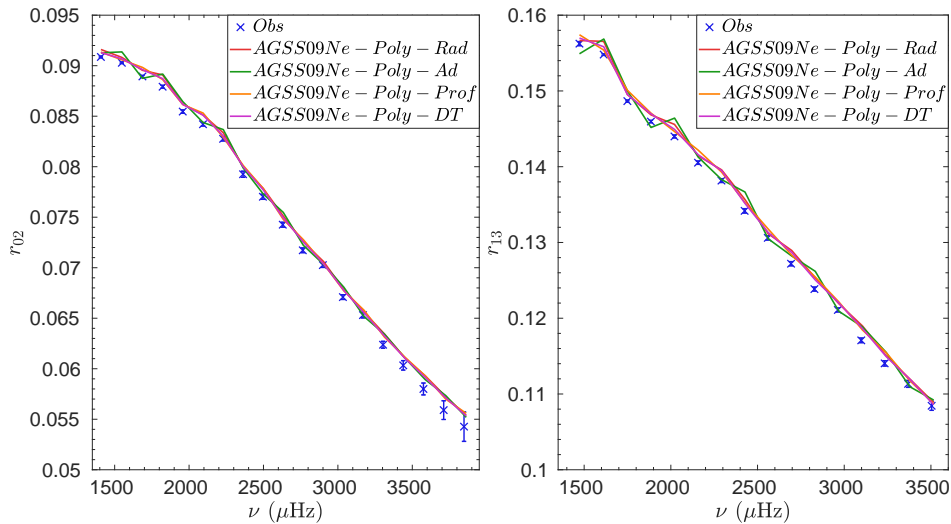}
\end{center}
\caption{Comparison between the observed frequency separation ratios and those of the solar models including modified opacity tables and additional macroscopic mixing.}\label{fig:RatiosOpacPoly}
\end{figure}

We illustrate in Fig. \ref{fig:OpacPolyInv} the results of the $c^{2}$, $S_{5/3}$ and $A$ inversions for these various modified models. In Fig. \ref{fig:RatiosOpacPoly}, we compare the frequency separation ratios of these models to those of BiSON data and in table \ref{tabMixModels}, we give the values of various parameters of these solar models of direct interest for helioseismology. We have used the following naming convention for the additional mixing at the base of the convective envelope: 'AGSS09Ne-Poly-DT' denotes a model where we used equation \ref{eqDTURB} with the values of $50$ and $2$ for the $D$ and $N$ coefficients respectively, whereas 'AGSS09Ne-Poly-Prof' denotes the use of the values $7500$ and $3$ for these coefficients. 'AGSS09Ne-Poly-Rad' denotes the uses of a step overshoot function of $0.3H_{P}$ using the radiative temperature gradient in the overshooting region and an instantaneous mixing of the chemical elements whereas 'AGSS09Ne-Poly-Ad' denotes the uses of the same step overshoot function but fixing the temperature gradient to the adiabatic gradient. 

\begin{table*}[t]
\caption{Parameters of the solar models with modified opacities and additional mixing used in this study}
\label{tabMixModels}
  \centering
\begin{tabular}{r | c | c | c | c | c | c }
\hline \hline
\textbf{Name}&\textbf{$\left(r/R\right)_{BCZ}$}&\textbf{$\left( m/M \right)_{CZ}$} &\textbf{$Y_{CZ}$}&\textbf{$Z_{CZ}$}&\textbf{$Y_{0}$}&\textbf{$Z_{0}$}\\ \hline
AGSS09Ne-Poly & $0.7122$&$0.9757$&$0.2416$&$0.01385$&$0.2692$& $0.01494$ \\
AGSS09Ne-Poly-DT & $0.7106$&$0.9762$&$0.2425$&$0.01383$& $0.2685$ &$0.01466$ \\
AGSS09Ne-Poly-Prof & $0.7121$&$0.9756$&$0.2460$&$0.01376$& $0.2696$ &$0.01500$ \\
AGSS09Ne-Poly-Rad & $0.7118$&$0.9757$&$0.2437$&$0.01381$& $0.2692$&$0.01495$ \\
AGSS09Ne-Poly-Ad & $0.6871$&$0.9757$&$0.2438$&$0.01381$& $0.2700$&$0.01506$ \\
\hline
\end{tabular}
\end{table*}

As can be seen from table \ref{tabMixModels}, the parameters of these models are in much better agreement with helioseismology. For nearly all models, the position of the base of the convective zone is in near perfect agreement with helioseismic constraints. The only exception being the model including adiabatic overshoot, which leads to an extension of the base of the convective zone far beyond what is expected from helioseismology and generates an glitch in the sound speed profile due to the too steep position of the transition in temperature gradients. This is perfectly illustrated in the $c^{2}$ and $A$ profiles (in red in Fig. \ref{fig:OpacPolyInv}) which show large deviations in the transition region.

All models also present a significant increase in the helium abundance in the convective zone. This is a direct consequence of the extended region over which the opacity is increased, which leads to a reduced initial hydrogen abundance and thus a higher initial helium abundance. However, the values still remain slightly lower than the helioseismic value\footnote{We consider here a conservative interval between $0.245$ and $0.26$ in agreement with recent studies \citep{Vorontsov13}.}, implying that, while the base of the convective zone is placed at the right position, something is still amiss in the solar models. This is confirmed by a closer analysis of the inversion results and the frequency separation ratios. 

From the upper-left panel of Fig. \ref{fig:OpacPolyInv}, we can see that the sound speed profile is in very good agreement with helioseismic results. However, considering the amount of ingredients that have been fine-tuned, the presence of very significant deviations below $0.6$ solar radii indicate that something still needs to be corrected in these models. Similarly, the height of the entropy plateau is still off by more than $1\%$ for all models, except the model including adiabatic overshoot which strongly disagrees with the base of the convective zone. Such disagreements give weight to the hypothesis that, beyond corrections to the radiative opacities, the modelling of the transition in temperature gradient at the base of the convective region will have a strong impact on the inversion results. This is a well-known fact, which has been analyzed and discussed by \citet{JCDOV,JCD18}. However, the current results show in a new way the potential of helioseismic data to provide very stringent constraints on the solar structure. 

The Ledoux discriminant inversions also illustrate the full potential of these diagnostics. In the lower panel of Fig. \ref{fig:OpacPolyInv}, we can see that the inversions clearly show the different behaviours of various types of chemical mixing unlike the results illustrated in Fig. \ref{fig:ASTD}. The reason why these different behaviours were not visible in Fig. \ref{fig:ASTD} is due to the fact that the mixed region was actually compared to a region that is fully mixed in the Sun. It seems that any type of mixing could provide a significant improvement in the Ledoux discriminant, especially the adiabatic overshooting. In the modified models, since the fully mixed region in the models more closely resembles that of the Sun, the Ledoux discriminant is far more useful in disentangling the various types of macroscopic mixing occuring below the transition in temperature gradient. 

In Fig \ref{fig:RatiosOpacPoly}, we compare the values of the frequency separation ratios from our modified theoretical models to observations. From a comparison between Figs. \ref{fig:ratiosSTD} and \ref{fig:RatiosOpacPoly}, we can see that the opacity modification induces a slight improvement in comparison with the standard AGSS09Ne model, especially at lower frequencies. However, it appears that the model built with the GS98 abundances still performs better. This also advocates for change in opacity over a wider range of temperature, which could be linked to a revision of the equation of state used in the opacity computations in such regimes. In practice, opacity computations are expected to be more robust at higher temperatures, as less transitions come into play. The equation of state used by different groups to compute the tables may explain some of the differences\footnote{This statement is however, difficult to assess, as the equation of state used for opacity computations is often not available, at least for the recent OPAS and OPLIB tables.}. However, the large differences observed in the \citet{Bailey} experiment results may also be linked to other issues in current opacity computations \citep{Nahar,Krief2016,Pain2019}. 

Indeed, changes of even a few percent at higher temperatures could significantly affect the frequency separation ratios, as well as the agreement with the helioseismic helium abundance in the convective envelope. Modifications of such amplitude are within the uncertainties of the opacity tables \citep{Guzik05,Guzik06} and thus do not imply significant revisions of the physics in opacity computations, unlike the modifications required at the base of the solar convective zone. Amongst the modified models, the addition of macroscopic mixing of the chemical elements does not have a significant impact on the ratios, with the exception of the adiabatic overshoot, which adds an oscillatory signal due to the large mismatch in the position of the base of the convective zone in this model. 

\begin{figure}[h!]
\begin{center}
\includegraphics[width=12cm]{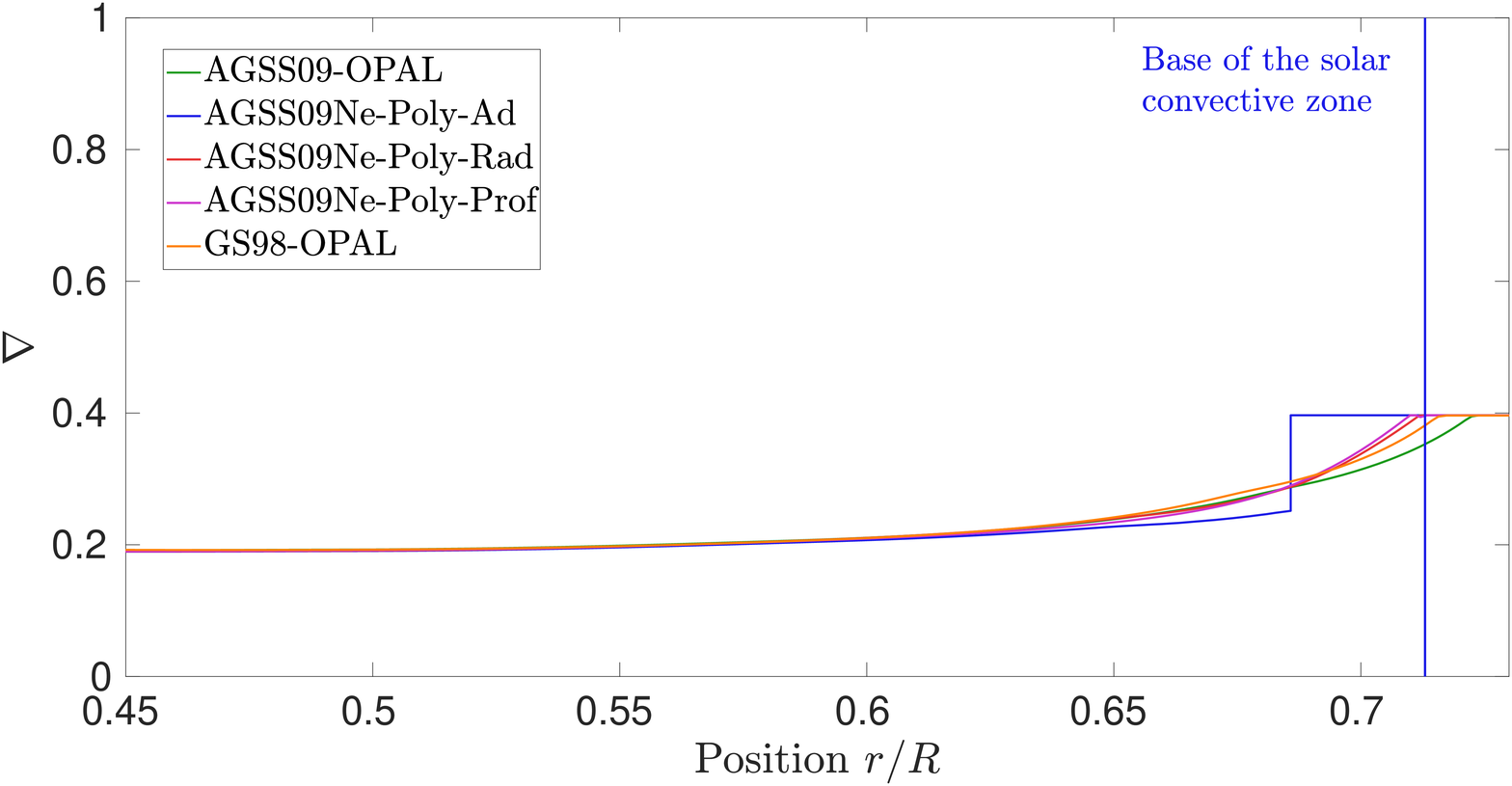}
\end{center}
\caption{Gradient of the natural logarithm of temperature with respect to the natural logarithm of pressure for the modified solar models considered in this study.}\label{fig:NablaTPoly}
\end{figure}

In Fig. \ref{fig:NablaTPoly}, we illustrate the gradient of the natural logarithm of temperature with respect to the natural logarithm of pressure $(\nabla = \frac{d \ln T}{d \ln P})$ for the various modified solar models considered in our study. As can be seen, the combination of both opacity modifications and chemical mixing allows to place the base of the convective zone in very good agreement with helioseismology, with the exception of course of the adiabatic overshoot. One can also note the slight differences in temperature gradient near the base of the convective zone for the various mixing. As the mean molecular weight gradients will be also very different, it can also be easily understood why the Ledoux discriminant inversions offer a great opportunity to probe chemical mixing just below the base of the convective zone. In Sect. \ref{sec:AppendixOne} , we discuss how the Ledoux discriminant can be separated in its chemical and thermal components and how additional insights could be gained from these inversions. It also appears that the temperature gradient quickly follows a very similar behaviour at $0.5$ solar radii for all models, as expected from the small amplitude of the considered opacity modifications at higher temperatures. These small shifts are however of constant sign over the whole radiative layers and thus still impact the initial hydrogen abundance of the calibrated model and its present-day helium abundance in the convective zone. 
\section{Impact of solar model modifications on the 16Cyg binary system}\label{Sec:Cygni}
\subsection{From global helioseismology to asteroseismology}\label{Sec:CygIntro}
With the advent of the CoRoT \citep{Baglin} and \textit{Kepler} missions \citep{Borucki}, asteroseismology of solar-like oscillators has become the golden path to characterize other stars than the Sun. Today, asteroseismic modelling is considered a standard tool to derive precise values of stellar fundamental parameters of stars, namely mass, radius and age which are of particular interest for fields such as exoplanetology and Galactic archaeology. While the high precision of these determinations is undisputable, as they result from the high precision of the seismic data, their accuracy will of course depend on the actual accuracy of the underlying stellar evolution models. 

Consequently, efforts have recently been made to quantify the impact of physical ingredients on the determination of these fundamental parameters. In parallel, the wealth of seismic and non-seismic data led to the development of sophisticated modelling tools \citep{Bazot2012, Gruberbauer, Rendle} and new analyses techniques \citep[see for example][]{Verma2014,Roxburgh2016,Farnir}. Most notably, the advent of space-based photometric data allowed the extension of seismic inversion techniques to other targets than the Sun. The use of these methods had been discussed with artificial data in a few pioneering works \citep[see for example][]{Gough1993,Gough1993b,Roxburgh1996,Roxburgh2002}. 

From a seismic point of view, the first obvious targets to attempt seismic inversions are low-mass main-sequence solar-like oscillating stars observed by \textit{Kepler} during the whole duration of the nominal mission. These stars have been assembled in a single catalog, called the \textit{Kepler} LEGACY sample \citep{Lund,Silva2017}. Amongst these stars, the most constrained targets are the components of the $16$Cyg binary systems. In addition to high-quality seismic data, interferometric, photometric and spectrocopic constraints are also available, providing an unprecedented dataset for such solar twins. Various studies have been dedicated to their modelling using forward and inverse approaches.

Given its extensive datasets, the $16$Cyg binary systems offers an excellent opportunity to test the ingredients of stellar models to a degree of sophistication similar to helioseismic investigations. From a physical point of view, one can consider the targets of the \textit{Kepler} LEGACY sample as additional experimental points to understand the solar modelling problem. In this section, we will carry out the academic exercise of considering the impact of the solar modelling problem on the seismic constraints of the $16$Cyg binary system. 

In our exercise, we computed different sets of models for the $16$Cyg binary systems using the same initial conditions, summarized in table \ref{tab16CygMODEL}, but various physical ingredients. These values have been taken from a preliminary modelling of the $16$Cyg binary system presented in \citet{Farnir}\footnote{Namely, the results are only presented for the A component in table $2$, second column of \citet{Farnir}, but the modelling was carried out for both components.}. The first set of models is composed of standard models of both stars built using the AGSS09 abundances, the FreeEOS equation of state, the OPAL opacities, following the diffusion formalism of \citet{Thoul}, the classical mixing-length theory of convection and using an Eddington grey atmosphere. First, we test opacity modifications, considering that a re-investigation of the $16$Cyg binary system would be required should updated opacity tables be made available\footnote{\citet{Buldgen2016B} also investigated the use of the OPAS opacities in their modelling, but this should be done more thoroughly.}. Hence, we computed a second set of models of $16$CygA$\&$B that includes the polynomial opacity modification that we considered for model ``AGSS09Ne-Poly'' of Sect. \ref{Sec:ModifiedModels}, represented in figure \ref{fig:kappaModif}. Finally, the third set of models considers both this opacity modification and the parametric macroscopic mixing of \citet{Proffitt} as in model ``AGSS09Ne-Poly-Prof'' of Sect. \ref{Sec:ModifiedModels}. The properties of the models of the various sets are summarized in table \ref{tab16Cyg}. The models have been calibrated by evolving them until they reach the radius values determined using interferometry \citep{White2013MNRAS}. Hence, each has a different age, as a result of the differences in their physical ingredients. All ages are however consistent between the components of the binary system. 

\begin{table*}[t]
\caption{Parameters of the $16$Cyg models with modified opacities and additional mixing used in this study}
\label{tab16CygMODEL}
  \centering
\begin{tabular}{r | c | c }
\hline \hline
&\textbf{$16$Cyg A}&\textbf{$16$Cyg B}\\ \hline
$M$ $(M_{\odot})$ & $1.06$ & $1.02$\\
$R$ $(R_{\odot})$ & $1.22$ & $1.12$\\
$X_{0}$ & $0.700$ & $0.700$\\
$Z_{0}$ & $0.022$ & $0.022$\\
$\alpha_{MLT}$ & $1.82$ & $1.82$\\
\hline
\end{tabular}
\end{table*}

The goal of this exercise is fairly simple: to illustrate the impact a potential solution to the solar modelling problem could have on the determination of stellar fundamental parameters such as mass, radius and age for stars other than the Sun. Hence, it also serves the purpose of reminding the model-dependence of asteroseismic investigations, but also how the high-quality asteroseismic data can help us better understand the physical processes acting inside stars by providing other experimental conditions to those at hand in helioseismology. A good illustration of the limitations of stellar models, and thus a central point for which a revision of their ingredients could have a significant impact is found for example in the current discussions related to the transport of angular momentum on both the main-sequence and the red giant branch \citep{Eggenberger2017,Benomar,Ouazzani18,Eggenberger2019}. In those cases, the sensitivity of the proposed mechanisms to the chemical composition gradients, and thus their validity, could be influenced by revisions of some of the physical ingredients at play when studying the solar modelling problem.

\subsection{Impact of the solar problem on classical seismic indices}\label{Sec:AsteroClassic}

Forward seismic modelling techniques can vary quite extensively depending on the quality of the data. The crudest approach uses the so-called scaling laws to infer stellar properties, whereas the most sophisticated techniques use various combinations of the individual frequencies. Amongst them, the use of the $r_{0,2}$ and $r_{1,3}$ ratios defined by \citet{RoxburghRatios} and defined in Eqs. \ref{eqratios1} and \ref{eqratios2} allow to infer the internal structure without too much dependency on the upper layers. Indeed, the direct use of the individual frequencies is not optimal for solar-like oscillators, as they are strongly influenced by the ``surface effect'' problem and lead to unrealistic precisions on stellar parameters\footnote{Indeed, early studies already discussed the fact that individual frequencies did not constitue independent constraints on stellar structure and should not be used directly as inputs of stellar forward modelling.}. 

\begin{figure}[h!]
\begin{center}
\includegraphics[width=17cm]{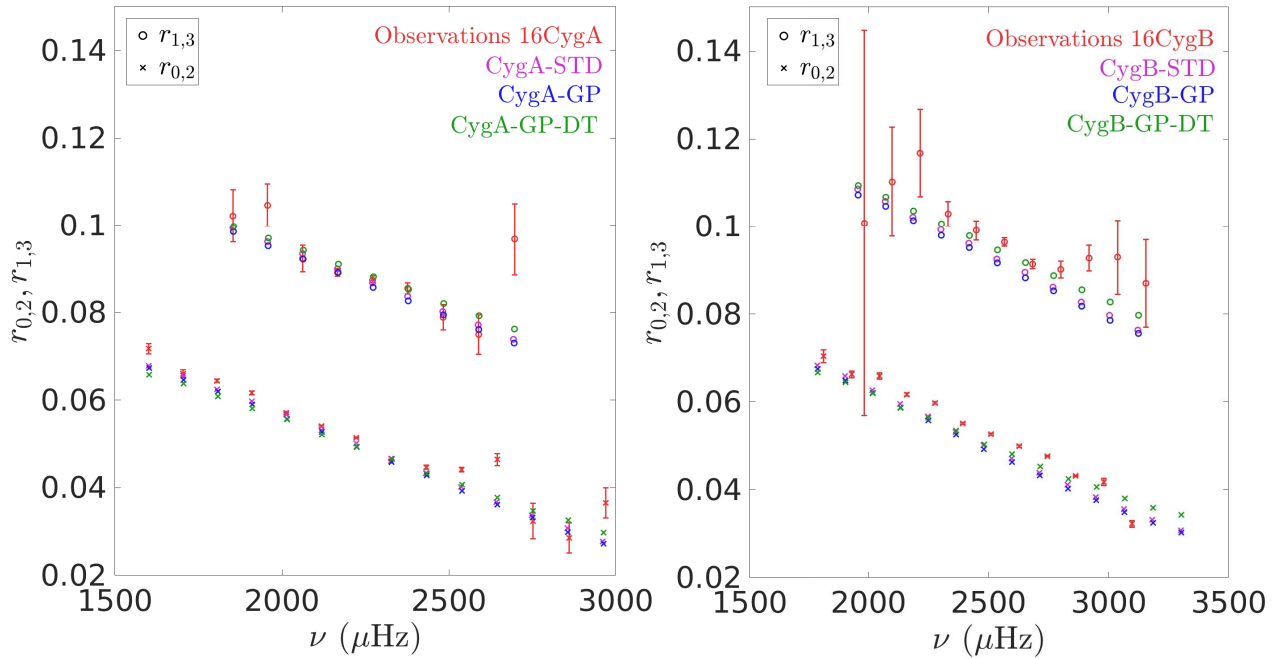}
\end{center}
\caption{Comparison between the observed frequency ratios and those of the $16$Cyg A and B models considered in this study.}\label{fig:ratCyg}
\end{figure}

In figure \ref{fig:ratCyg}, we compare the frequency separation ratios of the various models with respect to the observations. As can be seen, none of the models fit very well the seismic data at hand, although the agreement is not catastrophic either. This is not surprising, as they have not been fitted to the individual frequency ratios. However, what is more striking is that the variations between the theoretical models that are induced by the opacity modifications and the additional turbulent mixing is significant with respect to the observational uncertainties for most of the data. This is not really surprising but emphasizes the model-dependence of seismic modelling results. As expected, the variations observed here are also reflected in the fundamental parameters. However, this behaviour should be inspected for a given fit of the seismic constraints to be certain that the accuracy of the inferences is so significantly affected.
\begin{table*}[t]
\caption{Parameters of the $16$Cyg models with modified opacities and additional mixing used in this study}
\label{tab16Cyg}
  \centering
\begin{tabular}{r | c | c | c | c | c}
\hline \hline
\textbf{Name}&\textbf{$\left(r/R\right)_{BCZ}$}&\textbf{$\left( m/M \right)_{CZ}$} &\textbf{$Y_{CZ}$}&\textbf{$Z_{CZ}$}& \textbf{Age}\\ \hline
CygA-Std & $0.7070$&$0.9793$&$0.2300$&$0.01863$& $7.18$ Gy\\
CygB-Std & $0.6979$&$0.9739$&$0.2332$&$0.01887$& $7.48$ Gy\\
CygA-Poly & $0.7000$&$0.9771$&$0.2312$&$0.01876$& $7.30$ Gy\\
CygB-Poly & $0.6.850$&$0.9691$&$0.2348$&$0.01898$& $7.61$Gy\\
CygA-Poly-Prof & $0.7000$&$0.9765$&$0.2454$&$0.01950$& $7.48$ Gy\\
CygB-Poly-Prof & $0.6860$&$0.9699$&$0.2438$&$0.01958$& $7.76$ Gy\\
\hline
\end{tabular}
\end{table*}

Unsurprisingly, the most affected parameter is the age, for which variations between the standard models and the models including opacity modifications and additional mixing can reach around $4\%$, whereas including only the opacity modifications only induces a variation of almost $2\%$. These variations are of course quite small. However, they are as large as the uncertainties on the fundamental parameters derived from seismic modelling studies using the whole \textit{Kepler} dataset \citep[see for example][]{Metcalfe2015ApJ,Silva2017} and are of course only indicative of the impact of the changes of given ingredients for a given set of initial conditions for the evolutionary models. Taking into account the uncertainties on the other fundamental parameters for a given set of constraints could lead to a larger spread in age. 

In \citet{Buldgen2016A}, we demonstrated that a similar spread in age could also be seen by altering the efficiency of microscopic diffusion. Hence, we can state that modifying the formalism of microscopic diffusion, using the \citet{Paquette} approach and considering partial ionization when computing microscopic diffusion would also cause a change in age of the order of one per cent. Consequently, we can confirm that for the current best \textit{Kepler} targets \citep{Borucki}, as well as for future TESS and PLATO targets \citep{PLATO, Ricker}, the main contributors to the fundamental parameters will not be the propagation of the observational uncertainties onto the inferred parameters, but the physical ingredients of the underlying grids of evolutionary models. In such a context, very high precision results, for example as those of \citet{Metcalfe2015ApJ} or \citep{Buldgen2016A,Buldgen2016B} for the $16$Cyg binary system, should only be taken as valid for a given set of physical ingredients. 

\begin{figure}[h!]
\begin{center}
\includegraphics[width=17cm]{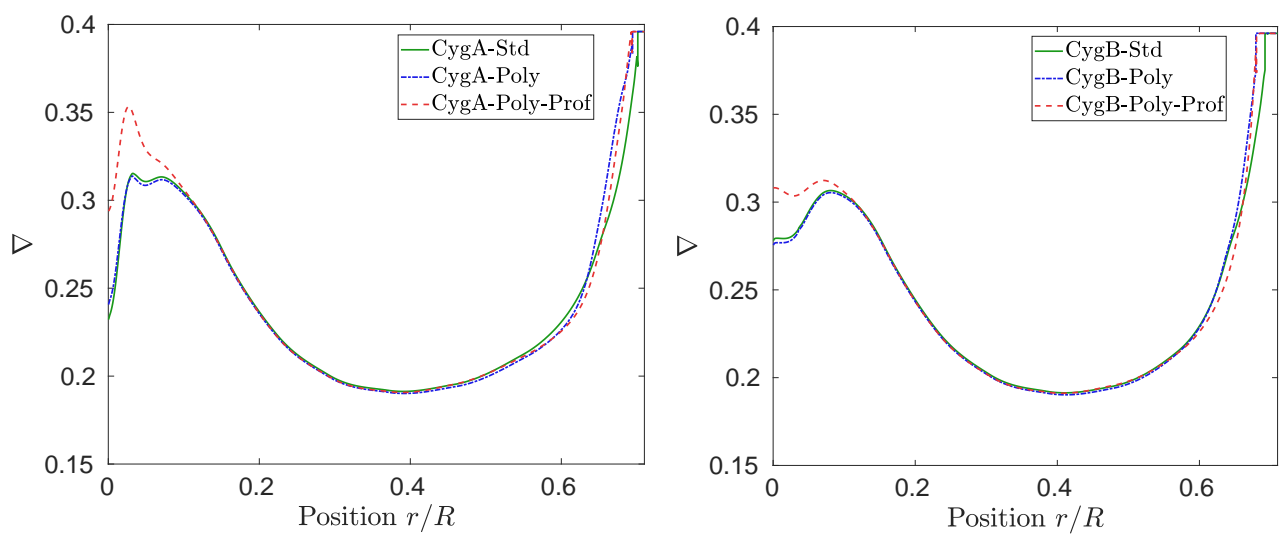}
\end{center}
\caption{Temperature gradient profiles as of function of $r/R$ for the $16$Cyg A and B models considered in this study.}\label{fig:NablaCyg}
\end{figure}

To illustrate some of the differences between the various models considered here, we show in Fig. \ref{fig:NablaCyg} the temperature gradients inside the models. As can be seen, the modification of the mean Rosseland opacity does not induce any significant variations in the deep layers. The main variation is unsurprisingly located at the base of the convective zone. A small modification is also seen in luminosity but well below the observational error bars. Interestingly, the inclusion of turbulent diffusion has altered the deep layers of the model. The variations are actually due to a change of the hydrogen abundance in the central layers. Indeed, $X_{C}$ is $0.038$ in the model including turbulent diffusion and $0.029$ in the standard model. This could be due to the inhibition of microscopic diffusion that is induced by the turbulent mixing.

\begin{figure}[h!]
\begin{center}
\includegraphics[width=17cm]{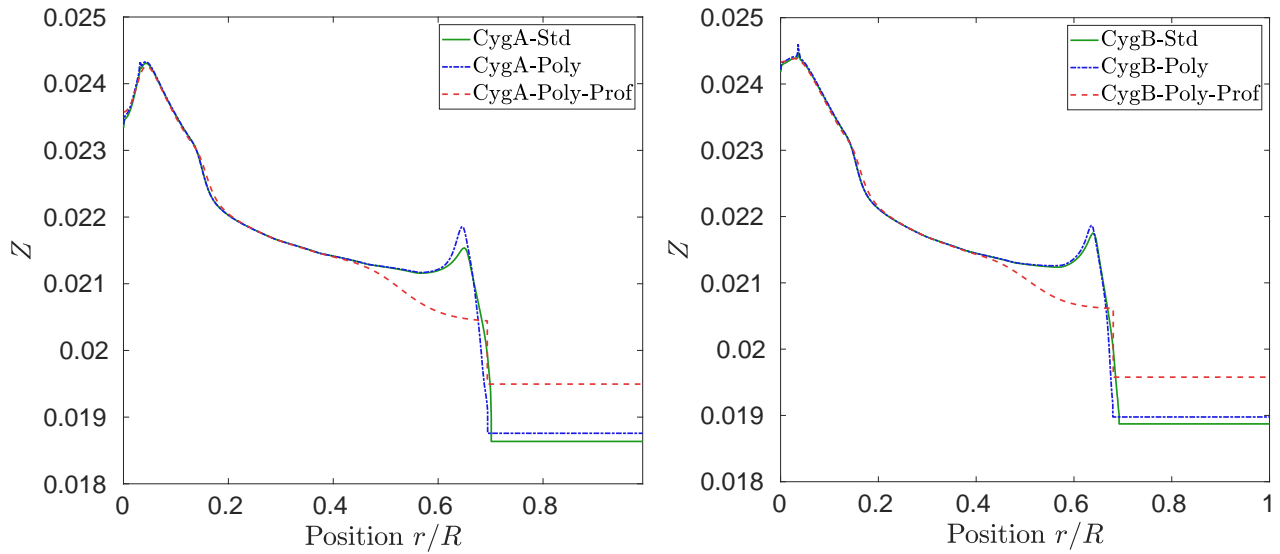}
\end{center}
\caption{Metallicity profiles as of function of $r/R$ for the $16$Cyg A and B models considered in this study.}\label{fig:ZCyg}
\end{figure}

The extreme impact of the \citet{Proffitt} parametric approach to turbulent diffusion can also be seen in Fig. \ref{fig:ZCyg}, where we illustrate the metallicity profile of the models of the $16$Cyg binary system. One can see the influence of turbulent diffusion on the surface abundance of metals that is significantly higher than in the standard model and the model including modified opacity. It is also interesting to note the slight differences between these two models. We emphasize here that there is no modification to the mixing of chemicals. However, there is an indirect impact of the opacity modification on chemicals through the modification of the position of the base of the convective envelope. Here, the higher opacity leads to a larger convective envelope, extending at higher temperatures. This implies that microscopic diffusion will be slightly less efficient and thus, that a higher metallicity will be found in the envelope but also just below the envelope, where metals tend to accumulate over the duration of the evolution. This is particularly well seen in the left panel of Fig. \ref{fig:ZCyg} in the case of $16$CygA.

Overall, the modifications we see in the models of both components remain quite small. They could, however, be modified by the seismic optimization procedure which will alter the initial conditions of the evolutionary sequence. Therefore, some variations seen in the models assuming the same initial parameters but different approaches for the mixing of the chemical elements might be erased at the expense of a change of fundamental parameters such as mass, radius and age, as was noted in \citet{Buldgen2016B}. On this matter, the case of $16$Cyg is particularly interesting and promising, as both stars form a binary system. This adds another level of constraint on their initial composition and their age, further reducing the amplitude of the changes one can make to the models. 

\subsection{Impact of the solar problem on indicator inversions}\label{Sec:IndicInversions}

In addition to classical seismic forward modelling, \citet{Reese2012}, \citet{Buldgen2015tau}, \citet{Buldgen2015tu} and \citet{Buldgen2018} developed inversions of so-called structural indicators, defined as integrated quantities, which can offer additional constraints beyond the use of classical seismic indices. In this section, we briefly discuss the potential variations in these seismic indicators that can be expected from the modifications of the physical ingredients of the $16$Cyg binary system models. It should however be noted that these results are preliminary and that the true diagnostic potential of the inversions might be further improved. For example, the use of non-linear inversions, following the formalism of \citet{Roxburgh2003Vor} may provide an excellent complement, less sensitive to surface effects, to the classical formalism used in global helioseismology. 

Here, we limit ourselves to a brief discussion on the diagnostic potential of structural indicators, namely the $t_{u}$ indicator from \citet{Buldgen2015tu}, defined as
\begin{align}
t_{u}=\int_{0}^{R}f(r) \left(\frac{du}{dr}\right)^{2}dr,
\end{align} 
with $u=P/\rho$, $R$ the stellar radius and $f(r)$ a suitably chosen parametric weight function \citep[see][for details]{Buldgen2015tu}. 

\begin{figure}[h!]
\begin{center}
\includegraphics[width=14cm]{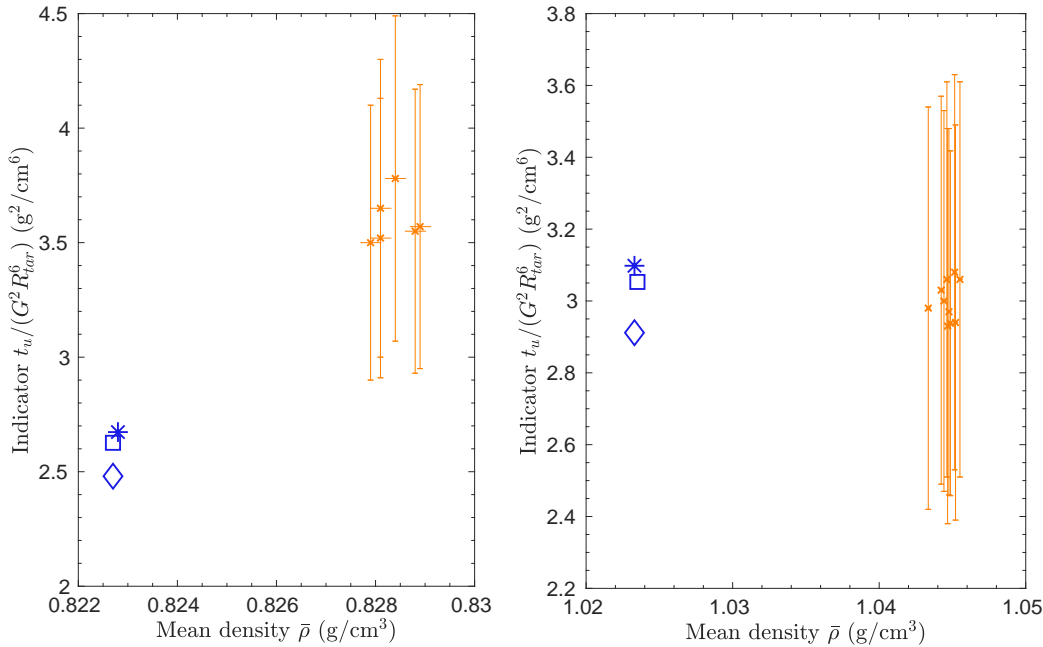}
\end{center}
\caption{Comparison of $t_{u}$ and $\bar{\rho}$ inversions for $16$CygA and B for the models considered in this study. The blue symbols refer to the reference models: $*$ for the standard models, $\square$ for the models with modified opacity and $\lozenge$ for the models including both the opacity modification and turbulent diffusion.}\label{fig:tuCyg}
\end{figure}

The impact of the changes in physics on the $t_{u}$ indicator  can be seen in Fig \ref{fig:tuCyg}, where we recreate the figures from \citet{Buldgen2016A} and \citet{Buldgen2016B} presenting the inversion results. As can be seen, the variations are quite small compared to the uncertainties of the inversions. However, the variations of the fundamental parameters will strongly affect the values of the structural indicators. As the $t_{u}$ indicator scales with $M^{2}$, its value is also strongly dependent on the mass and radii inferred from the forward modelling procedure. Including a small mixing at the base of the envelope of $16$CygB, \citet{Buldgen2016B} found a variation of the $t_{u}$ indicator between $12\%$ and $20\%$ between some models. However, it is very unlikely that this variation only results from changes in the structure of the models, but rather stems also from inaccuracies in the stellar fundamental parameters. This is illustrated in figure \ref{fig:tuCyg}, since the the models are built using the same mass and evolved until they have the same radius. From these tests, we can see that the maximum variations in $t_{u}$ at a given mass and radius are of approximately $6\%$; this is well below the uncertainties of the $t_{u}$ indicator that can reach values around $16\%$. On a sidenote, it also appears that the mean density value is not well reproduced for $16$CygB, with a difference of around $2\%$ between the reference models and the inverted value. The $t_{u}$ value seems however to be in very good agreement with the inversions, but it remains to be seen whether one can obtain a good agreement for all inverted quantities. In addition to the $t_{u}$ and $\bar{\rho}$ inversions, additional indicators presented in \citet{Buldgen2018} can be used to constrain the internal structure of the star. Moreover, taking into account the lithium and beryllium abundances, whenever measured, are also key additional observations to accurately depict the evolution of solar-like stars \citep{Deal2015, Thevenin17}. In the case of the $16$Cyg binary system, the lithium abundance for both stars has been determined \citep{King1997,Tucci2014}, finding the lithium abundance to be more depleted in the B component by approximately a factor $5$. \citet{Deal2015} suggested that those differences could have been provoked by an accretion of planetary matter on $16$CygB which would have triggered thermohaline mixing. Following these results, the differences observed in both the indicator inversions and the lithium abundance motivated the study of \citet{Buldgen2016B} who assessed the impact of extra-mixing on the inversion results. However, since various physical ingredients could affect the inversion result, \citet{Buldgen2016B} concluded that the problem might be degenerate and required a careful re-study. Moreover, it is still unclear whether the process leading to the far more significant lithium depletion in the B component would still leave a mark on its present-day structure. 

Results for some \textit{Kepler} LEGACY stars have already been presented in \citet{BuldgenLegacy}, showing the diagnostic potential of these additional indicators. However, \citet{Appourchaux15} demonstrated the potential of the method of \citet{Roxburgh2003Vor} by providing an inversion of the whole hydrostatic structure of a \textit{Kepler} target. There is no doubt that the use of such an approach on the targets of the \textit{Kepler} LEGACY sample in the Gaia era will provide invaluable information for stellar modellers, allowing to test with unprecedented thoroughness our depiction of stellar structure in a much more model-independent way than what is achievable with linear asteroseismic inversions. 

\section{Prospects and discussion}\label{Sec:Discussion}

In the previous sections, we presented the current state of the solar modelling problem, with a strong emphasis on helioseismic diagnostics and their capabilities. While it is obvious that global helioseismology is an essential tool to probe the internal structure of the Sun, it does not imply that other fields cannot also reshape the picture of the current solar issue. As mentioned earlier, constraints provided by helioseismic inversions are somewhat degenerate. Indeed, they do not give direct constraints on the temperature gradients inside the Sun as they also probe variables related to a combination of temperature and mean molecular weight. 

For example, measurements of neutrinos fluxes also provide stringent complementary constraints on the temperature of the most central regions of the solar core, probing a zone inaccessible to global helioseismology. Recent simultaneous measurements of all neutrinos of the pp-chain \citep{Borexino} provide a very complete picture of the solar core. In the future, measurements of the CNO neutrinos could provide more direct constraints on the chemical composition of the solar core \citep[see][for a recent discussion]{Gough2019}, in particular, its oxygen abundance, offering strong constraints on the chemical mixing during the evolution of the Sun. However, it is also clear that the neutrino fluxes measurements could also be significantly affected by a revision of the electronic screening formulas used in stellar models \citep{Mussack2011}, as mentioned by \citet{Vinyoles}. 

Similarly, the solar lithium and beryllium abundances also play a key role in understanding the evolution of the Sun \citep{Richard96Sun, Piau2001}. They are closely linked to the intensity and the extent of the mixing at the base of the convective zone, thus constraining the physical processes that can be at play in this narrow region. As such, a key point for the future of solar modelling is understanding the nature and impact of so-called ``non-standard'' processes often treated using ad-hoc prescriptions. The stakes of the solar modelling problem are not so much to validate a value of the solar metallicity, but to trigger the development of new generations of stellar models. 

In this perspective, the advent of space-based photometry missions and the rapid development of asteroseismology offers an unprecedented opportunity for stellar modellers. Today, we can use seismology to precisely probe the interior of thousands of stars, providing stellar modellers with additional experimental measurements to refine their understanding of the theory of stellar structure and evolution, from the microscopic scales of nuclear reactions and radiative transfer to the large scale of turbulent hydrodynamical motions. 

In a provocative way, one could state that stellar physics is far from being reduced to an optimization problem and that the main concerns of asteroseismic modellers should not be on providing extremely precise stellar fundamental parameters. Indeed, those will always be model-dependent. Thus, their precision will always be overestimated, as the systematic differences that can result from inaccurate physical ingredients are difficult to estimate. On the contrary, stellar and solar seismologists should focus on the quality and relevance of their inferences and the connection between their data and the actual physical constraints that they contain. With this mindset, asteroseismology will truly fulfill its role of complementary, ``experimental'' domain of theoretical stellar physics. 

Of course, the progress of stellar physics will require a strong effort on the modelling side. Improvements of the physical ingredients of stellar models are the keys to the solution of the solar modelling problem. From a macrophysical point of view for example, the development of hydrodynamical simulations also offers great potential for our understanding of turbulence in stellar conditions \citep[see for example][for an application in the solar case]{Jorgensen2018}. Linking these simulations to a formalism that can be used in stellar evolution codes is one of the key challenges of the coming years, especially for our depiction of the evolution of convective cores. From a microphysical point of view, further improvements of radiative opacities, microscopic diffusion, or the equation of state will also lead to revolutions in the field and will certainly play a key role for the solar problem. 

A first step in this direction is to compare various evolution codes to separate the numerical contributions to the uncertainties to those that clearly result from physical inaccuracies. This approach, although time-consuming and not very rewarding, is also crucial to motivate further developments and improvements of stellar evolution codes from a numerical point of view. In that respect, it is of course pointless to claim the superiority of one code over others, as much as it is useless to use them as blackboxes. It is clear that the numerical development of some codes has been focused on implementing thoroughly specific aspects (e.g. rotation, magnetic instabilities and internal gravity waves for the GENEC code \citep{Eggenberger2008GENEC}, microscopic diffusion including a complete treatment of radiative acceleration for the Montréal-Montpellier code \citep{Turcotte, Richer2000, Richard2001} and the Toulouse-Geneva evolution code  \citep{Theado12}, or the consistent evolution of convective boundary and quality of the models for seismology of the Liège code \citep{ScuflaireCles}). 

\section{Conclusion}\label{Sec:Conclusion}

This paper has focused on providing a brief review of the solar modelling problem, mainly from a helioseismic perspective. We have discussed in Sect. \ref{Sec:solarproblem} the various contributors to the current issue. Unsurprisingly, the opacity remains the usual suspect and probably the most significant contributor to the disagreements between standard solar models and helioseismic constraints. Beside the opacities, the mixing of the chemical element and the equation of state are the other usual suspects who could have a significant impact on the solar structure. In Sect. \ref{Sec:SolarPMS}, we also briefly presented some additional processes that could impact the present-day solar structure and thus the current discrepancies. While they are not commonly presented in the litterature, they should perhaps not be totally dismissed.

In Sect. \ref{Sec:Inversions}, we presented inversion results, frequency separation ratios and convective envelope properties of a sample of solar models built with various physical ingredients. The constraining nature of combining this entire set of information into one consistent study is very clear, as it allows to isolate the effect of the various contributors to the solar problem. We showed the impact of an extended modification of the mean Rosseland opacity, for which the largest amplitude of the correction lay in the conditions of the iron opacity peak at $log T=6.35$. We also show that the increased neon abundance found by \citet{Landi} and \citet{Young} significantly reduces the discrepancies of the low-metallicity solar models. In addition, we show that the combined inversions could provide stringent constraints on the type of mixing at the base of the solar convective zone. Further extensions of this study using a non-linear inversion technique and/or the phase shift of the mode frequencies to properly reproduce the transition in temperature gradients will provide key constraints for the physical implementation of overshooting at the base of stellar envelopes. 

In Sect. \ref{Sec:Cygni}, we have briefly discussed the impact of the solar modelling problem on the structure of the best \textit{Kepler} targets, the components of the $16$Cyg binary system. To do so, we have computed models with a given set of physical ingredients using the standard solar model framework, including the opacity modification we used for our modified solar models or including both the opacity modification and turbulent diffusion using the parametric approach of \citet{Proffitt}. We demonstrated that the impact of such modifications would be significant at the level of precision required from asteroseismic investigations. In that respect, improving the current seismic inference techniques is crucial to better exploit the constraints on these uncertain processes and, by providing more stringent analyses, to ensure the success of future space missions such as PLATO and to bring theoretical stellar physics to a new level of accuracy. Improving indicator inversions but also generalizing the use of non-linear inversions can be foreseen as the most promising way to fully exploit the data. However, new approaches to treat the seismic information in forward modelling methods also provide important insights in the limitations of seismic information and define the necessary reference models for seismic inversions \citep{Farnir}. 

Ultimately, the extension of such advanced modelling strategies further away from the solar conditions will allow to truly probe the limitations of the current state of theoretical stellar physics. In conclusion, the future of asteroseismology is deeply rooted in its history and the early developments of helioseismology. From these solid grounds, asteroseismologists can further develop this young and successful research field. This requires to solve the solar modelling problem, to promote synergies between stellar physicists, seismic modellers and experts in hydrodynamical simulations without perhaps falling into the trap of a race to precision of stellar parameters that are intrisically model-dependent. 

\section*{Conflict of Interest Statement}

The authors declare that the research was conducted in the absence of any commercial or financial relationships that could be construed as a potential conflict of interest.

\section*{Author Contributions}

The Author Contributions section is mandatory for all articles, including articles by sole authors. If an appropriate statement is not provided on submission, a standard one will be inserted during the production process. The Author Contributions statement must describe the contributions of individual authors referred to by their initials and, in doing so, all authors agree to be accountable for the content of the work. Please see  \href{http://home.frontiersin.org/about/author-guidelines#AuthorandContributors}{here} for full authorship criteria.

\section*{Funding}
G.B. acknowledges support from the ERC Consolidator Grant funding scheme ({\em project ASTEROCHRONOMETRY}, G.A. n. 772293). This work is sponsored the Swiss National Science Foundation (project number 200020-172505). S.J.A.J.S. is funded by the Wallonia-Brussels Federation ARC grant for Concerted Research Actions.

\bibliographystyle{frontiersinSCNS_ENG_HUMS} 
\bibliography{ArticleFrontiers}

\begin{thebibliography}{260}
\providecommand{\natexlab}[1]{#1}
\expandafter\ifx\csname urlstyle\endcsname\relax
  \providecommand{\doi}[1]{doi:\discretionary{}{}{}#1}\else
  \providecommand{\doi}{doi:\discretionary{}{}{}\begingroup
  \urlstyle{rm}\Url}\fi
\providecommand{\selectlanguage}[1]{\relax}
\providecommand{\bibAnnoteFile}[1]{%
  \IfFileExists{#1}{\begin{quotation}\noindent\textsc{Key:} #1\\
  \textsc{Annotation:}\ \input{#1}\end{quotation}}{}}
\providecommand{\bibAnnote}[2]{%
  \begin{quotation}\noindent\textsc{Key:} #1\\
  \textsc{Annotation:}\ #2\end{quotation}}

\bibitem[{{Adelberger} et~al.(2011){Adelberger}, {Garc{\'{\i}}a}, {Robertson},
  {Snover}, {Balantekin}, {Heeger} et~al.}]{Adelberger}
{Adelberger}, E.~G., {Garc{\'{\i}}a}, A., {Robertson}, R.~G.~H., {Snover},
  K.~A., {Balantekin}, A.~B., {Heeger}, K., et~al. (2011).
\newblock {Solar fusion cross sections. II. The pp chain and CNO cycles}.
\newblock \emph{Reviews of Modern Physics} 83, 195--246.
\newblock \doi{10.1103/RevModPhys.83.195}
\bibAnnoteFile{Adelberger}

\bibitem[{{Ahmad} et~al.(2002){Ahmad}, {Allen}, {Andersen}, {Anglin}, {Barton},
  {Beier} et~al.}]{Ahmad2002}
{Ahmad}, Q.~R., {Allen}, R.~C., {Andersen}, T.~C., {Anglin}, J.~D., {Barton},
  J.~C., {Beier}, E.~W., et~al. (2002).
\newblock {Direct Evidence for Neutrino Flavor Transformation from
  Neutral-Current Interactions in the Sudbury Neutrino Observatory}.
\newblock \emph{Physical Review Letters} 89, 011301.
\newblock \doi{10.1103/PhysRevLett.89.011301}
\bibAnnoteFile{Ahmad2002}

\bibitem[{{Airapetian} et~al.(2016){Airapetian}, {Glocer}, {Gronoff},
  {H{\'e}brard}, and {Danchi}}]{Airapetian2016}
{Airapetian}, V.~S., {Glocer}, A., {Gronoff}, G., {H{\'e}brard}, E., and
  {Danchi}, W. (2016).
\newblock {Prebiotic chemistry and atmospheric warming of early Earth by an
  active young Sun}.
\newblock \emph{Nature Geoscience} 9, 452--455.
\newblock \doi{10.1038/ngeo2719}
\bibAnnoteFile{Airapetian2016}

\bibitem[{{Alecian} and {LeBlanc}(2002)}]{Alecian2002}
{Alecian}, G. and {LeBlanc}, F. (2002).
\newblock {New approximate formulae for radiative accelerations in stars}.
\newblock \emph{MNRAS} 332, 891--900.
\newblock \doi{10.1046/j.1365-8711.2002.05352.x}
\bibAnnoteFile{Alecian2002}

\bibitem[{{Allende Prieto} et~al.(2001){Allende Prieto}, {Lambert}, and
  {Asplund}}]{Allende2001}
{Allende Prieto}, C., {Lambert}, D.~L., and {Asplund}, M. (2001).
\newblock {The Forbidden Abundance of Oxygen in the Sun}.
\newblock \emph{ApJL} 556, L63--L66.
\newblock \doi{10.1086/322874}
\bibAnnoteFile{Allende2001}

\bibitem[{{Antia} and {Basu}(1994)}]{Antia94}
{Antia}, H.~M. and {Basu}, S. (1994).
\newblock {Nonasymptotic helioseismic inversion for solar structure.}
\newblock \emph{A\&Aps} 107, 421--444
\bibAnnoteFile{Antia94}

\bibitem[{{Antia} and {Basu}(2005)}]{Antia05}
{Antia}, H.~M. and {Basu}, S. (2005).
\newblock {The Discrepancy between Solar Abundances and Helioseismology}.
\newblock \emph{ApJL} 620, L129--L132.
\newblock \doi{10.1086/428652}
\bibAnnoteFile{Antia05}

\bibitem[{{Antia} and {Basu}(2006)}]{Antia2006}
{Antia}, H.~M. and {Basu}, S. (2006).
\newblock {Determining Solar Abundances Using Helioseismology}.
\newblock \emph{ApJ} 644, 1292--1298.
\newblock \doi{10.1086/503707}
\bibAnnoteFile{Antia2006}

\bibitem[{{Appourchaux} et~al.(2015){Appourchaux}, {Antia}, {Ball}, {Creevey},
  {Lebreton}, {Verma} et~al.}]{Appourchaux15}
{Appourchaux}, T., {Antia}, H.~M., {Ball}, W., {Creevey}, O., {Lebreton}, Y.,
  {Verma}, K., et~al. (2015).
\newblock {A seismic and gravitationally bound double star observed by Kepler.
  Implication for the presence of a convective core}.
\newblock \emph{A\&A} 582, A25.
\newblock \doi{10.1051/0004-6361/201526610}
\bibAnnoteFile{Appourchaux15}

\bibitem[{{Appourchaux} et~al.(2010){Appourchaux}, {Belkacem}, {Broomhall},
  {Chaplin}, {Gough}, {Houdek} et~al.}]{Appourchaux2010}
{Appourchaux}, T., {Belkacem}, K., {Broomhall}, A.-M., {Chaplin}, W.~J.,
  {Gough}, D.~O., {Houdek}, G., et~al. (2010).
\newblock {The quest for the solar g modes}.
\newblock \emph{Astron. Astrophys. Rev.} 18, 197--277.
\newblock \doi{10.1007/s00159-009-0027-z}
\bibAnnoteFile{Appourchaux2010}

\bibitem[{{Asplund} et~al.(2005{\natexlab{a}}){Asplund}, {Grevesse}, and
  {Sauval}}]{AGS05}
{Asplund}, M., {Grevesse}, N., and {Sauval}, A.~J. (2005{\natexlab{a}}).
\newblock {The Solar Chemical Composition}.
\newblock In \emph{Cosmic Abundances as Records of Stellar Evolution and
  Nucleosynthesis}, eds. T.~G. {Barnes}, III and F.~N. {Bash}. vol. 336 of
  \emph{Astronomical Society of the Pacific Conference Series}, 25
\bibAnnoteFile{AGS05}

\bibitem[{{Asplund} et~al.(2005{\natexlab{b}}){Asplund}, {Grevesse}, {Sauval},
  {Allende Prieto}, and {Blomme}}]{AGS05C}
{Asplund}, M., {Grevesse}, N., {Sauval}, A.~J., {Allende Prieto}, C., and
  {Blomme}, R. (2005{\natexlab{b}}).
\newblock {Line formation in solar granulation. VI. [C I], C I, CH and C$_{2}$
  lines and the photospheric C abundance}.
\newblock \emph{A\&Ap} 431, 693--705
\bibAnnoteFile{AGS05C}

\bibitem[{{Asplund} et~al.(2004){Asplund}, {Grevesse}, {Sauval}, {Allende
  Prieto}, and {Kiselman}}]{AGS04O}
{Asplund}, M., {Grevesse}, N., {Sauval}, A.~J., {Allende Prieto}, C., and
  {Kiselman}, D. (2004).
\newblock {Line formation in solar granulation. IV. [O I], O I and OH lines and
  the photospheric O abundance}.
\newblock \emph{A\&Ap} 417, 751--768
\bibAnnoteFile{AGS04O}

\bibitem[{{Asplund} et~al.(2009){Asplund}, {Grevesse}, {Sauval}, and
  {Scott}}]{AGSS09}
{Asplund}, M., {Grevesse}, N., {Sauval}, A.~J., and {Scott}, P. (2009).
\newblock {The Chemical Composition of the Sun}.
\newblock \emph{ARA\&A} 47, 481--522
\bibAnnoteFile{AGSS09}

\bibitem[{{Ayukov} and {Baturin}(2011)}]{Ayukov2011}
{Ayukov}, S.~V. and {Baturin}, V.~A. (2011).
\newblock {Low-Z solar model: Sound speed profile under the convection zone}.
\newblock In \emph{Journal of Physics Conference Series}. vol. 271 of
  \emph{Journal of Physics Conference Series}, 012033.
\newblock \doi{10.1088/1742-6596/271/1/012033}
\bibAnnoteFile{Ayukov2011}

\bibitem[{{Ayukov} and {Baturin}(2017)}]{Ayukov2017}
{Ayukov}, S.~V. and {Baturin}, V.~A. (2017).
\newblock {Helioseismic models of the sun with a low heavy element abundance}.
\newblock \emph{Astronomy Reports} 61, 901--913.
\newblock \doi{10.1134/S1063772917100018}
\bibAnnoteFile{Ayukov2017}

\bibitem[{{Badnell} et~al.(2005){Badnell}, {Bautista}, {Butler}, {Delahaye},
  {Mendoza}, {Palmeri} et~al.}]{Badnell}
{Badnell}, N.~R., {Bautista}, M.~A., {Butler}, K., {Delahaye}, F., {Mendoza},
  C., {Palmeri}, P., et~al. (2005).
\newblock {Updated opacities from the Opacity Project}.
\newblock \emph{MNRAS} 360, 458--464.
\newblock \doi{10.1111/j.1365-2966.2005.08991.x}
\bibAnnoteFile{Badnell}

\bibitem[{{Baglin} et~al.(2009){Baglin}, {Auvergne}, {Barge}, {Deleuil},
  {Michel}, and {CoRoT Exoplanet Science Team}}]{Baglin}
{Baglin}, A., {Auvergne}, M., {Barge}, P., {Deleuil}, M., {Michel}, E., and
  {CoRoT Exoplanet Science Team} (2009).
\newblock {CoRoT: Description of the Mission and Early Results}.
\newblock In \emph{Transiting Planets}, eds. F.~{Pont}, D.~{Sasselov}, and
  M.~J. {Holman}. vol. 253 of \emph{IAU Symposium}, 71--81.
\newblock \doi{10.1017/S1743921308026252}
\bibAnnoteFile{Baglin}

\bibitem[{{Bahcall} et~al.(2005{\natexlab{a}}){Bahcall}, {Basu},
  {Pinsonneault}, and {Serenelli}}]{Bahcall2005a}
{Bahcall}, J.~N., {Basu}, S., {Pinsonneault}, M., and {Serenelli}, A.~M.
  (2005{\natexlab{a}}).
\newblock {Helioseismological Implications of Recent Solar Abundance
  Determinations}.
\newblock \emph{ApJ} 618, 1049--1056.
\newblock \doi{10.1086/426070}
\bibAnnoteFile{Bahcall2005a}

\bibitem[{{Bahcall} et~al.(2005{\natexlab{b}}){Bahcall}, {Basu}, and
  {Serenelli}}]{Bahcall05}
{Bahcall}, J.~N., {Basu}, S., and {Serenelli}, A.~M. (2005{\natexlab{b}}).
\newblock {What Is the Neon Abundance of the Sun?}
\newblock \emph{ApJ} 631, 1281--1285.
\newblock \doi{10.1086/431926}
\bibAnnoteFile{Bahcall05}

\bibitem[{{Bahcall} et~al.(1982){Bahcall}, {Huebner}, {Lubow}, {Parker}, and
  {Ulrich}}]{Bahcall82}
{Bahcall}, J.~N., {Huebner}, W.~F., {Lubow}, S.~H., {Parker}, P.~D., and
  {Ulrich}, R.~K. (1982).
\newblock {Standard solar models and the uncertainties in predicted capture
  rates of solar neutrinos}.
\newblock \emph{Reviews of Modern Physics} 54, 767--799.
\newblock \doi{10.1103/RevModPhys.54.767}
\bibAnnoteFile{Bahcall82}

\bibitem[{{Bahcall} and {Pe{\~n}a-Garay}(2004)}]{BahcallNeutrino}
{Bahcall}, J.~N. and {Pe{\~n}a-Garay}, C. (2004).
\newblock {Solar models and solar neutrino oscillations}.
\newblock \emph{New Journal of Physics} 6, 63
\bibAnnoteFile{BahcallNeutrino}

\bibitem[{{Bahcall} et~al.(2005{\natexlab{c}}){Bahcall}, {Serenelli}, and
  {Basu}}]{Bahcall2005b}
{Bahcall}, J.~N., {Serenelli}, A.~M., and {Basu}, S. (2005{\natexlab{c}}).
\newblock {New Solar Opacities, Abundances, Helioseismology, and Neutrino
  Fluxes}.
\newblock \emph{ApJL} 621, L85--L88.
\newblock \doi{10.1086/428929}
\bibAnnoteFile{Bahcall2005b}

\bibitem[{{Bahcall} et~al.(2006){Bahcall}, {Serenelli}, and
  {Basu}}]{Bahcall2006}
{Bahcall}, J.~N., {Serenelli}, A.~M., and {Basu}, S. (2006).
\newblock {10,000 Standard Solar Models: A Monte Carlo Simulation}.
\newblock \emph{ApJS} 165, 400--431.
\newblock \doi{10.1086/504043}
\bibAnnoteFile{Bahcall2006}

\bibitem[{{Bailey} et~al.(2015){Bailey}, {Nagayama}, {Loisel}, {Rochau},
  {Blancard}, {Colgan} et~al.}]{Bailey}
{Bailey}, J.~E., {Nagayama}, T., {Loisel}, G.~P., {Rochau}, G.~A., {Blancard},
  C., {Colgan}, J., et~al. (2015).
\newblock {A higher-than-predicted measurement of iron opacity at solar
  interior temperatures}.
\newblock \emph{Nature} 517, 3
\bibAnnoteFile{Bailey}

\bibitem[{{Basu} and {Antia}(1994)}]{Basu1994}
{Basu}, S. and {Antia}, H.~M. (1994).
\newblock {Effects of Diffusion on the Extent of Overshoot Below the Solar
  Convection Zone}.
\newblock \emph{MNRAS} 269, 1137.
\newblock \doi{10.1093/mnras/269.4.1137}
\bibAnnoteFile{Basu1994}

\bibitem[{{Basu} and {Antia}(1995)}]{BasuYSun}
{Basu}, S. and {Antia}, H.~M. (1995).
\newblock {Helium abundance in the solar envelope}.
\newblock \emph{MNRAS} 276, 1402--1408
\bibAnnoteFile{BasuYSun}

\bibitem[{{Basu} and {Antia}(1997)}]{Basu97BCZ}
{Basu}, S. and {Antia}, H.~M. (1997).
\newblock {Seismic measurement of the depth of the solar convection zone}.
\newblock \emph{MNRAS} 287, 189--198
\bibAnnoteFile{Basu97BCZ}

\bibitem[{{Basu} and {Antia}(2004)}]{Basu2004}
{Basu}, S. and {Antia}, H.~M. (2004).
\newblock {Constraining Solar Abundances Using Helioseismology}.
\newblock \emph{ApJL} 606, L85--L88.
\newblock \doi{10.1086/421110}
\bibAnnoteFile{Basu2004}

\bibitem[{{Basu} and {Antia}(2008)}]{Basu08}
{Basu}, S. and {Antia}, H.~M. (2008).
\newblock {Helioseismology and solar abundances}.
\newblock \emph{Phys. Rep.} 457, 217--283.
\newblock \doi{10.1016/j.physrep.2007.12.002}
\bibAnnoteFile{Basu08}

\bibitem[{{Basu} et~al.(2009){Basu}, {Chaplin}, {Elsworth}, {New}, and
  {Serenelli}}]{BasuSun}
{Basu}, S., {Chaplin}, W.~J., {Elsworth}, Y., {New}, R., and {Serenelli}, A.~M.
  (2009).
\newblock {Fresh Insights on the Structure of the Solar Core}.
\newblock \emph{ApJ} 699, 1403--1417
\bibAnnoteFile{BasuSun}

\bibitem[{{Basu} and {Christensen-Dalsgaard}(1997)}]{Basu1997}
{Basu}, S. and {Christensen-Dalsgaard}, J. (1997).
\newblock {Equation of state and helioseismic inversions.}
\newblock \emph{A\&A} 322, L5--L8
\bibAnnoteFile{Basu1997}

\bibitem[{{Basu} et~al.(1996){Basu}, {Christensen-Dalsgaard}, {Schou},
  {Thompson}, and {Tomczyk}}]{Basu1996}
{Basu}, S., {Christensen-Dalsgaard}, J., {Schou}, J., {Thompson}, M.~J., and
  {Tomczyk}, S. (1996).
\newblock {Solar structure as revealed by 1 year LOWL data}.
\newblock \emph{Bulletin of the Astronomical Society of India} 24, 147
\bibAnnoteFile{Basu1996}

\bibitem[{{Baturin} et~al.(2013){Baturin}, {Ayukov}, {Gryaznov}, {Iosilevskiy},
  {Fortov}, and {Starostin}}]{Baturin}
{Baturin}, V.~A., {Ayukov}, S.~V., {Gryaznov}, V.~K., {Iosilevskiy}, I.~L.,
  {Fortov}, V.~E., and {Starostin}, A.~N. (2013).
\newblock {The Current Version of the SAHA-S Equation of State: Improvement and
  Perspective}.
\newblock In \emph{Progress in Physics of the Sun and Stars: A New Era in
  Helio- and Asteroseismology}, eds. H.~{Shibahashi} and A.~E. {Lynas-Gray}.
  vol. 479 of \emph{Astronomical Society of the Pacific Conference Series}, 11
\bibAnnoteFile{Baturin}

\bibitem[{{Baturin} et~al.(2015){Baturin}, {Gorshkov}, and
  {Oreshina}}]{Baturin2015}
{Baturin}, V.~A., {Gorshkov}, A.~B., and {Oreshina}, A.~V. (2015).
\newblock {Formation of a chemical-composition gradient beneath the convection
  zone and the early evolution of the sun}.
\newblock \emph{Astronomy Reports} 59, 46--57.
\newblock \doi{10.1134/S1063772915010023}
\bibAnnoteFile{Baturin2015}

\bibitem[{{Bazot} et~al.(2012){Bazot}, {Bourguignon}, and
  {Christensen-Dalsgaard}}]{Bazot2012}
{Bazot}, M., {Bourguignon}, S., and {Christensen-Dalsgaard}, J. (2012).
\newblock {A Bayesian approach to the modelling of {$\alpha$} Cen A}.
\newblock \emph{MNRAS} 427, 1847--1866.
\newblock \doi{10.1111/j.1365-2966.2012.21818.x}
\bibAnnoteFile{Bazot2012}

\bibitem[{{Benomar} et~al.(2018){Benomar}, {Bazot}, {Nielsen}, {Gizon},
  {Sekii}, {Takata} et~al.}]{Benomar}
{Benomar}, O., {Bazot}, M., {Nielsen}, M.~B., {Gizon}, L., {Sekii}, T.,
  {Takata}, M., et~al. (2018).
\newblock {Asteroseismic detection of latitudinal differential rotation in 13
  Sun-like stars}.
\newblock \emph{Science} 361, 1231--1234.
\newblock \doi{10.1126/science.aao6571}
\bibAnnoteFile{Benomar}

\bibitem[{{Blancard} et~al.(2016){Blancard}, {Colgan}, {Coss{\'e}},
  {Faussurier}, {Fontes}, {Gilleron} et~al.}]{Blancard2016}
{Blancard}, C., {Colgan}, J., {Coss{\'e}}, P., {Faussurier}, G., {Fontes},
  C.~J., {Gilleron}, F., et~al. (2016).
\newblock {Comment on ``Large Enhancement in High-Energy Photoionization of Fe
  XVII and Missing Continuum Plasma Opacity''}.
\newblock \emph{Physical Review Letters} 117, 249501.
\newblock \doi{10.1103/PhysRevLett.117.249501}
\bibAnnoteFile{Blancard2016}

\bibitem[{{Blancard} et~al.(2012){Blancard}, {Coss{\'e}}, and
  {Faussurier}}]{BlancardOpacDetail}
{Blancard}, C., {Coss{\'e}}, P., and {Faussurier}, G. (2012).
\newblock {Solar Mixture Opacity Calculations Using Detailed Configuration and
  Level Accounting Treatments}.
\newblock \emph{ApJ} 745, 10
\bibAnnoteFile{BlancardOpacDetail}

\bibitem[{{B{\"o}hm-Vitense}(1958)}]{Bohm}
{B{\"o}hm-Vitense}, E. (1958).
\newblock {{\"U}ber die Wasserstoffkonvektionszone in Sternen verschiedener
  Effektivtemperaturen und Leuchtkr{\"a}fte. Mit 5 Textabbildungen}.
\newblock \emph{Zeitschrift für Astrophysik} 46, 108
\bibAnnoteFile{Bohm}

\bibitem[{{Boothroyd} and {Sackmann}(2003)}]{Boothroyd03}
{Boothroyd}, A.~I. and {Sackmann}, I.-J. (2003).
\newblock {Our Sun. IV. The Standard Model and Helioseismology: Consequences of
  Uncertainties in Input Physics and in Observed Solar Parameters}.
\newblock \emph{ApJ} 583, 1004--1023.
\newblock \doi{10.1086/345407}
\bibAnnoteFile{Boothroyd03}

\bibitem[{{Borexino Collaboration} et~al.(2018){Borexino Collaboration},
  {Agostini}, {Altenm{\"u}ller}, {Appel}, {Atroshchenko}, {Bagdasarian}
  et~al.}]{Borexino}
{Borexino Collaboration}, {Agostini}, M., {Altenm{\"u}ller}, K., {Appel}, S.,
  {Atroshchenko}, V., {Bagdasarian}, Z., et~al. (2018).
\newblock {Comprehensive measurement of pp-chain solar neutrinos}.
\newblock \emph{Nature} 562, 505--510.
\newblock \doi{10.1038/s41586-018-0624-y}
\bibAnnoteFile{Borexino}

\bibitem[{{Borucki} et~al.(2010){Borucki}, {Koch}, {Basri}, {Batalha}, {Brown},
  {Caldwell} et~al.}]{Borucki}
{Borucki}, W.~J., {Koch}, D., {Basri}, G., {Batalha}, N., {Brown}, T.,
  {Caldwell}, D., et~al. (2010).
\newblock {Kepler Planet-Detection Mission: Introduction and First Results}.
\newblock \emph{Science} 327, 977.
\newblock \doi{10.1126/science.1185402}
\bibAnnoteFile{Borucki}

\bibitem[{{Boury} et~al.(1975){Boury}, {Gabriel}, {Noels}, {Scuflaire}, and
  {Ledoux}}]{Boury1975}
{Boury}, A., {Gabriel}, M., {Noels}, A., {Scuflaire}, R., and {Ledoux}, P.
  (1975).
\newblock {Vibrational instability of a 1 solar mass star towards non-radial
  oscillations}.
\newblock \emph{A\&Ap} 41, 279--285
\bibAnnoteFile{Boury1975}

\bibitem[{{Bristow} et~al.(2017){Bristow}, {Haberle}, {Blake}, {Des Marais},
  {Eigenbrode}, {Fair{\'e}n} et~al.}]{Bristow2017}
{Bristow}, T.~F., {Haberle}, R.~M., {Blake}, D.~F., {Des Marais}, D.~J.,
  {Eigenbrode}, J.~L., {Fair{\'e}n}, A.~G., et~al. (2017).
\newblock {Low Hesperian PCO2 constrained from in situ mineralogical analysis
  at Gale Crater, Mars}.
\newblock \emph{Proceedings of the National Academy of Science} 114,
  2166--2170.
\newblock \doi{10.1073/pnas.1616649114}
\bibAnnoteFile{Bristow2017}

\bibitem[{{Brookes} et~al.(1978){Brookes}, {Isaak}, and {van der
  Raay}}]{Brookes1978}
{Brookes}, J.~R., {Isaak}, G.~R., and {van der Raay}, H.~B. (1978).
\newblock {A resonant-scattering solar spectrometer}.
\newblock \emph{MNRAS} 185, 1--18.
\newblock \doi{10.1093/mnras/185.1.1}
\bibAnnoteFile{Brookes1978}

\bibitem[{{Brown} and {Morrow}(1987)}]{BrownRota}
{Brown}, T.~M. and {Morrow}, C.~A. (1987).
\newblock {Observations of solar p-mode rotational splittings}.
\newblock In \emph{The Internal Solar Angular Velocity}, eds. B.~R. {Durney}
  and S.~{Sofia}. vol. 137 of \emph{Astrophysics and Space Science Library},
  7--17.
\newblock \doi{10.1007/978-94-009-3903-5_2}
\bibAnnoteFile{BrownRota}

\bibitem[{{Brun} et~al.(2002){Brun}, {Antia}, {Chitre}, and {Zahn}}]{Brun02}
{Brun}, A.~S., {Antia}, H.~M., {Chitre}, S.~M., and {Zahn}, J.-P. (2002).
\newblock {Seismic tests for solar models with tachocline mixing}.
\newblock \emph{A\&A} 391, 725--739.
\newblock \doi{10.1051/0004-6361:20020837}
\bibAnnoteFile{Brun02}

\bibitem[{{Buldgen} et~al.(2017{\natexlab{a}}){Buldgen}, {Reese}, and
  {Dupret}}]{BuldgenLegacy}
{Buldgen}, G., {Reese}, D., and {Dupret}, M.-A. (2017{\natexlab{a}}).
\newblock {Asteroseismic inversions in the Kepler era: application to the
  Kepler Legacy sample}.
\newblock In \emph{European Physical Journal Web of Conferences}. vol. 160 of
  \emph{European Physical Journal Web of Conferences}, 03005.
\newblock \doi{10.1051/epjconf/201716003005}
\bibAnnoteFile{BuldgenLegacy}

\bibitem[{{Buldgen} et~al.(2015{\natexlab{a}}){Buldgen}, {Reese}, and
  {Dupret}}]{Buldgen2015tu}
{Buldgen}, G., {Reese}, D.~R., and {Dupret}, M.~A. (2015{\natexlab{a}}).
\newblock {Using seismic inversions to obtain an indicator of internal mixing
  processes in main-sequence solar-like stars}.
\newblock \emph{A\&A} 583, A62.
\newblock \doi{10.1051/0004-6361/201526390}
\bibAnnoteFile{Buldgen2015tu}

\bibitem[{{Buldgen} et~al.(2016{\natexlab{a}}){Buldgen}, {Reese}, and
  {Dupret}}]{Buldgen2016A}
{Buldgen}, G., {Reese}, D.~R., and {Dupret}, M.~A. (2016{\natexlab{a}}).
\newblock {Constraints on the structure of 16 Cygni A and 16 Cygni B using
  inversion techniques}.
\newblock \emph{A\&A} 585, A109.
\newblock \doi{10.1051/0004-6361/201527032}
\bibAnnoteFile{Buldgen2016A}

\bibitem[{{Buldgen} et~al.(2017{\natexlab{b}}){Buldgen}, {Reese}, and
  {Dupret}}]{BuldgenKer}
{Buldgen}, G., {Reese}, D.~R., and {Dupret}, M.~A. (2017{\natexlab{b}}).
\newblock {Analysis of the linear approximation of seismic inversions for
  various structural pairs}.
\newblock \emph{A\&A} 598, A21.
\newblock \doi{10.1051/0004-6361/201629485}
\bibAnnoteFile{BuldgenKer}

\bibitem[{{Buldgen} et~al.(2018){Buldgen}, {Reese}, and {Dupret}}]{Buldgen2018}
{Buldgen}, G., {Reese}, D.~R., and {Dupret}, M.~A. (2018).
\newblock {Constraining convective regions with asteroseismic linear structural
  inversions}.
\newblock \emph{A\&A} 609, A95.
\newblock \doi{10.1051/0004-6361/201730693}
\bibAnnoteFile{Buldgen2018}

\bibitem[{{Buldgen} et~al.(2015{\natexlab{b}}){Buldgen}, {Reese}, {Dupret}, and
  {Samadi}}]{Buldgen2015tau}
{Buldgen}, G., {Reese}, D.~R., {Dupret}, M.~A., and {Samadi}, R.
  (2015{\natexlab{b}}).
\newblock {Stellar acoustic radii, mean densities, and ages from seismic
  inversion techniques}.
\newblock \emph{A\&A} 574, A42.
\newblock \doi{10.1051/0004-6361/201424613}
\bibAnnoteFile{Buldgen2015tau}

\bibitem[{{Buldgen} et~al.(2017{\natexlab{c}}){Buldgen}, {Salmon}, {Godart},
  {Noels}, {Scuflaire}, {Dupret} et~al.}]{BuldgenA}
{Buldgen}, G., {Salmon}, S.~J.~A.~J., {Godart}, M., {Noels}, A., {Scuflaire},
  R., {Dupret}, M.~A., et~al. (2017{\natexlab{c}}).
\newblock {Inversions of the Ledoux discriminant: a closer look at the
  tachocline}.
\newblock \emph{MNRAS} 472, L70--L74.
\newblock \doi{10.1093/mnrasl/slx139}
\bibAnnoteFile{BuldgenA}

\bibitem[{{Buldgen} et~al.(2017{\natexlab{d}}){Buldgen}, {Salmon}, {Noels},
  {Scuflaire}, {Dupret}, and {Reese}}]{BuldgenZ}
{Buldgen}, G., {Salmon}, S.~J.~A.~J., {Noels}, A., {Scuflaire}, R., {Dupret},
  M.~A., and {Reese}, D.~R. (2017{\natexlab{d}}).
\newblock {Determining the metallicity of the solar envelope using seismic
  inversion techniques}.
\newblock \emph{MNRAS} 472, 751--764.
\newblock \doi{10.1093/mnras/stx2057}
\bibAnnoteFile{BuldgenZ}

\bibitem[{{Buldgen} et~al.(2017{\natexlab{e}}){Buldgen}, {Salmon}, {Noels},
  {Scuflaire}, {Reese}, {Dupret} et~al.}]{BuldgenS}
{Buldgen}, G., {Salmon}, S.~J.~A.~J., {Noels}, A., {Scuflaire}, R., {Reese},
  D.~R., {Dupret}, M.-A., et~al. (2017{\natexlab{e}}).
\newblock {Seismic inversion of the solar entropy. A case for improving the
  standard solar model}.
\newblock \emph{A\&A} 607, A58.
\newblock \doi{10.1051/0004-6361/201731354}
\bibAnnoteFile{BuldgenS}

\bibitem[{{Buldgen} et~al.(2016{\natexlab{b}}){Buldgen}, {Salmon}, {Reese}, and
  {Dupret}}]{Buldgen2016B}
{Buldgen}, G., {Salmon}, S.~J.~A.~J., {Reese}, D.~R., and {Dupret}, M.~A.
  (2016{\natexlab{b}}).
\newblock {In-depth study of 16CygB using inversion techniques}.
\newblock \emph{A\&A} 596, A73.
\newblock \doi{10.1051/0004-6361/201628773}
\bibAnnoteFile{Buldgen2016B}

\bibitem[{{Burgers}(1969)}]{Burgers1969}
{Burgers}, J.~M. (1969).
\newblock \emph{{Flow Equations for Composite Gases}}
\bibAnnoteFile{Burgers1969}

\bibitem[{{Caffau} et~al.(2011){Caffau}, {Ludwig}, {Steffen}, {Freytag}, and
  {Bonifacio}}]{Caffau}
{Caffau}, E., {Ludwig}, H.-G., {Steffen}, M., {Freytag}, B., and {Bonifacio},
  P. (2011).
\newblock {Solar Chemical Abundances Determined with a CO5BOLD 3D Model
  Atmosphere}.
\newblock \emph{Sol.Phys.} 268, 255--269.
\newblock \doi{10.1007/s11207-010-9541-4}
\bibAnnoteFile{Caffau}

\bibitem[{{Castro} et~al.(2007){Castro}, {Vauclair}, and
  {Richard}}]{Castro2007}
{Castro}, M., {Vauclair}, S., and {Richard}, O. (2007).
\newblock {Low abundances of heavy elements in the solar outer layers:
  comparisons of solar models with helioseismic inversions}.
\newblock \emph{A\&A} 463, 755--758.
\newblock \doi{10.1051/0004-6361:20066327}
\bibAnnoteFile{Castro2007}

\bibitem[{{Chapman} and {Cowling}(1970)}]{Chapman1970}
{Chapman}, S. and {Cowling}, T.~G. (1970).
\newblock \emph{{The mathematical theory of non-uniform gases. an account of
  the kinetic theory of viscosity, thermal conduction and diffusion in gases}}
\bibAnnoteFile{Chapman1970}

\bibitem[{{Charbonnel} and {Talon}(2005)}]{Charbonnel}
{Charbonnel}, C. and {Talon}, S. (2005).
\newblock {Influence of Gravity Waves on the Internal Rotation and Li Abundance
  of Solar-Type Stars}.
\newblock \emph{Science} 309, 2189--2191.
\newblock \doi{10.1126/science.1116849}
\bibAnnoteFile{Charbonnel}

\bibitem[{{Charnay} et~al.(2017){Charnay}, {Le Hir}, {Fluteau}, {Forget}, and
  {Catling}}]{Charnay2017}
{Charnay}, B., {Le Hir}, G., {Fluteau}, F., {Forget}, F., and {Catling}, D.~C.
  (2017).
\newblock {A warm or a cold early Earth? New insights from a 3-D climate-carbon
  model}.
\newblock \emph{Earth and Planetary Science Letters} 474, 97--109.
\newblock \doi{10.1016/j.epsl.2017.06.029}
\bibAnnoteFile{Charnay2017}

\bibitem[{{Christensen-Dalsgaard} and {D\"appen}(1992)}]{CEFF}
{Christensen-Dalsgaard}, J. and {D\"appen}, W. (1992).
\newblock {Solar oscillations and the equation of state}.
\newblock \emph{Astron. Astrophys. Rev.} 4, 267--361.
\newblock \doi{10.1007/BF00872687}
\bibAnnoteFile{CEFF}

\bibitem[{{Christensen-Dalsgaard} et~al.(1996){Christensen-Dalsgaard},
  {D\"appen}, {Ajukov}, {Anderson}, {Antia}, {Basu} et~al.}]{MODELS}
{Christensen-Dalsgaard}, J., {D\"appen}, W., {Ajukov}, S.~V., {Anderson},
  E.~R., {Antia}, H.~M., {Basu}, S., et~al. (1996).
\newblock {The Current State of Solar Modeling}.
\newblock \emph{Science} 272, 1286--1292.
\newblock \doi{10.1126/science.272.5266.1286}
\bibAnnoteFile{MODELS}

\bibitem[{{Christensen-Dalsgaard} et~al.(2009){Christensen-Dalsgaard}, {di
  Mauro}, {Houdek}, and {Pijpers}}]{JCD2009}
{Christensen-Dalsgaard}, J., {di Mauro}, M.~P., {Houdek}, G., and {Pijpers}, F.
  (2009).
\newblock {On the opacity change required to compensate for the revised solar
  composition}.
\newblock \emph{A\&A} 494, 205--208.
\newblock \doi{10.1051/0004-6361:200810170}
\bibAnnoteFile{JCD2009}

\bibitem[{{Christensen-Dalsgaard} et~al.(1974){Christensen-Dalsgaard}, {Dilke},
  and {Gough}}]{JCD1974}
{Christensen-Dalsgaard}, J., {Dilke}, F.~W.~W., and {Gough}, D.~O. (1974).
\newblock {The stability of a solar model to non-radial oscillations}.
\newblock \emph{MNRAS} 169, 429--445.
\newblock \doi{10.1093/mnras/169.3.429}
\bibAnnoteFile{JCD1974}

\bibitem[{{Christensen-Dalsgaard} and {Gough}(1975)}]{JCD1975}
{Christensen-Dalsgaard}, J. and {Gough}, D.~O. (1975).
\newblock {Nonadiabatic nonradial oscillations of a solar model}.
\newblock \emph{Memoires of the Societe Royale des Sciences de Liege} 8,
  309--316
\bibAnnoteFile{JCD1975}

\bibitem[{{Christensen-Dalsgaard} et~al.(2018){Christensen-Dalsgaard}, {Gough},
  and {Knudstrup}}]{JCD18}
{Christensen-Dalsgaard}, J., {Gough}, D.~O., and {Knudstrup}, E. (2018).
\newblock {On the hydrostatic stratification of the solar tachocline}.
\newblock \emph{MNRAS} 477, 3845--3852.
\newblock \doi{10.1093/mnras/sty752}
\bibAnnoteFile{JCD18}

\bibitem[{{Christensen-Dalsgaard} et~al.(1991){Christensen-Dalsgaard}, {Gough},
  and {Thompson}}]{JCD91Conv}
{Christensen-Dalsgaard}, J., {Gough}, D.~O., and {Thompson}, M.~J. (1991).
\newblock {The depth of the solar convection zone}.
\newblock \emph{ApJ} 378, 413--437.
\newblock \doi{10.1086/170441}
\bibAnnoteFile{JCD91Conv}

\bibitem[{{Christensen-Dalsgaard} and {Houdek}(2010)}]{JCD2010}
{Christensen-Dalsgaard}, J. and {Houdek}, G. (2010).
\newblock {Prospects for asteroseismology}.
\newblock \emph{APSS} 328, 51--66.
\newblock \doi{10.1007/s10509-009-0227-z}
\bibAnnoteFile{JCD2010}

\bibitem[{{Christensen-Dalsgaard} et~al.(2011){Christensen-Dalsgaard},
  {Monteiro}, {Rempel}, and {Thompson}}]{JCDOV}
{Christensen-Dalsgaard}, J., {Monteiro}, M.~J.~P.~F.~G., {Rempel}, M., and
  {Thompson}, M.~J. (2011).
\newblock {A more realistic representation of overshoot at the base of the
  solar convective envelope as seen by helioseismology}.
\newblock \emph{MNRAS} 414, 1158--1174.
\newblock \doi{10.1111/j.1365-2966.2011.18460.x}
\bibAnnoteFile{JCDOV}

\bibitem[{{Christensen-Dalsgaard} et~al.(1993){Christensen-Dalsgaard},
  {Proffitt}, and {Thompson}}]{JCD1993Diff}
{Christensen-Dalsgaard}, J., {Proffitt}, C.~R., and {Thompson}, M.~J. (1993).
\newblock {Effects of diffusion on solar models and their oscillation
  frequencies}.
\newblock \emph{ApJL} 403, L75--L78.
\newblock \doi{10.1086/186725}
\bibAnnoteFile{JCD1993Diff}

\bibitem[{{Christensen-Dalsgaard} and {Thompson}(2007)}]{JCD2007}
{Christensen-Dalsgaard}, J. and {Thompson}, M.~J. (2007).
\newblock {Observational results and issues concerning the tachocline}.
\newblock In \emph{The Solar Tachocline}, eds. D.~W. {Hughes}, R.~{Rosner}, and
  N.~O. {Weiss}. 53
\bibAnnoteFile{JCD2007}

\bibitem[{{Colgan} et~al.(2016){Colgan}, {Kilcrease}, {Magee}, {Sherrill},
  {Abdallah}, {Hakel} et~al.}]{Colgan}
{Colgan}, J., {Kilcrease}, D.~P., {Magee}, N.~H., {Sherrill}, M.~E.,
  {Abdallah}, J., Jr., {Hakel}, P., et~al. (2016).
\newblock {A New Generation of Los Alamos Opacity Tables}.
\newblock \emph{ApJ} 817, 116
\bibAnnoteFile{Colgan}

\bibitem[{{Corbard} et~al.(1999){Corbard}, {Blanc-F{\'e}raud}, {Berthomieu},
  and {Provost}}]{corbard99}
{Corbard}, T., {Blanc-F{\'e}raud}, L., {Berthomieu}, G., and {Provost}, J.
  (1999).
\newblock {Non linear regularization for helioseismic inversions. Application
  for the study of the solar tachocline}.
\newblock \emph{A\&A} 344, 696--708
\bibAnnoteFile{corbard99}

\bibitem[{{Cox} and {Giuli}(1968)}]{Cox}
{Cox}, J.~P. and {Giuli}, R.~T. (1968).
\newblock \emph{{Principles of stellar structure }}
\bibAnnoteFile{Cox}

\bibitem[{{D\"appen} et~al.(1988){D\"appen}, {Mihalas}, {Hummer}, and
  {Mihalas}}]{MHDIII}
{D\"appen}, W., {Mihalas}, D., {Hummer}, D.~G., and {Mihalas}, B.~W. (1988).
\newblock {The equation of state for stellar envelopes. III - Thermodynamic
  quantities}.
\newblock \emph{ApJ} 332, 261--270.
\newblock \doi{10.1086/166650}
\bibAnnoteFile{MHDIII}

\bibitem[{{Davies} et~al.(2014){Davies}, {Broomhall}, {Chaplin}, {Elsworth},
  and {Hale}}]{Davies}
{Davies}, G.~R., {Broomhall}, A.~M., {Chaplin}, W.~J., {Elsworth}, Y., and
  {Hale}, S.~J. (2014).
\newblock {Low-frequency, low-degree solar p-mode properties from 22 years of
  Birmingham Solar Oscillations Network data}.
\newblock \emph{MNRAS} 439, 2025--2032.
\newblock \doi{10.1093/mnras/stu080}
\bibAnnoteFile{Davies}

\bibitem[{{Davis} et~al.(1968){Davis}, {Harmer}, and {Hoffman}}]{Davis1968}
{Davis}, R., {Harmer}, D.~S., and {Hoffman}, K.~C. (1968).
\newblock {Search for Neutrinos from the Sun}.
\newblock \emph{Physical Review Letters} 20, 1205--1209.
\newblock \doi{10.1103/PhysRevLett.20.1205}
\bibAnnoteFile{Davis1968}

\bibitem[{{Deal} et~al.(2018){Deal}, {Alecian}, {Lebreton}, {Goupil},
  {Marques}, {LeBlanc} et~al.}]{Deal2018}
{Deal}, M., {Alecian}, G., {Lebreton}, Y., {Goupil}, M.~J., {Marques}, J.~P.,
  {LeBlanc}, F., et~al. (2018).
\newblock {Impacts of radiative accelerations on solar-like oscillating
  main-sequence stars}.
\newblock \emph{A\&A} 618, A10.
\newblock \doi{10.1051/0004-6361/201833361}
\bibAnnoteFile{Deal2018}

\bibitem[{{Deal} et~al.(2015){Deal}, {Richard}, and {Vauclair}}]{Deal2015}
{Deal}, M., {Richard}, O., and {Vauclair}, S. (2015).
\newblock {Accretion of planetary matter and the lithium problem in the 16
  Cygni stellar system}.
\newblock \emph{A\&A} 584, A105.
\newblock \doi{10.1051/0004-6361/201526917}
\bibAnnoteFile{Deal2015}

\bibitem[{{Defouw}(1970)}]{Defouw1970}
{Defouw}, R.~J. (1970).
\newblock {Thermal-Convective Instability}.
\newblock \emph{ApJ} 160, 659.
\newblock \doi{10.1086/150460}
\bibAnnoteFile{Defouw1970}

\bibitem[{{Delahaye} and {Pinsonneault}(2006)}]{Delahaye2006}
{Delahaye}, F. and {Pinsonneault}, M.~H. (2006).
\newblock {The Solar Heavy-Element Abundances. I. Constraints from Stellar
  Interiors}.
\newblock \emph{ApJ} 649, 529--540.
\newblock \doi{10.1086/505260}
\bibAnnoteFile{Delahaye2006}

\bibitem[{{Dilke} and {Gough}(1972)}]{Dilke1972}
{Dilke}, F.~W.~W. and {Gough}, D.~O. (1972).
\newblock {The Solar Spoon}.
\newblock \emph{Nature} 240, 262--294.
\newblock \doi{10.1038/240262a0}
\bibAnnoteFile{Dilke1972}

\bibitem[{{Domingo} et~al.(1995){Domingo}, {Fleck}, and {Poland}}]{Domingo1995}
{Domingo}, V., {Fleck}, B., and {Poland}, A.~I. (1995).
\newblock {The SOHO Mission: an Overview}.
\newblock \emph{Sol.Phys.} 162, 1--37.
\newblock \doi{10.1007/BF00733425}
\bibAnnoteFile{Domingo1995}

\bibitem[{{Dziembowski}(1982)}]{Dziembowski1982}
{Dziembowski}, W. (1982).
\newblock {Nonlinear mode coupling in oscillating stars. I - Second order
  theory of the coherent mode coupling}.
\newblock \emph{Acta Astron.} 32, 147--171
\bibAnnoteFile{Dziembowski1982}

\bibitem[{{Dziembowski}(1983)}]{Dziembowski1983}
{Dziembowski}, W. (1983).
\newblock {Resonant coupling between solar gravity modes}.
\newblock \emph{Sol.Phys.} 82, 259--266.
\newblock \doi{10.1007/BF00145568}
\bibAnnoteFile{Dziembowski1983}

\bibitem[{{Dziembowski} et~al.(1990){Dziembowski}, {Pamyatnykh}, and
  {Sienkiewicz}}]{Dziemboswki90}
{Dziembowski}, W.~A., {Pamyatnykh}, A.~A., and {Sienkiewicz}, R. (1990).
\newblock {Solar model from helioseismology and the neutrino flux problem}.
\newblock \emph{MNRAS} 244, 542--550
\bibAnnoteFile{Dziemboswki90}

\bibitem[{{Eggenberger} et~al.(2019){Eggenberger}, {Deheuvels}, {Miglio},
  {Ekstr{\"o}m}, {Georgy}, {Meynet} et~al.}]{Eggenberger2019}
{Eggenberger}, P., {Deheuvels}, S., {Miglio}, A., {Ekstr{\"o}m}, S., {Georgy},
  C., {Meynet}, G., et~al. (2019).
\newblock {Asteroseismology of evolved stars to constrain the internal
  transport of angular momentum. I. Efficiency of transport during the subgiant
  phase}.
\newblock \emph{A\&A} 621, A66.
\newblock \doi{10.1051/0004-6361/201833447}
\bibAnnoteFile{Eggenberger2019}

\bibitem[{{Eggenberger} et~al.(2017){Eggenberger}, {Lagarde}, {Miglio},
  {Montalb{\'a}n}, {Ekstr{\"o}m}, {Georgy} et~al.}]{Eggenberger2017}
{Eggenberger}, P., {Lagarde}, N., {Miglio}, A., {Montalb{\'a}n}, J.,
  {Ekstr{\"o}m}, S., {Georgy}, C., et~al. (2017).
\newblock {Constraining the efficiency of angular momentum transport with
  asteroseismology of red giants: the effect of stellar mass}.
\newblock \emph{A\&A} 599, A18.
\newblock \doi{10.1051/0004-6361/201629459}
\bibAnnoteFile{Eggenberger2017}

\bibitem[{{Eggenberger} et~al.(2005){Eggenberger}, {Maeder}, and
  {Meynet}}]{Eggenberger}
{Eggenberger}, P., {Maeder}, A., and {Meynet}, G. (2005).
\newblock {Stellar evolution with rotation and magnetic fields. IV. The solar
  rotation profile}.
\newblock \emph{A\&A} 440, L9--L12.
\newblock \doi{10.1051/0004-6361:200500156}
\bibAnnoteFile{Eggenberger}

\bibitem[{{Eggenberger} et~al.(2008){Eggenberger}, {Meynet}, {Maeder},
  {Hirschi}, {Charbonnel}, {Talon} et~al.}]{Eggenberger2008GENEC}
{Eggenberger}, P., {Meynet}, G., {Maeder}, A., {Hirschi}, R., {Charbonnel}, C.,
  {Talon}, S., et~al. (2008).
\newblock {The Geneva stellar evolution code}.
\newblock \emph{Ap\&SS} 316, 43--54.
\newblock \doi{10.1007/s10509-007-9511-y}
\bibAnnoteFile{Eggenberger2008GENEC}

\bibitem[{{Eguchi} et~al.(2003){Eguchi}, {Enomoto}, {Furuno}, {Goldman},
  {Hanada}, {Ikeda} et~al.}]{Eguchi2003}
{Eguchi}, K., {Enomoto}, S., {Furuno}, K., {Goldman}, J., {Hanada}, H.,
  {Ikeda}, H., et~al. (2003).
\newblock {First Results from KamLAND: Evidence for Reactor Antineutrino
  Disappearance}.
\newblock \emph{Physical Review Letters} 90, 021802.
\newblock \doi{10.1103/PhysRevLett.90.021802}
\bibAnnoteFile{Eguchi2003}

\bibitem[{{Elliott}(1996)}]{Elliott1996}
{Elliott}, J.~R. (1996).
\newblock {Equation of state in the solar convection zone and the implications
  of helioseismology}.
\newblock \emph{MNRAS} 280, 1244--1256.
\newblock \doi{10.1093/mnras/280.4.1244}
\bibAnnoteFile{Elliott1996}

\bibitem[{{Elliott} and {Gough}(1999)}]{Elliott99Tacho}
{Elliott}, J.~R. and {Gough}, D.~O. (1999).
\newblock {Calibration of the Thickness of the Solar Tachocline}.
\newblock \emph{ApJ} 516, 475--481
\bibAnnoteFile{Elliott99Tacho}

\bibitem[{{Elsworth} et~al.(1990){Elsworth}, {Howe}, {Isaak}, {McLeod}, and
  {New}}]{Elsworth1990}
{Elsworth}, Y., {Howe}, R., {Isaak}, G.~R., {McLeod}, C.~P., and {New}, R.
  (1990).
\newblock {Evidence from solar seismology against non-standard solar-core
  models}.
\newblock \emph{Nature} 347, 536--539.
\newblock \doi{10.1038/347536a0}
\bibAnnoteFile{Elsworth1990}

\bibitem[{{Farnir} et~al.(2019){Farnir}, {Dupret}, {Salmon}, {Noels}, and
  {Buldgen}}]{Farnir}
{Farnir}, M., {Dupret}, M.-A., {Salmon}, S.~J.~A.~J., {Noels}, A., and
  {Buldgen}, G. (2019).
\newblock {Comprehensive stellar seismic analysis. New method exploiting the
  glitches information in solar-like pulsators}.
\newblock \emph{A\&A} 622, A98.
\newblock \doi{10.1051/0004-6361/201834044}
\bibAnnoteFile{Farnir}

\bibitem[{Ferziger and Kaper(1972)}]{Ferziger1972}
Ferziger, J.~H. and Kaper, H.~G. (1972).
\newblock {\selectlanguage{English}\emph{Mathematical theory of transport
  processes in gases. [By] J. H. Ferziger and H. G. Kaper}} (North-Holland Pub.
  Co Amsterdam)
\bibAnnoteFile{Ferziger1972}

\bibitem[{{Forget} et~al.(2013){Forget}, {Wordsworth}, {Millour}, {Madeleine},
  {Kerber}, {Leconte} et~al.}]{Forget2013}
{Forget}, F., {Wordsworth}, R., {Millour}, E., {Madeleine}, J.-B., {Kerber},
  L., {Leconte}, J., et~al. (2013).
\newblock {3D modelling of the early martian climate under a denser CO$_{2}$
  atmosphere: Temperatures and CO$_{2}$ ice clouds}.
\newblock \emph{Icarus} 222, 81--99.
\newblock \doi{10.1016/j.icarus.2012.10.019}
\bibAnnoteFile{Forget2013}

\bibitem[{{Fossat} et~al.(2017){Fossat}, {Boumier}, {Corbard}, {Provost},
  {Salabert}, {Schmider} et~al.}]{Fossat2017}
{Fossat}, E., {Boumier}, P., {Corbard}, T., {Provost}, J., {Salabert}, D.,
  {Schmider}, F.~X., et~al. (2017).
\newblock {Asymptotic g modes: Evidence for a rapid rotation of the solar
  core}.
\newblock \emph{A\&A} 604, A40.
\newblock \doi{10.1051/0004-6361/201730460}
\bibAnnoteFile{Fossat2017}

\bibitem[{{Fukuda} et~al.(1999){Fukuda}, {Hayakawa}, {Ichihara}, {Inoue},
  {Ishihara}, {Ishino} et~al.}]{Fukuda1999}
{Fukuda}, Y., {Hayakawa}, T., {Ichihara}, E., {Inoue}, K., {Ishihara}, K.,
  {Ishino}, H., et~al. (1999).
\newblock {Constraints on Neutrino Oscillation Parameters from the Measurement
  of Day-Night Solar Neutrino Fluxes at Super-Kamiokande}.
\newblock \emph{Physical Review Letters} 82, 1810--1814.
\newblock \doi{10.1103/PhysRevLett.82.1810}
\bibAnnoteFile{Fukuda1999}

\bibitem[{{Gabriel}(1997)}]{Gabriel97}
{Gabriel}, M. (1997).
\newblock {Influence of heavy element and rotationally induced diffusions on
  the solar models.}
\newblock \emph{A\&A} 327, 771--778
\bibAnnoteFile{Gabriel97}

\bibitem[{{Gabriel} et~al.(1976){Gabriel}, {Noels}, {Scuflaire}, and
  {Boury}}]{Gabriel1976}
{Gabriel}, M., {Noels}, A., {Scuflaire}, R., and {Boury}, A. (1976).
\newblock {On the evolution of a one-solar-mass star with a periodically mixed
  core}.
\newblock \emph{A\&Ap} 47, 137--141
\bibAnnoteFile{Gabriel1976}

\bibitem[{{Garc{\'{\i}}a} et~al.(2007){Garc{\'{\i}}a}, {Turck-Chi{\`e}ze},
  {Jim{\'e}nez-Reyes}, {Ballot}, {Pall{\'e}}, {Eff-Darwich}
  et~al.}]{Garcia2007}
{Garc{\'{\i}}a}, R.~A., {Turck-Chi{\`e}ze}, S., {Jim{\'e}nez-Reyes}, S.~J.,
  {Ballot}, J., {Pall{\'e}}, P.~L., {Eff-Darwich}, A., et~al. (2007).
\newblock {Tracking Solar Gravity Modes: The Dynamics of the Solar Core}.
\newblock \emph{Science} 316, 1591.
\newblock \doi{10.1126/science.1140598}
\bibAnnoteFile{Garcia2007}

\bibitem[{{Gorshkov} and {Baturin}(2008)}]{Gorshkov2008}
{Gorshkov}, A.~B. and {Baturin}, V.~A. (2008).
\newblock {Diffusion settling of heavy elements in the solar interior}.
\newblock \emph{Astronomy Reports} 52, 760--771.
\newblock \doi{10.1134/S1063772908090072}
\bibAnnoteFile{Gorshkov2008}

\bibitem[{{Gorshkov} and {Baturin}(2010)}]{Gorshkov10}
{Gorshkov}, A.~B. and {Baturin}, V.~A. (2010).
\newblock {Elemental diffusion and segregation processes in partially ionized
  solar plasma}.
\newblock \emph{APSS} 328, 171--174.
\newblock \doi{10.1007/s10509-010-0277-2}
\bibAnnoteFile{Gorshkov10}

\bibitem[{{Gough}(2015)}]{Gough2015}
{Gough}, D.~O. (2015).
\newblock {Some Glimpses from Helioseismology at the Dynamics of the Deep Solar
  Interior}.
\newblock \emph{Space Sci. Rev.} 196, 15--47.
\newblock \doi{10.1007/s11214-015-0159-6}
\bibAnnoteFile{Gough2015}

\bibitem[{{Gough}(2019)}]{Gough2019}
{Gough}, D.~O. (2019).
\newblock {Anticipating the Sun's heavy-element abundance}.
\newblock \emph{MNRAS} 485, L114--L115.
\newblock \doi{10.1093/mnrasl/slz044}
\bibAnnoteFile{Gough2019}

\bibitem[{{Gough} and {Kosovichev}(1993{\natexlab{a}})}]{Gough1993b}
{Gough}, D.~O. and {Kosovichev}, A.~G. (1993{\natexlab{a}}).
\newblock {Initial asteroseismic inversions}.
\newblock In \emph{IAU Colloq. 137: Inside the Stars}, eds. W.~W. {Weiss} and
  A.~{Baglin}. vol.~40 of \emph{Astronomical Society of the Pacific Conference
  Series}, 541
\bibAnnoteFile{Gough1993b}

\bibitem[{{Gough} and {Kosovichev}(1993{\natexlab{b}})}]{Gough1993}
{Gough}, D.~O. and {Kosovichev}, A.~G. (1993{\natexlab{b}}).
\newblock {The Influence of Low-Degree P-Mode Frequencies on the Determination
  of the Structure of the Solar Interior}.
\newblock \emph{MNRAS} 264, 522.
\newblock \doi{10.1093/mnras/264.2.522}
\bibAnnoteFile{Gough1993}

\bibitem[{{Gough} and {McIntyre}(1998)}]{Gough1998}
{Gough}, D.~O. and {McIntyre}, M.~E. (1998).
\newblock {Inevitability of a magnetic field in the Sun's radiative interior}.
\newblock \emph{Nature} 394, 755--757.
\newblock \doi{10.1038/29472}
\bibAnnoteFile{Gough1998}

\bibitem[{{Gough} and {Thompson}(1991)}]{Gough1991}
{Gough}, D.~O. and {Thompson}, M.~J. (1991).
\newblock \emph{{The inversion problem}}.
\newblock 519--561
\bibAnnoteFile{Gough1991}

\bibitem[{{Graedel} et~al.(1991){Graedel}, {Sackmann}, and
  {Boothroyd}}]{Graedel1991}
{Graedel}, T.~E., {Sackmann}, I.-J., and {Boothroyd}, A.~I. (1991).
\newblock {Early solar mass loss - A potential solution to the weak sun
  paradox}.
\newblock \emph{Geophys. Res. Lett.} 18, 1881--1884.
\newblock \doi{10.1029/91GL02314}
\bibAnnoteFile{Graedel1991}

\bibitem[{{Grevesse} and {Noels}(1993)}]{GrevNoels}
{Grevesse}, N. and {Noels}, A. (1993).
\newblock {Cosmic abundances of the elements.}
\newblock In \emph{Origin and Evolution of the Elements}, eds. N.~{Prantzos},
  E.~{Vangioni-Flam}, and M.~{Casse}. 15--25
\bibAnnoteFile{GrevNoels}

\bibitem[{{Grevesse} and {Sauval}(1998)}]{GreSauv}
{Grevesse}, N. and {Sauval}, A.~J. (1998).
\newblock {Standard Solar Composition}.
\newblock \emph{Space Science Reviews} 85, 161--174
\bibAnnoteFile{GreSauv}

\bibitem[{{Grevesse} et~al.(2015){Grevesse}, {Scott}, {Asplund}, and
  {Sauval}}]{Grevesse2015}
{Grevesse}, N., {Scott}, P., {Asplund}, M., and {Sauval}, A.~J. (2015).
\newblock {The elemental composition of the Sun. III. The heavy elements Cu to
  Th}.
\newblock \emph{A\&A} 573, A27.
\newblock \doi{10.1051/0004-6361/201424111}
\bibAnnoteFile{Grevesse2015}

\bibitem[{{Gruberbauer} et~al.(2012){Gruberbauer}, {Guenther}, and
  {Kallinger}}]{Gruberbauer}
{Gruberbauer}, M., {Guenther}, D.~B., and {Kallinger}, T. (2012).
\newblock {Toward a New Kind of Asteroseismic Grid Fitting}.
\newblock \emph{ApJ} 749, 109.
\newblock \doi{10.1088/0004-637X/749/2/109}
\bibAnnoteFile{Gruberbauer}

\bibitem[{Gryaznov et~al.(2013)Gryaznov, Iosilevskiy, Fortov, Starostin,
  Roerich, Baturin et~al.}]{Gryaznov13}
Gryaznov, V., Iosilevskiy, I., Fortov, V., Starostin, A., Roerich, V., Baturin,
  V.~A., et~al. (2013).
\newblock Saha-s thermodynamic model of solar plasma.
\newblock \emph{Contributions to Plasma Physics} 53, 392--396.
\newblock \doi{10.1002/ctpp.201200109}
\bibAnnoteFile{Gryaznov13}

\bibitem[{{Gryaznov} et~al.(2004){Gryaznov}, {Ayukov}, {Baturin},
  {Iosilevskiy}, {Starostin}, and {Fortov}}]{Gryaznov04}
{Gryaznov}, V.~K., {Ayukov}, S.~V., {Baturin}, V.~A., {Iosilevskiy}, I.~L.,
  {Starostin}, A.~N., and {Fortov}, V.~E. (2004).
\newblock {SAHA-S model: Equation of State and Thermodynamic Functions of Solar
  Plasma}.
\newblock In \emph{Equation-of-State and Phase-Transition in Models of Ordinary
  Astrophysical Matter}, eds. V.~{Celebonovic}, D.~{Gough}, and
  W.~{D{\"a}ppen}. vol. 731 of \emph{American Institute of Physics Conference
  Series}, 147--161.
\newblock \doi{10.1063/1.1828400}
\bibAnnoteFile{Gryaznov04}

\bibitem[{Gryaznov et~al.(2006)Gryaznov, Ayukov, Baturin, Iosilevskiy,
  Starostin, and Fortov}]{Gryaznov06}
Gryaznov, V.~K., Ayukov, S.~V., Baturin, V.~A., Iosilevskiy, I.~L., Starostin,
  A.~N., and Fortov, V.~E. (2006).
\newblock Solar plasma: calculation of thermodynamic functions and equation of
  state.
\newblock \emph{Journal of Physics A: Mathematical and General} 39, 4459
\bibAnnoteFile{Gryaznov06}

\bibitem[{{Guzik}(2008)}]{Guzik2008}
{Guzik}, J.~A. (2008).
\newblock {Problems for the standard solar model arising from the new solar
  mixture.}
\newblock \emph{MmSAI} 79, 481
\bibAnnoteFile{Guzik2008}

\bibitem[{{Guzik} et~al.(2016){Guzik}, {Fontes}, {Walczak}, {Wood}, {Mussack},
  and {Farag}}]{Guzik16}
{Guzik}, J.~A., {Fontes}, C.~J., {Walczak}, P., {Wood}, S.~R., {Mussack}, K.,
  and {Farag}, E. (2016).
\newblock {Sound speed and oscillation frequencies for solar models evolved
  with Los Alamos ATOMIC opacities}.
\newblock \emph{IAU Focus Meeting} 29, 532--535.
\newblock \doi{10.1017/S1743921316006062}
\bibAnnoteFile{Guzik16}

\bibitem[{{Guzik} and {Mussack}(2010)}]{Guzik2010}
{Guzik}, J.~A. and {Mussack}, K. (2010).
\newblock {Exploring Mass Loss, Low-Z Accretion, and Convective Overshoot in
  Solar Models to Mitigate the Solar Abundance Problem}.
\newblock \emph{ApJ} 713, 1108--1119.
\newblock \doi{10.1088/0004-637X/713/2/1108}
\bibAnnoteFile{Guzik2010}

\bibitem[{{Guzik} et~al.(2005){Guzik}, {Watson}, and {Cox}}]{Guzik05}
{Guzik}, J.~A., {Watson}, L.~S., and {Cox}, A.~N. (2005).
\newblock {Can Enhanced Diffusion Improve Helioseismic Agreement for Solar
  Models with Revised Abundances?}
\newblock \emph{ApJ} 627, 1049--1056.
\newblock \doi{10.1086/430438}
\bibAnnoteFile{Guzik05}

\bibitem[{{Guzik} et~al.(2006){Guzik}, {Watson}, and {Cox}}]{Guzik06}
{Guzik}, J.~A., {Watson}, L.~S., and {Cox}, A.~N. (2006).
\newblock {Implications of revised solar abundances for helioseismology.}
\newblock \emph{MmSAI} 77, 389
\bibAnnoteFile{Guzik06}

\bibitem[{{Guzik} et~al.(1987){Guzik}, {Willson}, and {Brunish}}]{Guzik1987}
{Guzik}, J.~A., {Willson}, L.~A., and {Brunish}, W.~M. (1987).
\newblock {A comparison between mass-losing and standard solar models}.
\newblock \emph{ApJ} 319, 957--965.
\newblock \doi{10.1086/165512}
\bibAnnoteFile{Guzik1987}

\bibitem[{{Harvey} et~al.(1988){Harvey}, {Abdel-Gawad}, {Ball}, {Boxum},
  {Bull}, {Cole} et~al.}]{Harvey1988}
{Harvey}, J., {Abdel-Gawad}, K., {Ball}, W., {Boxum}, B., {Bull}, F., {Cole},
  J., et~al. (1988).
\newblock {The GONG instrument}.
\newblock In \emph{Seismology of the Sun and Sun-Like Stars}, ed. E.~J.
  {Rolfe}. vol. 286 of \emph{ESA Special Publication}
\bibAnnoteFile{Harvey1988}

\bibitem[{{Haxton} et~al.(2013){Haxton}, {Hamish Robertson}, and
  {Serenelli}}]{Haxton2013}
{Haxton}, W.~C., {Hamish Robertson}, R.~G., and {Serenelli}, A.~M. (2013).
\newblock {Solar Neutrinos: Status and Prospects}.
\newblock \emph{Annu. Rev. Astron. Astrophys} 51, 21--61.
\newblock \doi{10.1146/annurev-astro-081811-125539}
\bibAnnoteFile{Haxton2013}

\bibitem[{{Holweger} and {Mueller}(1974)}]{Holweger1974}
{Holweger}, H. and {Mueller}, E.~A. (1974).
\newblock {The photospheric barium spectrum - Solar abundance and collision
  broadening of BA II lines by hydrogen}.
\newblock \emph{Sol.Phys.} 39, 19--30.
\newblock \doi{10.1007/BF00154968}
\bibAnnoteFile{Holweger1974}

\bibitem[{{Hotta}(2017)}]{Hotta}
{Hotta}, H. (2017).
\newblock {Solar Overshoot Region and Small-scale Dynamo with Realistic Energy
  Flux}.
\newblock \emph{ApJ} 843, 52.
\newblock \doi{10.3847/1538-4357/aa784b}
\bibAnnoteFile{Hotta}

\bibitem[{{Houdek} and {Gough}(2011)}]{Houdek2011}
{Houdek}, G. and {Gough}, D.~O. (2011).
\newblock {On the seismic age and heavy-element abundance of the Sun}.
\newblock \emph{MNRAS} 418, 1217--1230.
\newblock \doi{10.1111/j.1365-2966.2011.19572.x}
\bibAnnoteFile{Houdek2011}

\bibitem[{{Houdek} and {Rogl}(1996)}]{Houdek1996}
{Houdek}, G. and {Rogl}, J. (1996).
\newblock {On the accuracy of opacity interpolation schemes}.
\newblock \emph{Bulletin of the Astronomical Society of India} 24, 317--320
\bibAnnoteFile{Houdek1996}

\bibitem[{{Hummer} and {Mihalas}(1988)}]{MHDI}
{Hummer}, D.~G. and {Mihalas}, D. (1988).
\newblock {The equation of state for stellar envelopes. I - an occupation
  probability formalism for the truncation of internal partition functions}.
\newblock \emph{ApJ} 331, 794--814.
\newblock \doi{10.1086/166600}
\bibAnnoteFile{MHDI}

\bibitem[{{Iglesias}(2015)}]{Iglesias2015}
{Iglesias}, C.~A. (2015).
\newblock {Iron-group opacities for B stars}.
\newblock \emph{MNRAS} 450, 2--9.
\newblock \doi{10.1093/mnras/stv591}
\bibAnnoteFile{Iglesias2015}

\bibitem[{{Iglesias} and {Hansen}(2017)}]{Iglesias2017}
{Iglesias}, C.~A. and {Hansen}, S.~B. (2017).
\newblock {Fe XVII Opacity at Solar Interior Conditions}.
\newblock \emph{ApJ} 835, 284.
\newblock \doi{10.3847/1538-4357/835/2/284}
\bibAnnoteFile{Iglesias2017}

\bibitem[{{Iglesias} and {Rogers}(1996)}]{OPAL}
{Iglesias}, C.~A. and {Rogers}, F.~J. (1996).
\newblock {Updated Opal Opacities}.
\newblock \emph{ApJ} 464, 943
\bibAnnoteFile{OPAL}

\bibitem[{{Irwin}(2012)}]{Irwin}
[Dataset] {Irwin}, A.~W. (2012).
\newblock {FreeEOS: Equation of State for stellar interiors calculations}.
\newblock Astrophysics Source Code Library
\bibAnnoteFile{Irwin}

\bibitem[{{Isaak} et~al.(1989){Isaak}, {McLeod}, {Palle}, {van der Raay}, and
  {Roca Cortes}}]{Isaak1989}
{Isaak}, G.~R., {McLeod}, C.~P., {Palle}, P.~L., {van der Raay}, H.~B., and
  {Roca Cortes}, T. (1989).
\newblock {Solar oscillations as seen in the NaI and KI absorption lines}.
\newblock \emph{A\&A} 208, 297--302
\bibAnnoteFile{Isaak1989}

\bibitem[{{J{\o}rgensen} et~al.(2018){J{\o}rgensen}, {Mosumgaard}, {Weiss},
  {Silva Aguirre}, and {Christensen-Dalsgaard}}]{Jorgensen2018}
{J{\o}rgensen}, A.~C.~S., {Mosumgaard}, J.~R., {Weiss}, A., {Silva Aguirre},
  V., and {Christensen-Dalsgaard}, J. (2018).
\newblock {Coupling 1D stellar evolution with 3D-hydrodynamical simulations on
  the fly - I. A new standard solar model}.
\newblock \emph{MNRAS} 481, L35--L39.
\newblock \doi{10.1093/mnrasl/sly152}
\bibAnnoteFile{Jorgensen2018}

\bibitem[{{King} et~al.(1997){King}, {Deliyannis}, {Hiltgen}, {Stephens},
  {Cunha}, and {Boesgaard}}]{King1997}
{King}, J.~R., {Deliyannis}, C.~P., {Hiltgen}, D.~D., {Stephens}, A., {Cunha},
  K., and {Boesgaard}, A.~M. (1997).
\newblock {Lithium Abundances in the Solar Twins 16 CYG A and B and the Solar
  Analog alpha CEN A, Calibration of the 6707 Angstrom Li Region Linelist, and
  Implications}.
\newblock \emph{AJ} 113, 1871.
\newblock \doi{10.1086/118399}
\bibAnnoteFile{King1997}

\bibitem[{{Kosovichev}(1988)}]{KosovichevRota}
{Kosovichev}, A.~G. (1988).
\newblock {The Internal Rotation of the Sun from Helioseismological Data}.
\newblock \emph{Soviet Astronomy Letters} 14, 145
\bibAnnoteFile{KosovichevRota}

\bibitem[{{Kosovichev}(1999)}]{Kosovichev}
{Kosovichev}, A.~G. (1999).
\newblock {Inversion methods in helioseismology and solar tomography.}
\newblock \emph{Journal of Computational and Applied Mathematics} 109, 1--39
\bibAnnoteFile{Kosovichev}

\bibitem[{{Kosovichev} and {Fedorova}(1991)}]{KosovBCZ}
{Kosovichev}, A.~G. and {Fedorova}, A.~V. (1991).
\newblock {Construction of a Seismic Model of the Sun}.
\newblock \emph{Sov.Astron.} 35, 507
\bibAnnoteFile{KosovBCZ}

\bibitem[{{Kosovichev} and {Severnyi}(1985)}]{Kosovichev1985}
{Kosovichev}, A.~G. and {Severnyi}, A.~B. (1985).
\newblock {The influence of chemical composition on the stability of solar
  gravity-mode oscillations}.
\newblock \emph{Izvestiya Ordena Trudovogo Krasnogo Znameni Krymskoj
  Astrofizicheskoj Observatorii} 72, 188--198
\bibAnnoteFile{Kosovichev1985}

\bibitem[{{Krief} et~al.(2016){Krief}, {Feigel}, and {Gazit}}]{Krief2016}
{Krief}, M., {Feigel}, A., and {Gazit}, D. (2016).
\newblock {Line Broadening and the Solar Opacity Problem}.
\newblock \emph{ApJ} 824, 98.
\newblock \doi{10.3847/0004-637X/824/2/98}
\bibAnnoteFile{Krief2016}

\bibitem[{{Landi} and {Testa}(2015)}]{Landi}
{Landi}, E. and {Testa}, P. (2015).
\newblock {Neon and Oxygen Abundances and Abundance Ratio in the Solar Corona}.
\newblock \emph{ApJ} 800, 110.
\newblock \doi{10.1088/0004-637X/800/2/110}
\bibAnnoteFile{Landi}

\bibitem[{{Le Pennec} et~al.(2015){Le Pennec}, {Turck-Chi{\`e}ze}, {Salmon},
  {Blancard}, {Coss{\'e}}, {Faussurier} et~al.}]{LePennec}
{Le Pennec}, M., {Turck-Chi{\`e}ze}, S., {Salmon}, S., {Blancard}, C.,
  {Coss{\'e}}, P., {Faussurier}, G., et~al. (2015).
\newblock {First New Solar Models with OPAS Opacity Tables}.
\newblock \emph{ApJL} 813, L42.
\newblock \doi{10.1088/2041-8205/813/2/L42}
\bibAnnoteFile{LePennec}

\bibitem[{{Lebreton} et~al.(2007){Lebreton}, {Montalb{\'a}n},
  {Christensen-Dalsgaard}, {Th{\'e}ado}, {Hui-Bon-Hoa}, {Monteiro}
  et~al.}]{ESTA1}
{Lebreton}, Y., {Montalb{\'a}n}, J., {Christensen-Dalsgaard}, J., {Th{\'e}ado},
  S., {Hui-Bon-Hoa}, A., {Monteiro}, M.~J.~P.~F.~G., et~al. (2007).
\newblock {Microscopic Diffusion in Stellar Evolution Codes: First Comparisons
  Results of ESTA-Task 3}.
\newblock In \emph{EAS Publications Series}, eds. C.~W. {Straka},
  Y.~{Lebreton}, and M.~J.~P.~F.~G. {Monteiro}. vol.~26 of \emph{EAS
  Publications Series}, 155--165.
\newblock \doi{10.1051/eas:2007134}
\bibAnnoteFile{ESTA1}

\bibitem[{{Ledoux}(1974)}]{Ledoux1974}
{Ledoux}, P. (1974).
\newblock {Non-radial oscillations}.
\newblock In \emph{Stellar Instability and Evolution}, eds. P.~{Ledoux},
  A.~{Noels}, and A.~W. {Rodgers}. vol.~59 of \emph{IAU Symposium}, 135--173
\bibAnnoteFile{Ledoux1974}

\bibitem[{{Li} and {Yang}(2007)}]{YangI}
{Li}, Y. and {Yang}, J.~Y. (2007).
\newblock {Testing turbulent convection theory in solar models - I. Structure
  of the solar convection zone}.
\newblock \emph{MNRAS} 375, 388--402.
\newblock \doi{10.1111/j.1365-2966.2006.11319.x}
\bibAnnoteFile{YangI}

\bibitem[{{Lund} et~al.(2017){Lund}, {Silva Aguirre}, {Davies}, {Chaplin},
  {Christensen-Dalsgaard}, {Houdek} et~al.}]{Lund}
{Lund}, M.~N., {Silva Aguirre}, V., {Davies}, G.~R., {Chaplin}, W.~J.,
  {Christensen-Dalsgaard}, J., {Houdek}, G., et~al. (2017).
\newblock {Standing on the Shoulders of Dwarfs: the Kepler Asteroseismic LEGACY
  Sample. I. Oscillation Mode Parameters}.
\newblock \emph{ApJ} 835, 172.
\newblock \doi{10.3847/1538-4357/835/2/172}
\bibAnnoteFile{Lund}

\bibitem[{{Marchenkov} et~al.(2000){Marchenkov}, {Roxburgh}, and
  {Vorontsov}}]{Marchenkov}
{Marchenkov}, K., {Roxburgh}, I., and {Vorontsov}, S. (2000).
\newblock {Non-linear inversion for the hydrostatic structure of the solar
  interior}.
\newblock \emph{MNRAS} 312, 39--50.
\newblock \doi{10.1046/j.1365-8711.2000.03059.x}
\bibAnnoteFile{Marchenkov}

\bibitem[{{Metcalfe} et~al.(2015){Metcalfe}, {Creevey}, and
  {Davies}}]{Metcalfe2015ApJ}
{Metcalfe}, T.~S., {Creevey}, O.~L., and {Davies}, G.~R. (2015).
\newblock {Asteroseismic Modeling of 16 Cyg A {\&} B using the Complete Kepler
  Data Set}.
\newblock \emph{ApJL} 811, L37.
\newblock \doi{10.1088/2041-8205/811/2/L37}
\bibAnnoteFile{Metcalfe2015ApJ}

\bibitem[{{Michaud} et~al.(2015){Michaud}, {Alecian}, and
  {Richer}}]{Michaud2015}
{Michaud}, G., {Alecian}, G., and {Richer}, J. (2015).
\newblock \emph{{Atomic Diffusion in Stars}}.
\newblock \doi{10.1007/978-3-319-19854-5}
\bibAnnoteFile{Michaud2015}

\bibitem[{{Michaud} et~al.(1976){Michaud}, {Charland}, {Vauclair}, and
  {Vauclair}}]{Michaud1976}
{Michaud}, G., {Charland}, Y., {Vauclair}, S., and {Vauclair}, G. (1976).
\newblock {Diffusion in main-sequence stars - Radiation forces, time scales,
  anomalies}.
\newblock \emph{ApJ} 210, 447--465.
\newblock \doi{10.1086/154848}
\bibAnnoteFile{Michaud1976}

\bibitem[{{Michaud} and {Proffitt}(1993)}]{Michaud1993}
{Michaud}, G. and {Proffitt}, C.~R. (1993).
\newblock {Particle transport processes}.
\newblock In \emph{IAU Colloq. 137: Inside the Stars}, eds. W.~W. {Weiss} and
  A.~{Baglin}. vol.~40 of \emph{Astronomical Society of the Pacific Conference
  Series}, 246--259
\bibAnnoteFile{Michaud1993}

\bibitem[{{Michaud} and {Richer.}(2008)}]{Michaud2008}
{Michaud}, G. and {Richer.}, J. (2008).
\newblock {Radiative accelerations in stellar evolution}.
\newblock \emph{MMSAI} 79, 592
\bibAnnoteFile{Michaud2008}

\bibitem[{{Mihalas} et~al.(1988){Mihalas}, {D\"appen}, and {Hummer}}]{MHDII}
{Mihalas}, D., {D\"appen}, W., and {Hummer}, D.~G. (1988).
\newblock {The equation of state for stellar envelopes. II - Algorithm and
  selected results}.
\newblock \emph{ApJ} 331, 815--825.
\newblock \doi{10.1086/166601}
\bibAnnoteFile{MHDII}

\bibitem[{{Mihalas} et~al.(1990){Mihalas}, {Hummer}, {Mihalas}, and
  {D\"appen}}]{MHDIV}
{Mihalas}, D., {Hummer}, D.~G., {Mihalas}, B.~W., and {D\"appen}, W. (1990).
\newblock {The equation of state for stellar envelopes. IV - Thermodynamic
  quantities and selected ionization fractions for six elemental mixes}.
\newblock \emph{ApJ} 350, 300--308.
\newblock \doi{10.1086/168383}
\bibAnnoteFile{MHDIV}

\bibitem[{{Minton} and {Malhotra}(2007)}]{Minton2007}
{Minton}, D.~A. and {Malhotra}, R. (2007).
\newblock {Assessing the Massive Young Sun Hypothesis to Solve the Warm Young
  Earth Puzzle}.
\newblock \emph{ApJ} 660, 1700--1706.
\newblock \doi{10.1086/514331}
\bibAnnoteFile{Minton2007}

\bibitem[{{Mondet} et~al.(2015){Mondet}, {Blancard}, {Coss{\'e}}, and
  {Faussurier}}]{Mondet}
{Mondet}, G., {Blancard}, C., {Coss{\'e}}, P., and {Faussurier}, G. (2015).
\newblock {Opacity Calculations for Solar Mixtures}.
\newblock \emph{ApJs} 220, 2
\bibAnnoteFile{Mondet}

\bibitem[{{Montalban} et~al.(2006){Montalban}, {Miglio}, {Theado}, {Noels}, and
  {Grevesse}}]{Montalban06}
{Montalban}, J., {Miglio}, A., {Theado}, S., {Noels}, A., and {Grevesse}, N.
  (2006).
\newblock {The new solar abundances - Part II: the crisis and possible
  solutions}.
\newblock \emph{Communications in Asteroseismology} 147, 80--84.
\newblock \doi{10.1553/cia147s80}
\bibAnnoteFile{Montalban06}

\bibitem[{{Montalb{\'a}n} et~al.(2007){Montalb{\'a}n}, {Th{\'e}ado}, and
  {Lebreton}}]{ESTA2}
{Montalb{\'a}n}, J., {Th{\'e}ado}, S., and {Lebreton}, Y. (2007).
\newblock {Comparisons for ESTA-Task3: CLES and CESAM}.
\newblock In \emph{EAS Publications Series}, eds. C.~W. {Straka},
  Y.~{Lebreton}, and M.~J.~P.~F.~G. {Monteiro}. vol.~26 of \emph{EAS
  Publications Series}, 167--176.
\newblock \doi{10.1051/eas:2007135}
\bibAnnoteFile{ESTA2}

\bibitem[{{Monteiro} et~al.(1994){Monteiro}, {Christensen-Dalsgaard}, and
  {Thompson}}]{Monteiro94}
{Monteiro}, M.~J.~P.~F.~G., {Christensen-Dalsgaard}, J., and {Thompson}, M.~J.
  (1994).
\newblock {Seismic study of overshoot at the base of the solar convective
  envelope}.
\newblock \emph{A\&A} 283, 247--262
\bibAnnoteFile{Monteiro94}

\bibitem[{{Mussack} and {D{\"a}ppen}(2011)}]{Mussack2011}
{Mussack}, K. and {D{\"a}ppen}, W. (2011).
\newblock {Dynamic Screening Correction for Solar p-p Reaction Rates}.
\newblock \emph{ApJ} 729, 96.
\newblock \doi{10.1088/0004-637X/729/2/96}
\bibAnnoteFile{Mussack2011}

\bibitem[{{Nahar} and {Pradhan}(2016)}]{Nahar}
{Nahar}, S.~N. and {Pradhan}, A.~K. (2016).
\newblock {Large Enhancement in High-Energy Photoionization of Fe XVII and
  Missing Continuum Plasma Opacity}.
\newblock \emph{Physical Review Letters} 116, 235003
\bibAnnoteFile{Nahar}

\bibitem[{{Noels} et~al.(1976){Noels}, {Boury}, {Gabriel}, and
  {Scuflaire}}]{Noels1976}
{Noels}, A., {Boury}, A., {Gabriel}, M., and {Scuflaire}, R. (1976).
\newblock {Vibrational Stability towards Non-radial Oscillations during Central
  Hydrogen Burning}.
\newblock \emph{A\&Ap} 49, 103
\bibAnnoteFile{Noels1976}

\bibitem[{{Noerdlinger}(1977)}]{Noerdlinger1977}
{Noerdlinger}, P.~D. (1977).
\newblock {Diffusion of helium in the sun}.
\newblock \emph{A\&A} 57, 407--415
\bibAnnoteFile{Noerdlinger1977}

\bibitem[{{Ouazzani} et~al.(2018){Ouazzani}, {Marques}, {Goupil}, {Christophe},
  {Antoci}, and {Salmon}}]{Ouazzani18}
{Ouazzani}, R.-M., {Marques}, J.~P., {Goupil}, M., {Christophe}, S., {Antoci},
  V., and {Salmon}, S.~J.~A.~J. (2018).
\newblock {$\{$$\backslash$gamma$\}$ Doradus stars as test of angular momentum
  transport models}.
\newblock \emph{arXiv e-prints}
\bibAnnoteFile{Ouazzani18}

\bibitem[{{Pain} and {Gilleron}(2019)}]{Pain2019}
{Pain}, J.-C. and {Gilleron}, F. (2019).
\newblock {Opacity calculations for stellar astrophysics}.
\newblock \emph{arXiv e-prints}
\bibAnnoteFile{Pain2019}

\bibitem[{{Pain} et~al.(2018){Pain}, {Gilleron}, and {Comet}}]{Pain2018}
{Pain}, J.-C., {Gilleron}, F., and {Comet}, M. (2018).
\newblock {Detailed Opacity Calculations for Stellar Models}.
\newblock In \emph{Workshop on Astrophysical Opacities}. vol. 515 of
  \emph{Astronomical Society of the Pacific Conference Series}, 35
\bibAnnoteFile{Pain2018}

\bibitem[{{Paquette} et~al.(1986){Paquette}, {Pelletier}, {Fontaine}, and
  {Michaud}}]{Paquette}
{Paquette}, C., {Pelletier}, C., {Fontaine}, G., and {Michaud}, G. (1986).
\newblock {Diffusion coefficients for stellar plasmas}.
\newblock \emph{ApJS} 61, 177--195.
\newblock \doi{10.1086/191111}
\bibAnnoteFile{Paquette}

\bibitem[{{Piau} and {Turck-Chi{\`e}ze}(2001)}]{Piau2001}
{Piau}, L. and {Turck-Chi{\`e}ze}, S. (2001).
\newblock {Lithium Burning in the Early Evolution of the Sun and Sun-like
  Stars}.
\newblock In \emph{From Darkness to Light: Origin and Evolution of Young
  Stellar Clusters}, eds. T.~{Montmerle} and P.~{Andr{\'e}}. vol. 243 of
  \emph{Astronomical Society of the Pacific Conference Series}, 639
\bibAnnoteFile{Piau2001}

\bibitem[{{Pijpers} and {Thompson}(1994)}]{Pijpers}
{Pijpers}, F.~P. and {Thompson}, M.~J. (1994).
\newblock {The SOLA method for helioseismic inversion}.
\newblock \emph{A\&Ap} 281, 231--240
\bibAnnoteFile{Pijpers}

\bibitem[{{Pinsonneault} and {Delahaye}(2009)}]{Delahaye2009}
{Pinsonneault}, M.~H. and {Delahaye}, F. (2009).
\newblock {The Solar Heavy Element Abundances. II. Constraints from Stellar
  Atmospheres}.
\newblock \emph{ApJ} 704, 1174--1188.
\newblock \doi{10.1088/0004-637X/704/2/1174}
\bibAnnoteFile{Delahaye2009}

\bibitem[{{Pradhan} and {Nahar}(2018)}]{Pradhan}
{Pradhan}, A.~K. and {Nahar}, S.~N. (2018).
\newblock {Recalculation of Astrophysical Opacities: Overview, Methodology, and
  Atomic Calculations}.
\newblock In \emph{Workshop on Astrophysical Opacities}. vol. 515 of
  \emph{Astronomical Society of the Pacific Conference Series}, 79
\bibAnnoteFile{Pradhan}

\bibitem[{{Proffitt} and {Michaud}(1991)}]{Proffitt}
{Proffitt}, C.~R. and {Michaud}, G. (1991).
\newblock {Gravitational settling in solar models}.
\newblock \emph{ApJ} 380, 238--250.
\newblock \doi{10.1086/170580}
\bibAnnoteFile{Proffitt}

\bibitem[{{Rabello-Soares} et~al.(1999){Rabello-Soares}, {Basu}, and
  {Christensen-Dalsgaard}}]{RabelloParam}
{Rabello-Soares}, M.~C., {Basu}, S., and {Christensen-Dalsgaard}, J. (1999).
\newblock {On the choice of parameters in solar-structure inversion}.
\newblock \emph{MNRAS} 309, 35--47
\bibAnnoteFile{RabelloParam}

\bibitem[{{Rauer} et~al.(2014){Rauer}, {Catala}, {Aerts}, {Appourchaux},
  {Benz}, {Brandeker} et~al.}]{PLATO}
{Rauer}, H., {Catala}, C., {Aerts}, C., {Appourchaux}, T., {Benz}, W.,
  {Brandeker}, A., et~al. (2014).
\newblock {The PLATO 2.0 mission}.
\newblock \emph{Experimental Astronomy} 38, 249--330.
\newblock \doi{10.1007/s10686-014-9383-4}
\bibAnnoteFile{PLATO}

\bibitem[{{Reese} et~al.(2012{\natexlab{a}}){Reese}, {Marques}, {Goupil},
  {Thompson}, and {Deheuvels}}]{ReeseDens}
{Reese}, D.~R., {Marques}, J.~P., {Goupil}, M.~J., {Thompson}, M.~J., and
  {Deheuvels}, S. (2012{\natexlab{a}}).
\newblock {Estimating stellar mean density through seismic inversions}.
\newblock \emph{A\&Ap} 539, A63
\bibAnnoteFile{ReeseDens}

\bibitem[{{Reese} et~al.(2012{\natexlab{b}}){Reese}, {Marques}, {Goupil},
  {Thompson}, and {Deheuvels}}]{Reese2012}
{Reese}, D.~R., {Marques}, J.~P., {Goupil}, M.~J., {Thompson}, M.~J., and
  {Deheuvels}, S. (2012{\natexlab{b}}).
\newblock {Estimating stellar mean density through seismic inversions}.
\newblock \emph{A\&A} 539, A63.
\newblock \doi{10.1051/0004-6361/201118156}
\bibAnnoteFile{Reese2012}

\bibitem[{{Rempel}(2004)}]{Rempel04}
{Rempel}, M. (2004).
\newblock {Overshoot at the Base of the Solar Convection Zone: A Semianalytical
  Approach}.
\newblock \emph{ApJ} 607, 1046--1064.
\newblock \doi{10.1086/383605}
\bibAnnoteFile{Rempel04}

\bibitem[{{Rendle} et~al.(2019){Rendle}, {Buldgen}, {Miglio}, {Reese}, {Noels},
  {Davies} et~al.}]{Rendle}
{Rendle}, B.~M., {Buldgen}, G., {Miglio}, A., {Reese}, D., {Noels}, A.,
  {Davies}, G.~R., et~al. (2019).
\newblock {AIMS - a new tool for stellar parameter determinations using
  asteroseismic constraints}.
\newblock \emph{MNRAS} 484, 771--786.
\newblock \doi{10.1093/mnras/stz031}
\bibAnnoteFile{Rendle}

\bibitem[{{Richard} et~al.(1998){Richard}, {Dziembowski}, {Sienkiewicz}, and
  {Goode}}]{RichardY}
{Richard}, O., {Dziembowski}, W.~A., {Sienkiewicz}, R., and {Goode}, P.~R.
  (1998).
\newblock {On the accuracy of helioseismic determination of solar helium
  abundance}.
\newblock \emph{A\&A} 338, 756--760
\bibAnnoteFile{RichardY}

\bibitem[{{Richard} et~al.(2001){Richard}, {Michaud}, and
  {Richer}}]{Richard2001}
{Richard}, O., {Michaud}, G., and {Richer}, J. (2001).
\newblock {Iron Convection Zones in B, A, and F Stars}.
\newblock \emph{ApJ} 558, 377--391.
\newblock \doi{10.1086/322264}
\bibAnnoteFile{Richard2001}

\bibitem[{{Richard} et~al.(2002{\natexlab{a}}){Richard}, {Michaud}, and
  {Richer}}]{Richard2002b}
{Richard}, O., {Michaud}, G., and {Richer}, J. (2002{\natexlab{a}}).
\newblock {Models of Metal-poor Stars with Gravitational Settling and Radiative
  Accelerations. III. Metallicity Dependence}.
\newblock \emph{ApJ} 580, 1100--1117.
\newblock \doi{10.1086/343733}
\bibAnnoteFile{Richard2002b}

\bibitem[{{Richard} et~al.(2002{\natexlab{b}}){Richard}, {Michaud}, {Richer},
  {Turcotte}, {Turck-Chi{\`e}ze}, and {VandenBerg}}]{Richard2002a}
{Richard}, O., {Michaud}, G., {Richer}, J., {Turcotte}, S., {Turck-Chi{\`e}ze},
  S., and {VandenBerg}, D.~A. (2002{\natexlab{b}}).
\newblock {Models of Metal-poor Stars with Gravitational Settling and Radiative
  Accelerations. I. Evolution and Abundance Anomalies}.
\newblock \emph{ApJ} 568, 979--997.
\newblock \doi{10.1086/338952}
\bibAnnoteFile{Richard2002a}

\bibitem[{{Richard} et~al.(1996){Richard}, {Vauclair}, {Charbonnel}, and
  {Dziembowski}}]{Richard96Sun}
{Richard}, O., {Vauclair}, S., {Charbonnel}, C., and {Dziembowski}, W.~A.
  (1996).
\newblock {New solar models including helioseismological constraints and
  light-element depletion.}
\newblock \emph{A\&Ap} 312, 1000--1011
\bibAnnoteFile{Richard96Sun}

\bibitem[{{Richer} et~al.(2000){Richer}, {Michaud}, and
  {Turcotte}}]{Richer2000}
{Richer}, J., {Michaud}, G., and {Turcotte}, S. (2000).
\newblock {The Evolution of AMFM Stars, Abundance Anomalies, and Turbulent
  Transport}.
\newblock \emph{ApJ} 529, 338--356.
\newblock \doi{10.1086/308274}
\bibAnnoteFile{Richer2000}

\bibitem[{{Ricker} et~al.(2015){Ricker}, {Winn}, {Vanderspek}, {Latham},
  {Bakos}, {Bean} et~al.}]{Ricker}
{Ricker}, G.~R., {Winn}, J.~N., {Vanderspek}, R., {Latham}, D.~W., {Bakos},
  G.~{\'A}., {Bean}, J.~L., et~al. (2015).
\newblock {Transiting Exoplanet Survey Satellite (TESS)}.
\newblock \emph{Journal of Astronomical Telescopes, Instruments, and Systems}
  1, 014003.
\newblock \doi{10.1117/1.JATIS.1.1.014003}
\bibAnnoteFile{Ricker}

\bibitem[{{Rogers} and {Nayfonov}(2002)}]{Rogerseos}
{Rogers}, F.~J. and {Nayfonov}, A. (2002).
\newblock {Updated and Expanded OPAL Equation-of-State Tables: Implications for
  Helioseismology}.
\newblock \emph{ApJ} 576, 1064--1074
\bibAnnoteFile{Rogerseos}

\bibitem[{{Rogers} et~al.(1996){Rogers}, {Swenson}, and
  {Iglesias}}]{Rogers1996}
{Rogers}, F.~J., {Swenson}, F.~J., and {Iglesias}, C.~A. (1996).
\newblock {OPAL Equation-of-State Tables for Astrophysical Applications}.
\newblock \emph{ApJ} 456, 902.
\newblock \doi{10.1086/176705}
\bibAnnoteFile{Rogers1996}

\bibitem[{{Roxburgh} et~al.(1998){Roxburgh}, {Audard}, {Basu},
  {Christensen-Dalsgaard}, and {Vorontsov}}]{Roxburgh1996}
{Roxburgh}, I., {Audard}, N., {Basu}, S., {Christensen-Dalsgaard}, J., and
  {Vorontsov}, S. (1998).
\newblock {Inversion for the internal structure of an evolved small mass star
  using modes with l=0-3 in proceedings}.
\newblock In \emph{Sounding Solar and Stellar Interiors}, eds. F.~{Schmider}
  and J.~{Provost}. IAU Coll. 181 Nice, p.245
\bibAnnoteFile{Roxburgh1996}

\bibitem[{{Roxburgh} and {Vorontsov}(2003{\natexlab{a}})}]{Roxburgh2003Vor}
{Roxburgh}, I. and {Vorontsov}, S. (2003{\natexlab{a}}).
\newblock {Diagnostics of the Internal Structure of Stars using the
  Differential Response Technique}.
\newblock \emph{Ap\&SS} 284, 187--191
\bibAnnoteFile{Roxburgh2003Vor}

\bibitem[{{Roxburgh}(1976)}]{Roxburgh1976}
{Roxburgh}, I.~W. (1976).
\newblock {The Internal Structure of the Sun and Solar Type Stars}.
\newblock In \emph{Basic Mechanisms of Solar Activity}, eds. V.~{Bumba} and
  J.~{Kleczek}. vol.~71 of \emph{IAU Symposium}, 453
\bibAnnoteFile{Roxburgh1976}

\bibitem[{{Roxburgh}(1984)}]{Roxburgh1984}
{Roxburgh}, I.~W. (1984).
\newblock {On Turbulent Mixing}.
\newblock In \emph{Observational Tests of the Stellar Evolution Theory}, eds.
  A.~{Maeder} and A.~{Renzini}. vol. 105 of \emph{IAU Symposium}, 519
\bibAnnoteFile{Roxburgh1984}

\bibitem[{{Roxburgh}(2016)}]{Roxburgh2016}
{Roxburgh}, I.~W. (2016).
\newblock {Asteroseismic model fitting by comparing {$\epsilon$}$_{nl}$
  values}.
\newblock \emph{A\&A} 585, A63.
\newblock \doi{10.1051/0004-6361/201526593}
\bibAnnoteFile{Roxburgh2016}

\bibitem[{{Roxburgh} and {Vorontsov}(2002)}]{Roxburgh2002}
{Roxburgh}, I.~W. and {Vorontsov}, S.~V. (2002).
\newblock {Inversion for a 0.8 M$_{solar}$ star using differential-response
  technique}.
\newblock In \emph{Stellar Structure and Habitable Planet Finding}, eds.
  B.~{Battrick}, F.~{Favata}, I.~W. {Roxburgh}, and D.~{Galadi}. vol. 485 of
  \emph{ESA Special Publication}, 337--339
\bibAnnoteFile{Roxburgh2002}

\bibitem[{{Roxburgh} and {Vorontsov}(2003{\natexlab{b}})}]{RoxburghRatios}
{Roxburgh}, I.~W. and {Vorontsov}, S.~V. (2003{\natexlab{b}}).
\newblock {The ratio of small to large separations of acoustic oscillations as
  a diagnostic of the interior of solar-like stars}.
\newblock \emph{A\&A} 411, 215--220
\bibAnnoteFile{RoxburghRatios}

\bibitem[{{Sackmann} and {Boothroyd}(2003)}]{Sackmann2003}
{Sackmann}, I.-J. and {Boothroyd}, A.~I. (2003).
\newblock {Our Sun. V. A Bright Young Sun Consistent with Helioseismology and
  Warm Temperatures on Ancient Earth and Mars}.
\newblock \emph{ApJ} 583, 1024--1039.
\newblock \doi{10.1086/345408}
\bibAnnoteFile{Sackmann2003}

\bibitem[{{Saio}(1980)}]{Saio80}
{Saio}, H. (1980).
\newblock {Stability of nonradial g/+/-mode pulsations in 1 solar mass models}.
\newblock \emph{ApJ} 240, 685--692.
\newblock \doi{10.1086/158275}
\bibAnnoteFile{Saio80}

\bibitem[{{Schatten}(1973)}]{Schatten1973}
{Schatten}, K.~H. (1973).
\newblock {Magnetic Convection}.
\newblock \emph{Sol.Phys.} 33, 305--318.
\newblock \doi{10.1007/BF00152420}
\bibAnnoteFile{Schatten1973}

\bibitem[{{Schlattl}(2002)}]{Schlattl2002}
{Schlattl}, H. (2002).
\newblock {Microscopic diffusion of partly ionized metals in the Sun and
  metal-poor stars}.
\newblock \emph{A\&A} 395, 85--95.
\newblock \doi{10.1051/0004-6361:20021212}
\bibAnnoteFile{Schlattl2002}

\bibitem[{{Schlattl} and {Salaris}(2003)}]{Schlattl2003}
{Schlattl}, H. and {Salaris}, M. (2003).
\newblock {Quantum corrections to microscopic diffusion constants}.
\newblock \emph{A\&A} 402, 29--35.
\newblock \doi{10.1051/0004-6361:20030230}
\bibAnnoteFile{Schlattl2003}

\bibitem[{{Schou} et~al.(1998){Schou}, {Antia}, {Basu}, {Bogart}, {Bush},
  {Chitre} et~al.}]{SchouRota}
{Schou}, J., {Antia}, H.~M., {Basu}, S., {Bogart}, R.~S., {Bush}, R.~I.,
  {Chitre}, S.~M., et~al. (1998).
\newblock {Helioseismic Studies of Differential Rotation in the Solar Envelope
  by the Solar Oscillations Investigation Using the Michelson Doppler Imager}.
\newblock \emph{ApJ} 505, 390--417.
\newblock \doi{10.1086/306146}
\bibAnnoteFile{SchouRota}

\bibitem[{{Schunker} et~al.(2018){Schunker}, {Schou}, {Gaulme}, and
  {Gizon}}]{Schunker2018}
{Schunker}, H., {Schou}, J., {Gaulme}, P., and {Gizon}, L. (2018).
\newblock {Fragile Detection of Solar g-Modes by Fossat et al.}
\newblock \emph{Sol.Phys.} 293, 95.
\newblock \doi{10.1007/s11207-018-1313-6}
\bibAnnoteFile{Schunker2018}

\bibitem[{{Schwarzschild}(1906)}]{Schwarzschild1906}
{Schwarzschild}, K. (1906).
\newblock {On the equilibrium of the Sun's atmosphere}.
\newblock \emph{Nachrichten von der K{\"o}niglichen Gesellschaft der
  Wissenschaften zu G{\"o}ttingen.~Math.-phys.~Klasse, 195, p.~41-53} 195,
  41--53
\bibAnnoteFile{Schwarzschild1906}

\bibitem[{{Scott} et~al.(2015{\natexlab{a}}){Scott}, {Asplund}, {Grevesse},
  {Bergemann}, and {Sauval}}]{Scott2015II}
{Scott}, P., {Asplund}, M., {Grevesse}, N., {Bergemann}, M., and {Sauval},
  A.~J. (2015{\natexlab{a}}).
\newblock {The elemental composition of the Sun. II. The iron group elements Sc
  to Ni}.
\newblock \emph{A\&A} 573, A26.
\newblock \doi{10.1051/0004-6361/201424110}
\bibAnnoteFile{Scott2015II}

\bibitem[{{Scott} et~al.(2015{\natexlab{b}}){Scott}, {Grevesse}, {Asplund},
  {Sauval}, {Lind}, {Takeda} et~al.}]{Scott2015I}
{Scott}, P., {Grevesse}, N., {Asplund}, M., {Sauval}, A.~J., {Lind}, K.,
  {Takeda}, Y., et~al. (2015{\natexlab{b}}).
\newblock {The elemental composition of the Sun. I. The intermediate mass
  elements Na to Ca}.
\newblock \emph{A\&A} 573, A25.
\newblock \doi{10.1051/0004-6361/201424109}
\bibAnnoteFile{Scott2015I}

\bibitem[{{Scuflaire} et~al.(1975){Scuflaire}, {Gabriel}, {Noels}, and
  {Boury}}]{Scuflaire1975}
{Scuflaire}, R., {Gabriel}, M., {Noels}, A., and {Boury}, A. (1975).
\newblock {Oscillatory periods in the sun and theoretical models with or
  without mixing}.
\newblock \emph{A\&Ap} 45, 15--18
\bibAnnoteFile{Scuflaire1975}

\bibitem[{{Scuflaire} et~al.(2008{\natexlab{a}}){Scuflaire}, {Montalb{\'a}n},
  {Th{\'e}ado}, {Bourge}, {Miglio}, {Godart} et~al.}]{ScuflaireOsc}
{Scuflaire}, R., {Montalb{\'a}n}, J., {Th{\'e}ado}, S., {Bourge}, P.-O.,
  {Miglio}, A., {Godart}, M., et~al. (2008{\natexlab{a}}).
\newblock {The Li{\`e}ge Oscillation code}.
\newblock \emph{ApSS} 316, 149--154
\bibAnnoteFile{ScuflaireOsc}

\bibitem[{{Scuflaire} et~al.(2008{\natexlab{b}}){Scuflaire}, {Th{\'e}ado},
  {Montalb{\'a}n}, {Miglio}, {Bourge}, {Godart} et~al.}]{ScuflaireCles}
{Scuflaire}, R., {Th{\'e}ado}, S., {Montalb{\'a}n}, J., {Miglio}, A., {Bourge},
  P.-O., {Godart}, M., et~al. (2008{\natexlab{b}}).
\newblock {CL{\'E}S, Code Li{\'e}geois d'{\'E}volution Stellaire}.
\newblock \emph{ApSS} 316, 83--91
\bibAnnoteFile{ScuflaireCles}

\bibitem[{{Serenelli} et~al.(2009){Serenelli}, {Basu}, {Ferguson}, and
  {Asplund}}]{SerenelliComp}
{Serenelli}, A.~M., {Basu}, S., {Ferguson}, J.~W., and {Asplund}, M. (2009).
\newblock {New Solar Composition: The Problem with Solar Models Revisited}.
\newblock \emph{ApJl} 705, L123--L127
\bibAnnoteFile{SerenelliComp}

\bibitem[{{Serenelli} et~al.(2011){Serenelli}, {Haxton}, and
  {Pe{\~n}a-Garay}}]{Serenelli2011}
{Serenelli}, A.~M., {Haxton}, W.~C., and {Pe{\~n}a-Garay}, C. (2011).
\newblock {Solar Models with Accretion. I. Application to the Solar Abundance
  Problem}.
\newblock \emph{ApJ} 743, 24.
\newblock \doi{10.1088/0004-637X/743/1/24}
\bibAnnoteFile{Serenelli2011}

\bibitem[{{Shibahashi} et~al.(1975){Shibahashi}, {Osaki}, and
  {Unno}}]{Shibahashi1975}
{Shibahashi}, H., {Osaki}, Y., and {Unno}, W. (1975).
\newblock {Nonradial g-mode oscillations and the stability of the sun}.
\newblock \emph{PASJ} 27, 401--410
\bibAnnoteFile{Shibahashi1975}

\bibitem[{{Silva Aguirre} et~al.(2017){Silva Aguirre}, {Lund}, {Antia}, {Ball},
  {Basu}, {Christensen-Dalsgaard} et~al.}]{Silva2017}
{Silva Aguirre}, V., {Lund}, M.~N., {Antia}, H.~M., {Ball}, W.~H., {Basu}, S.,
  {Christensen-Dalsgaard}, J., et~al. (2017).
\newblock {Standing on the Shoulders of Dwarfs: the Kepler Asteroseismic LEGACY
  Sample. II.Radii, Masses, and Ages}.
\newblock \emph{ApJ} 835, 173.
\newblock \doi{10.3847/1538-4357/835/2/173}
\bibAnnoteFile{Silva2017}

\bibitem[{{Sonoi} and {Shibahashi}(2012{\natexlab{a}})}]{Sonoi2012b}
{Sonoi}, T. and {Shibahashi}, H. (2012{\natexlab{a}}).
\newblock {Dipole low-order g-mode instability of metal-poor low-mass
  main-sequence stars due to the {$\epsilon$} mechanism}.
\newblock \emph{MNRAS} 422, 2642--2647.
\newblock \doi{10.1111/j.1365-2966.2012.20827.x}
\bibAnnoteFile{Sonoi2012b}

\bibitem[{{Sonoi} and {Shibahashi}(2012{\natexlab{b}})}]{Sonoi2012a}
{Sonoi}, T. and {Shibahashi}, H. (2012{\natexlab{b}}).
\newblock {Fully Nonadiabatic Analysis of Vibrational Instability of Population
  III Stars due to the varepsilon-Mechanism}.
\newblock \emph{PASJ} 64, 2.
\newblock \doi{10.1093/pasj/64.1.2}
\bibAnnoteFile{Sonoi2012a}

\bibitem[{{Spalding} et~al.(2018){Spalding}, {Fischer}, and
  {Laughlin}}]{Spalding2018}
{Spalding}, C., {Fischer}, W.~W., and {Laughlin}, G. (2018).
\newblock {An Orbital Window into the Ancient Sun's Mass}.
\newblock \emph{ApJL} 869, L19.
\newblock \doi{10.3847/2041-8213/aaf219}
\bibAnnoteFile{Spalding2018}

\bibitem[{{Spiegel} and {Zahn}(1992)}]{SpiegelZahn1992}
{Spiegel}, E.~A. and {Zahn}, J.-P. (1992).
\newblock {The solar tachocline}.
\newblock \emph{A\&Ap} 265, 106--114
\bibAnnoteFile{SpiegelZahn1992}

\bibitem[{{Takata} and {Montgomery}(2002)}]{Takata2002}
{Takata}, M. and {Montgomery}, M.~H. (2002).
\newblock {Seismic Inversions for White Dwarf Stars}.
\newblock In \emph{IAU Colloq. 185: Radial and Nonradial Pulsationsn as Probes
  of Stellar Physics}, eds. C.~{Aerts}, T.~R. {Bedding}, and
  J.~{Christensen-Dalsgaard}. vol. 259 of \emph{Astronomical Society of the
  Pacific Conference Series}, 606
\bibAnnoteFile{Takata2002}

\bibitem[{{Takata} and {Shibahashi}(2001)}]{Takata2001}
{Takata}, M. and {Shibahashi}, H. (2001).
\newblock {Solar Metal Abundance Inferred from Helioseismology}.
\newblock In \emph{Recent Insights into the Physics of the Sun and Heliosphere:
  Highlights from SOHO and Other Space Missions}, eds. P.~{Brekke}, B.~{Fleck},
  and J.~B. {Gurman}. vol. 203 of \emph{IAU Symposium}, 43
\bibAnnoteFile{Takata2001}

\bibitem[{{Th{\'e}ado} et~al.(2012){Th{\'e}ado}, {Alecian}, {LeBlanc}, and
  {Vauclair}}]{Theado12}
{Th{\'e}ado}, S., {Alecian}, G., {LeBlanc}, F., and {Vauclair}, S. (2012).
\newblock {The new Toulouse-Geneva stellar evolution code including radiative
  accelerations of heavy elements}.
\newblock \emph{A\&A} 546, A100.
\newblock \doi{10.1051/0004-6361/201219610}
\bibAnnoteFile{Theado12}

\bibitem[{{Theado} and {Vauclair}(2010)}]{Theado2010}
{Theado}, S. and {Vauclair}, S. (2010).
\newblock {Radiative accelerations, accumulation of iron and thermohaline
  convection inside stars}.
\newblock \emph{Theado} 328, 209--212.
\newblock \doi{10.1007/s10509-009-0221-5}
\bibAnnoteFile{Theado2010}

\bibitem[{{Th{\'e}ado} et~al.(2005){Th{\'e}ado}, {Vauclair}, {Castro},
  {Charpinet}, and {Dolez}}]{Theado2005}
{Th{\'e}ado}, S., {Vauclair}, S., {Castro}, M., {Charpinet}, S., and {Dolez},
  N. (2005).
\newblock {Asteroseismic tests of element diffusion in solar type stars}.
\newblock \emph{A\&A} 437, 553--560.
\newblock \doi{10.1051/0004-6361:20042328}
\bibAnnoteFile{Theado2005}

\bibitem[{{Th{\'e}venin} et~al.(2017){Th{\'e}venin}, {Oreshina}, {Baturin},
  {Gorshkov}, {Morel}, and {Provost}}]{Thevenin17}
{Th{\'e}venin}, F., {Oreshina}, A.~V., {Baturin}, V.~A., {Gorshkov}, A.~B.,
  {Morel}, P., and {Provost}, J. (2017).
\newblock {Evolution of lithium abundance in the Sun and solar twins}.
\newblock \emph{A\&A} 598, A64.
\newblock \doi{10.1051/0004-6361/201629385}
\bibAnnoteFile{Thevenin17}

\bibitem[{{Thoul} et~al.(1994){Thoul}, {Bahcall}, and {Loeb}}]{Thoul}
{Thoul}, A.~A., {Bahcall}, J.~N., and {Loeb}, A. (1994).
\newblock {Element diffusion in the solar interior}.
\newblock \emph{ApJ} 421, 828--842.
\newblock \doi{10.1086/173695}
\bibAnnoteFile{Thoul}

\bibitem[{{Tucci Maia} et~al.(2014){Tucci Maia}, {Mel{\'e}ndez}, and
  {Ram{\'{\i}}rez}}]{Tucci2014}
{Tucci Maia}, M., {Mel{\'e}ndez}, J., and {Ram{\'{\i}}rez}, I. (2014).
\newblock {High Precision Abundances in the 16 Cyg Binary System: A Signature
  of the Rocky Core in the Giant Planet}.
\newblock \emph{ApJL} 790, L25.
\newblock \doi{10.1088/2041-8205/790/2/L25}
\bibAnnoteFile{Tucci2014}

\bibitem[{{Turbet} et~al.(2017){Turbet}, {Forget}, {Head}, and
  {Wordsworth}}]{Turbet2017}
{Turbet}, M., {Forget}, F., {Head}, J.~W., and {Wordsworth}, R. (2017).
\newblock {3D modelling of the climatic impact of outflow channel formation
  events on early Mars}.
\newblock \emph{Icarus} 288, 10--36.
\newblock \doi{10.1016/j.icarus.2017.01.024}
\bibAnnoteFile{Turbet2017}

\bibitem[{{Turck-Chi{\`e}ze}(2005)}]{TC2005}
{Turck-Chi{\`e}ze}, S. (2005).
\newblock {How does helioseismology constrain solar neutrino properties?}
\newblock \emph{Nuclear Physics B Proceedings Supplements} 143, 35--42.
\newblock \doi{10.1016/j.nuclphysbps.2005.01.085}
\bibAnnoteFile{TC2005}

\bibitem[{{Turck-Chieze} et~al.(1988){Turck-Chieze}, {Cahen}, {Casse}, and
  {Doom}}]{TC1988}
{Turck-Chieze}, S., {Cahen}, S., {Casse}, M., and {Doom}, C. (1988).
\newblock {Revisiting the standard solar model}.
\newblock \emph{ApJ} 335, 415--424.
\newblock \doi{10.1086/166936}
\bibAnnoteFile{TC1988}

\bibitem[{{Turck-Chi{\`e}ze} and {Couvidat}(2011)}]{TC2011}
{Turck-Chi{\`e}ze}, S. and {Couvidat}, S. (2011).
\newblock {Solar neutrinos, helioseismology and the solar internal dynamics}.
\newblock \emph{Reports on Progress in Physics} 74, 086901.
\newblock \doi{10.1088/0034-4885/74/8/086901}
\bibAnnoteFile{TC2011}

\bibitem[{{Turck-Chi{\`e}ze} et~al.(2004){Turck-Chi{\`e}ze}, {Couvidat},
  {Piau}, {Ferguson}, {Lambert}, {Ballot} et~al.}]{TC2004}
{Turck-Chi{\`e}ze}, S., {Couvidat}, S., {Piau}, L., {Ferguson}, J., {Lambert},
  P., {Ballot}, J., et~al. (2004).
\newblock {Surprising Sun: A New Step Towards a Complete Picture?}
\newblock \emph{Physical Review Letters} 93, 211102.
\newblock \doi{10.1103/PhysRevLett.93.211102}
\bibAnnoteFile{TC2004}

\bibitem[{{Turck-Chi{\`e}ze} et~al.(2011){Turck-Chi{\`e}ze}, {Piau}, and
  {Couvidat}}]{Turck2011}
{Turck-Chi{\`e}ze}, S., {Piau}, L., and {Couvidat}, S. (2011).
\newblock {The Solar Energetic Balance Revisited by Young Solar Analogs,
  Helioseismology, and Neutrinos}.
\newblock \emph{ApJL} 731, L29.
\newblock \doi{10.1088/2041-8205/731/2/L29}
\bibAnnoteFile{Turck2011}

\bibitem[{{Turcotte} et~al.(1998){Turcotte}, {Richer}, {Michaud}, {Iglesias},
  and {Rogers}}]{Turcotte}
{Turcotte}, S., {Richer}, J., {Michaud}, G., {Iglesias}, C.~A., and {Rogers},
  F.~J. (1998).
\newblock {Consistent Solar Evolution Model Including Diffusion and Radiative
  Acceleration Effects}.
\newblock \emph{ApJ} 504, 539--558.
\newblock \doi{10.1086/306055}
\bibAnnoteFile{Turcotte}

\bibitem[{{Ulrich}(1974)}]{Ulrich1974}
{Ulrich}, R.~K. (1974).
\newblock {Solar Models with Low Neutrino Fluxes}.
\newblock \emph{ApJ} 188, 369--378.
\newblock \doi{10.1086/152725}
\bibAnnoteFile{Ulrich1974}

\bibitem[{{Ulrich}(1975)}]{Ulrich1975}
{Ulrich}, R.~K. (1975).
\newblock {Solar neutrinos and variations in the solar luminosity}.
\newblock \emph{Science} 190, 619--624
\bibAnnoteFile{Ulrich1975}

\bibitem[{{Ulrich} and {Rood}(1973)}]{Ulrich1973}
{Ulrich}, R.~K. and {Rood}, R.~T. (1973).
\newblock {Mixing in Stellar Models}.
\newblock \emph{Nature Physical Science} 241, 111--112.
\newblock \doi{10.1038/physci241111a0}
\bibAnnoteFile{Ulrich1973}

\bibitem[{{Unno}(1975)}]{Unno1975}
{Unno}, W. (1975).
\newblock {On the stability of the solar core}.
\newblock \emph{PASJ} 27, 81--99
\bibAnnoteFile{Unno1975}

\bibitem[{{VandenBerg} et~al.(2002){VandenBerg}, {Richard}, {Michaud}, and
  {Richer}}]{VandenBerg2002}
{VandenBerg}, D.~A., {Richard}, O., {Michaud}, G., and {Richer}, J. (2002).
\newblock {Models of Metal-poor Stars with Gravitational Settling and Radiative
  Accelerations. II. The Age of the Oldest Stars}.
\newblock \emph{ApJ} 571, 487--500.
\newblock \doi{10.1086/339895}
\bibAnnoteFile{VandenBerg2002}

\bibitem[{{Verma} et~al.(2014){Verma}, {Faria}, {Antia}, {Basu}, {Mazumdar},
  {Monteiro} et~al.}]{Verma2014}
{Verma}, K., {Faria}, J.~P., {Antia}, H.~M., {Basu}, S., {Mazumdar}, A.,
  {Monteiro}, M.~J.~P.~F.~G., et~al. (2014).
\newblock {Asteroseismic Estimate of Helium Abundance of a Solar Analog Binary
  System}.
\newblock \emph{ApJ} 790, 138.
\newblock \doi{10.1088/0004-637X/790/2/138}
\bibAnnoteFile{Verma2014}

\bibitem[{{Vernazza} et~al.(1976){Vernazza}, {Avrett}, and
  {Loeser}}]{Vernazza1976}
{Vernazza}, J.~E., {Avrett}, E.~H., and {Loeser}, R. (1976).
\newblock {Structure of the solar chromosphere. II - The underlying photosphere
  and temperature-minimum region}.
\newblock \emph{ApJS} 30, 1--60.
\newblock \doi{10.1086/190356}
\bibAnnoteFile{Vernazza1976}

\bibitem[{{Viallet} et~al.(2015){Viallet}, {Meakin}, {Prat}, and
  {Arnett}}]{Viallet}
{Viallet}, M., {Meakin}, C., {Prat}, V., and {Arnett}, D. (2015).
\newblock {Toward a consistent use of overshooting parametrizations in 1D
  stellar evolution codes}.
\newblock \emph{A\&A} 580, A61.
\newblock \doi{10.1051/0004-6361/201526294}
\bibAnnoteFile{Viallet}

\bibitem[{{Vinyoles} et~al.(2017){Vinyoles}, {Serenelli}, {Villante}, {Basu},
  {Bergstr{\"o}m}, {Gonzalez-Garcia} et~al.}]{Vinyoles}
{Vinyoles}, N., {Serenelli}, A.~M., {Villante}, F.~L., {Basu}, S.,
  {Bergstr{\"o}m}, J., {Gonzalez-Garcia}, M.~C., et~al. (2017).
\newblock {A New Generation of Standard Solar Models}.
\newblock \emph{ApJ} 835, 202.
\newblock \doi{10.3847/1538-4357/835/2/202}
\bibAnnoteFile{Vinyoles}

\bibitem[{{Vorontsov} et~al.(2013){Vorontsov}, {Baturin}, {Ayukov}, and
  {Gryaznov}}]{Vorontsov13}
{Vorontsov}, S.~V., {Baturin}, V.~A., {Ayukov}, S.~V., and {Gryaznov}, V.~K.
  (2013).
\newblock {Helioseismic calibration of the equation of state and chemical
  composition in the solar convective envelope}.
\newblock \emph{MNRAS} 430, 1636--1652.
\newblock \doi{10.1093/mnras/sts701}
\bibAnnoteFile{Vorontsov13}

\bibitem[{{Vorontsov} et~al.(1991){Vorontsov}, {Baturin}, and
  {Pamiatnykh}}]{Vorontsov91}
{Vorontsov}, S.~V., {Baturin}, V.~A., and {Pamiatnykh}, A.~A. (1991).
\newblock {Seismological measurement of solar helium abundance}.
\newblock \emph{Nature} 349, 49--51.
\newblock \doi{10.1038/349049a0}
\bibAnnoteFile{Vorontsov91}

\bibitem[{{Weiss} and {Heners}(2013)}]{Weiss2013}
{Weiss}, A. and {Heners}, N. (2013).
\newblock {Low-mass stars: Open problems all along their evolution}.
\newblock In \emph{European Physical Journal Web of Conferences}. vol.~43 of
  \emph{European Physical Journal Web of Conferences}, 01002.
\newblock \doi{10.1051/epjconf/20134301002}
\bibAnnoteFile{Weiss2013}

\bibitem[{{White} et~al.(2013){White}, {Huber}, {Maestro}, {Bedding},
  {Ireland}, {Baron} et~al.}]{White2013MNRAS}
{White}, T.~R., {Huber}, D., {Maestro}, V., {Bedding}, T.~R., {Ireland}, M.~J.,
  {Baron}, F., et~al. (2013).
\newblock {Interferometric radii of bright Kepler stars with the CHARA Array:
  {$\theta$} Cygni and 16 Cygni A and B}.
\newblock \emph{MNRAS} 433, 1262--1270.
\newblock \doi{10.1093/mnras/stt802}
\bibAnnoteFile{White2013MNRAS}

\bibitem[{{Winnick} et~al.(2002){Winnick}, {Demarque}, {Basu}, and
  {Guenther}}]{Winnick2002}
{Winnick}, R.~A., {Demarque}, P., {Basu}, S., and {Guenther}, D.~B. (2002).
\newblock {Seismic Test of Solar Models, Solar Neutrinos, and Implications for
  Metal-rich Accretion}.
\newblock \emph{ApJ} 576, 1075--1084.
\newblock \doi{10.1086/341795}
\bibAnnoteFile{Winnick2002}

\bibitem[{{Wood} et~al.(2005){Wood}, {M{\"u}ller}, {Zank}, {Linsky}, and
  {Redfield}}]{Wood2005}
{Wood}, B.~E., {M{\"u}ller}, H.-R., {Zank}, G.~P., {Linsky}, J.~L., and
  {Redfield}, S. (2005).
\newblock {New Mass-Loss Measurements from Astrospheric Ly{$\alpha$}
  Absorption}.
\newblock \emph{ApJL} 628, L143--L146.
\newblock \doi{10.1086/432716}
\bibAnnoteFile{Wood2005}

\bibitem[{{Wood} et~al.(2018){Wood}, {Mussack}, and {Guzik}}]{Wood2018}
{Wood}, S.~R., {Mussack}, K., and {Guzik}, J.~A. (2018).
\newblock {Solar Models with Dynamic Screening and Early Mass Loss Tested by
  Helioseismic, Astrophysical, and Planetary Constraints}.
\newblock \emph{Sol.Phys.} 293, 111.
\newblock \doi{10.1007/s11207-018-1334-1}
\bibAnnoteFile{Wood2018}

\bibitem[{{Wordsworth}(2016)}]{Wordsworth2016}
{Wordsworth}, R.~D. (2016).
\newblock {The Climate of Early Mars}.
\newblock \emph{Annual Review of Earth and Planetary Sciences} 44, 381--408.
\newblock \doi{10.1146/annurev-earth-060115-012355}
\bibAnnoteFile{Wordsworth2016}

\bibitem[{{Xiong} and {Deng}(2001)}]{Xiong01}
{Xiong}, D.~R. and {Deng}, L. (2001).
\newblock {The structure of the solar convective overshooting zone}.
\newblock \emph{MNRAS} 327, 1137--1144.
\newblock \doi{10.1046/j.1365-8711.2001.04820.x}
\bibAnnoteFile{Xiong01}

\bibitem[{{Yang} and {Li}(2007)}]{YangII}
{Yang}, J.~Y. and {Li}, Y. (2007).
\newblock {Testing turbulent convection theory in solar models - II. Solar
  p-mode oscillations}.
\newblock \emph{MNRAS} 375, 403--414.
\newblock \doi{10.1111/j.1365-2966.2006.11320.x}
\bibAnnoteFile{YangII}

\bibitem[{{Young}(2018)}]{Young}
{Young}, P.~R. (2018).
\newblock {Element Abundance Ratios in the Quiet Sun Transition Region}.
\newblock \emph{ApJ} 855, 15.
\newblock \doi{10.3847/1538-4357/aaab48}
\bibAnnoteFile{Young}

\bibitem[{{Zaatri} et~al.(2007){Zaatri}, {Provost}, {Berthomieu}, {Morel}, and
  {Corbard}}]{Zaatri2007}
{Zaatri}, A., {Provost}, J., {Berthomieu}, G., {Morel}, P., and {Corbard}, T.
  (2007).
\newblock {Sensitivity of low degree oscillations to the change in solar
  abundances}.
\newblock \emph{A\&A} 469, 1145--1149.
\newblock \doi{10.1051/0004-6361:20077212}
\bibAnnoteFile{Zaatri2007}

\bibitem[{{Zhang}(2017)}]{Zhang2017}
{Zhang}, Q.~S. (2017).
\newblock {Numerical Integral of Resistance Coefficients in Diffusion}.
\newblock \emph{ApJ} 834, 132.
\newblock \doi{10.3847/1538-4357/834/2/132}
\bibAnnoteFile{Zhang2017}

\bibitem[{{Zhao} et~al.(2018){Zhao}, {Eissner}, {Nahar}, and {Pradhan}}]{Zhao}
{Zhao}, L., {Eissner}, W., {Nahar}, S.~N., and {Pradhan}, A.~K. (2018).
\newblock {Converged R-Matrix Calculations of the Photoionization of Fexvii in
  Astrophysical Plasmas: from Convergence to Completeness}.
\newblock In \emph{Workshop on Astrophysical Opacities}. vol. 515 of
  \emph{Astronomical Society of the Pacific Conference Series}, 89
\bibAnnoteFile{Zhao}

\end{thebibliography}


\section*{Appendix}\label{sec:AppendixOne}

\begin{figure}[h!]
\begin{center}
\includegraphics[width=8cm]{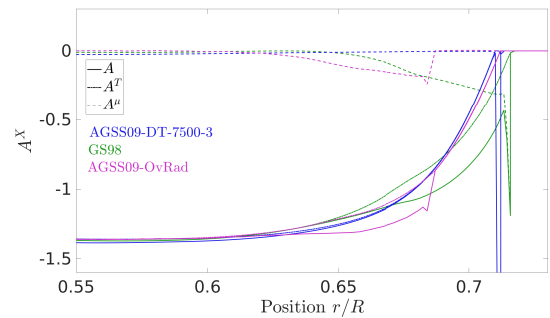}
\end{center}
\caption{Decomposition of the Ledoux discriminant in its thermal and chemical composition gradients contributions for a model built with the GS98 abundances, and other models including modified opacity tables and macroscopic chemical mixing.}\label{fig:DecompOpacPoly}
\end{figure}

To have a closer look at the impact of extra mixing at the base of the convective zone, we illustrate in Fig. \ref{fig:DecompOpacPoly} the various contributions to the Ledoux discriminant, A. We use the following definition 
\begin{align}
A=-\frac{r \delta}{H_{P}}\left( \nabla_{Ad} -\nabla + \frac{\phi}{\delta}\nabla_{\mu}\right),
\end{align}
with $H_{P}$ the pressure scale height, $\mu$ the mean molecular weight, $P$ the pressure and $\delta=-\left( \frac{\partial \ln \rho}{\partial \ln T} \right)_{P,\mu}$, $\phi=\left( \frac{\partial \ln \rho}{\partial \ln \mu}\right)_{P,T}$, $\nabla=\frac{d \ln T}{d \ln P}$, $\nabla_{Ad}=\left(\frac{\partial \ln T}{\partial \ln P}\right)_{S}$, with $S$ the entropy, $\nabla_{\mu}=\frac{d \ln \mu}{d \ln P}$. We define the thermal and chemical contributions to the Ledoux discriminant, $A^{T}$ and $A^{\mu}$ as 
\begin{align}
A^{\mu}=-\frac{r \phi}{H_{P}}\nabla_{\mu}, \\
A^{T}=-\frac{r \delta}{H_{P}}\left(\nabla_{Ad}-\nabla \right).
\end{align}

From Fig. \ref{fig:DecompOpacPoly}, we see that the thermal contribution largely dominates the behaviour of the Ledoux discriminant, with the exception of the last percent of the radiative regions. There is a clear difference between a fully mixed overshoot and that of turbulent diffusion. This opens up the possibility to combine seismic diagnostics to distinguish the form of the macroscopic mixing at the base of the solar convective zone and explains the observed differences in Fig. \ref{fig:OpacPolyInv}. Using non-linear inversions, as in \citet{corbard99}, may also help to further constrain the profile of the Ledoux discriminant between $0.67$ and $0.71$ solar radii. There also appears to be a clear difference in the temperature contribution, $A^{T}$,  between the various models, around $0.65$ solar radii. This difference is due to the too steep temperature gradient at this position,  seen for the GS98 model and the model built using radiative overshoot. A parametric modelling of the transition in both temperature and chemical gradients should shed new light the existing degeneracies and provide seismic constraints to works aiming at including hydrodynamical prescriptions in stellar evolutionary codes. To that end, a combination of the analysis of the phase-shift of frequencies, as carried out by \citet{JCDOV} or the use of a non-linear RLS method following the approach of \citet{corbard99} would allow to probe the sharp transition in Ledoux discriminant at the base of the convective zone. 

\end{document}